\documentclass[onefignum,onetabnum,letterpaper]{siamart251104}

\usepackage{lipsum}
\usepackage{amsfonts}
\usepackage{graphicx}
\usepackage{epstopdf}
\usepackage{algorithmic}
\ifpdf%
  \DeclareGraphicsExtensions{.eps,.pdf,.png,.jpg}
\else
  \DeclareGraphicsExtensions{.eps}
\fi

\makeatletter
\newcommand*{\emails}[2][@gmail.com]{%
    \def\@tempa{\@gobble}%
    \@for\qrr@email:=#2\do{%
        \edef\@tempb{\noexpand\href{mailto:\qrr@email #1}{\qrr@email}}%
        \edef\@tempa{\unexpanded\expandafter{\@tempa}{, }\unexpanded\expandafter{\@tempb}}}%
    \{\@tempa\}#1%
}
\makeatother

\usepackage{soul}

\newsiamremark{remark}{Remark}
\newsiamremark{hypothesis}{Hypothesis}
\crefname{hypothesis}{Hypothesis}{Hypotheses}
\newsiamthm{claim}{Claim}
\newsiamremark{fact}{Fact}
\crefname{fact}{Fact}{Facts}
\newsiamremark{assumption}{Assumption}
\crefname{assumption}{Assumption}{Assumptions}
\newsiamremark{constraint}{Constraint}
\crefname{constraint}{Constraint}{Constraints}

\headers{
    Sparse Mode Selecting Neural Net
}{T. Koike, P. Mohan, M. T. H. Frahan, E. Qian, J. Bessac}

\title{
    Sparse POD Mode Selection and Manifold Dimensionality Reduction with Neural Networks\thanks{Submitted to the editors \today. % Submission date
        \funding{
            The authors were supported by the Department of Energy Office of Science Advanced Scientific Computing Research, DOE Award DE-SC0024721. 
        }
    }
}

\author{
    Tomoki Koike\thanks{School of Aerospace Engineering, 
    Georgia Institute of Technology, Atlanta, GA 
    (\emails[@gatech.edu]{tkoike,eqian})} \and
    Prakash Mohan\thanks{Computational Science Center, National Laboratory of the Rockies (NLR), Golden, CO
    (\emails[@nlr.gov]{julie.bessac})} \and 
    Marc T. Henry de Frahan\footnotemark[3] \and 
    Elizabeth Qian\footnotemark[2] \and
    Julie Bessac\footnotemark[3]
}

\usepackage{amsopn}
\usepackage[acronym]{glossaries}
\setkeys{glslink}{hyper=false}
\usepackage{booktabs}
\usepackage{custom}                     % custom math expressions from "custom.sty"

\ifpdf%
\hypersetup{
  pdftitle={Sparse POD Mode Selection and Manifold Dimensionality Reduction with Neural Networks},
  pdfauthor={T. Koike, P. Mohan, M. T. H. Frahan, E. Qian, J. Bessac}
}
\fi

\makeatletter
\newcommand*{\addFileDependency}[1]{
\typeout{(#1)}
\@addtofilelist{#1}
\IfFileExists{#1}{}{\typeout{No file #1.}}
}
\makeatother

\newcommand*{\myexternaldocument}[1]{%
\externaldocument[][nocite]{#1}%
\addFileDependency{#1.tex}%
\addFileDependency{#1.aux}%
}
\myexternaldocument{supplement}

\usepackage{pdfpages} % include PDFs
\usepackage{pgffor} % for loops
\makeatletter
\AtBeginDocument{\let\LS@rot\@undefined}
\makeatother
\def\supplementfilename{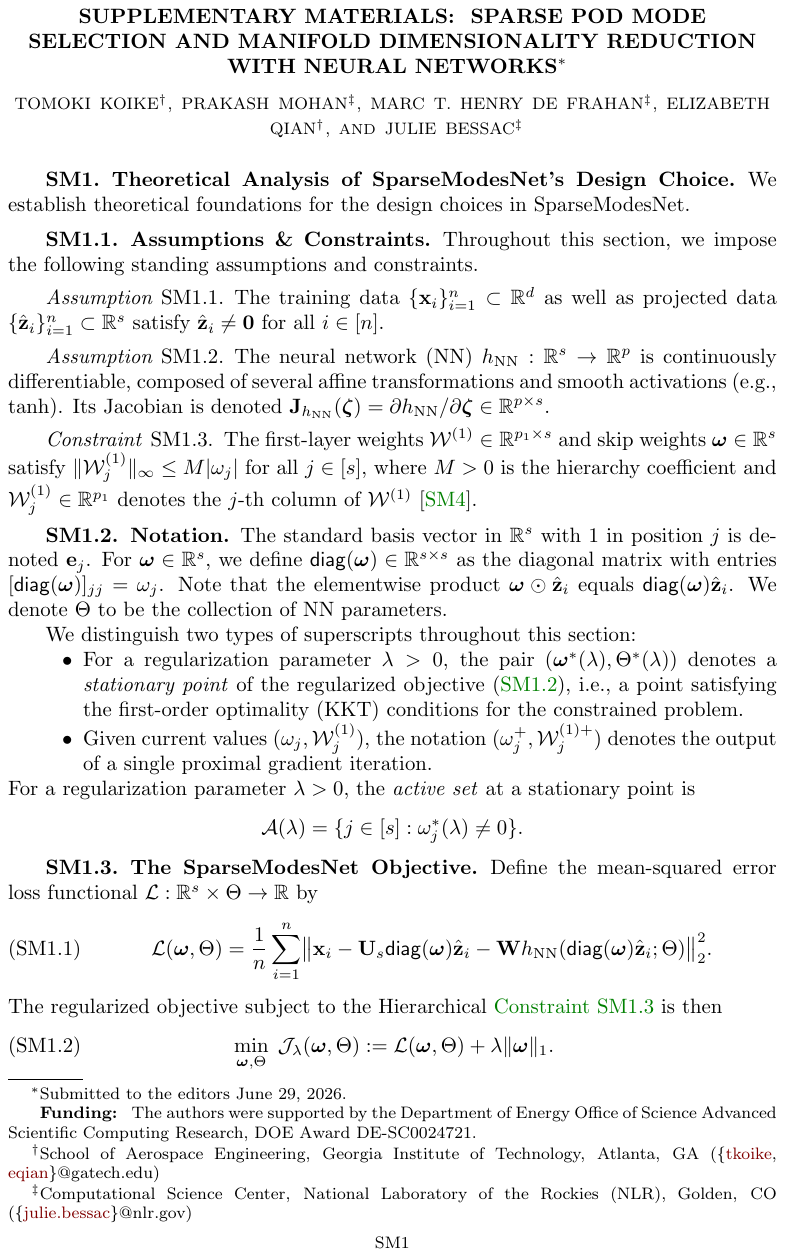}
\pdfximage{\supplementfilename}
\def\numbersupplementpages{\the\pdflastximagepages}
\newif\ifarXiv
\arXivtrue
\begin{document}

\maketitle

%==================================%
% REQUIRED Abstract
%==================================%
\begin{abstract}
    Linear dimensionality reduction methods such as proper orthogonal
    decomposition (POD) make high-dimensional data amenable to analysis by
    identifying the principal components, or modes, that capture the most
    variance, or energy, in the data and constructing a low-dimensional
    representation in the subspace they span. Such linear methods struggle,
    however, for data with slowly decaying Kolmogorov \(n\)-widths, such as
    advection-dominated and turbulent flows, which require many modes for
    accurate reconstruction; moreover, energy-based truncation can discard
    low-energy modes needed to capture small-scale features. Recent nonlinear
    manifold methods using polynomial mappings with alternating or greedy mode
    selection achieve better reconstruction with fewer modes, but fix the form
    of the nonlinear mapping a priori, limiting expressivity. In contrast,
    neural network (NN) manifolds offer greater expressivity yet employ
    energy-based selection. We present SparseModesNet, a dimensionality
    reduction framework that employs linear encoding and nonlinear NN decoding.
    The decoder leverages LassoNet, a method enforcing hierarchical sparsity
    through a residual connection with a linear skip layer, to simultaneously
    select informative modes and learn a nonlinear mapping that minimizes
    reconstruction error. On benchmark advection-dominated and chaotic flows,
    SparseModesNet matches or exceeds state-of-the-art performance. For
    turbulent channel flow at friction Reynolds number \(Re_\tau = 5200\), our
    method reduces reconstruction error by 51--78\% compared to existing
    polynomial manifold methods while maintaining interpretability through
    physically meaningful mode selection.
\end{abstract}

%==================================%
% REQUIRED keywords
%==================================%

%==================================%
% REQUIRED MSCcodes
%==================================%

%================%
%% MAIN SECTIONS
%================%
\section{Introduction}\label{sec:intro}

High-dimensional data are ubiquitous in computational science and engineering,
arising from numerical simulations of \glspl*{pde} as well as from experiments
and observations, and are frequently too large to analyze, store, or process
directly. Dimensionality reduction addresses this challenge by constructing a
low-dimensional representation that preserves the essential structure of the
data and enables efficient downstream tasks such as inverse problems, control, and optimization~\cite{benner2015Surveya,quarteroni2015reduced,kunisch2002galerkin,antoulas2005approximation}. The most widely used methods are
linear. For example, \gls{pod} and \gls{pca} computes via \gls{svd}
an orthonormal set of modes ordered by the variance, or energy, onto which the data can be projected to yield a low-dimensional representation~\cite{berkooz1993Proper,holmes2012turbulence,rowley2004Model}. 

However, \gls{pod}'s limitation to linear subspaces causes it to underperform for transport-dominated, advection-dominated, or turbulent systems. In such cases, slow spectral decay, known as the Kolmogorov \(n\)-width barrier, requires retaining prohibitively many modes for accuracy, undermining dimensionality reduction~\cite{cohen2016Kolmogorov}. Recent work has proposed addressing this barrier by moving beyond linear subspaces to nonlinear manifolds that allow nonlinear combinations of modes~\cite{ahmed2020Breaking,peherstorfer2022Breaking}.

Some researchers have explored nonlinear manifolds that augment linear \gls{pod} subspaces with quadratic~\cite{qiao2012Explicit,xia2014Explicit,jain2017Quadratic,rutzmoser2017Generalization} or higher-order polynomial mappings~\cite{geelen2023Learning}. These approaches have successfully demonstrated the potential to overcome the Kolmogorov barrier with fewer modes. However, they rely on heuristic energy-based mode selection, where modes are ordered by energy retention and then truncated at fixed thresholds. This selection can be suboptimal since low-energy modes may be critical to capture dynamics such as energy dissipation in turbulent flows~\cite{balajewicz2016Minimal,carlberg2017Galerkin}.

In~\cite{geelen2023Learning,geelen2024Learning}, the authors define a polynomial mapping that decodes a reduced representation and optimize both the linear basis and polynomial coefficients through an alternating minimization scheme, while~\cite{schwerdtner2024Greedy} proposes a greedy algorithm that iteratively selects modes based on their contribution to reducing state reconstruction error, later extended to sparse data in~\cite{schwerdtner2025Empirical}. While advancing beyond energy-based truncation, these methods face two limitations. First, to accurately capture increasingly complex physics requires higher-degree polynomials; including all cross-terms among the reduced variables then incurs combinatorial growth in the number of coefficients, creating a tradeoff between expressivity and computational tractability. Second, mode selection is performed separately from the mapping optimization rather than jointly, which may yield suboptimal mode-mapping combinations. Importantly, while the polynomial coefficients are optimized, the form of the mapping (e.g., quadratic) is prescribed a priori rather than learned from data, which limits expressivity.

\Glspl*{nn} offer an alternative to overcoming the Kolmogorov \(n\)-width barrier by learning nonlinear manifold representations directly from data. A prevalent architecture is the \gls{ae}, which consists of an encoder that compresses high-dimensional states into a low-dimensional latent representation and a decoder that reconstructs the original state from this representation. Unlike polynomial mappings with predetermined functional forms, \glspl*{ae} learn both the compression and reconstruction mappings from data, enabling greater flexibility in capturing complex nonlinear structure. The work~\cite{lee2020Model} first applied deep convolutional \glspl*{ae} to construct trial manifolds for \gls{mor}, demonstrating improved reconstruction over linear methods for advection-dominated problems. Subsequent developments have extended this framework in several directions:~\cite{romor2023Nonlinear} incorporates hyper-reduction techniques to lower online computational costs,~\cite{tencer2021Tailored} generalizes convolutional architectures to unstructured meshes common in finite element discretizations, and~\cite{chen2024Nonlinearmanifold} introduces adaptive decoder filters that evolve with the latent state. Despite these advances, \gls{ae}-based methods encode information into latent variables that lack direct physical meaning, defying \emph{interpretability} of the learned representations.

Interpretability can be considered the ability to understand which physical features or modes the reduced model captures and how they contribute to the reconstruction~\cite{chipman2005interpretable}. Several works have pursued this goal by imposing structure on the learned latent space according to physical considerations:~\cite{kadeethum2022Reduced} employs self-supervised learning to produce embeddings distinguishable according to physical parameters,~\cite{solera-rico2024VVariational} uses \( \beta \)-variational \glspl*{ae} to learn disentangled, near-orthogonal latent variables, and~\cite{zhang2023Nonlinear} encodes frequency information to yield latents corresponding to distinct temporal scales. While these methods endow the latent space with meaningful structure, the resulting representations remain dense---all latent coordinates contribute to every reconstruction, making it difficult to isolate which modes are essential for a given state. A complementary principle from interpretable machine learning is that \emph{sparsity}, the use of fewer active features, enhances interpretability by enabling direct examination of individual contributions~\cite{sudjianto2021Designing}. Sparse representations that activate only a subset of modes would allow practitioners to identify precisely which physical structures are necessary for accurate reconstruction.

This work bridges interpretable nonlinear manifolds with sparse mode selection and expressive \gls{nn} decoders. We adapt LassoNet~\cite{lemhadri2021Lassonet}, a framework enforcing hierarchical feature sparsity where features participate in \glspl*{nn} only if corresponding linear skip weights are active, to transform the combinatorial mode selection problem into a continuous regularization path promoting sparsity during \gls{nn} training. While LassoNet remains underexplored in scientific computing beyond limited applications such as physics-informed neural networks for inverse problems~\cite{ma2024incorporating}, we demonstrate its efficacy for mode selection with a \gls{nn} decoder, enabling automated, scalable selection of minimal sets of physically interpretable modes while leveraging \gls{nn} expressivity.

The contributions of this work are:
\begin{itemize}
    \item We propose SparseModesNet, a novel framework that combines the interpretability of \gls{pod} modes with \gls{nn} expressivity, enabling sparse mode selection during \gls{nn} training.
    \item We show SparseModesNet's effectiveness on dynamical data with slowly decaying Kolmogorov \(n\)-widths, including turbulent channel flow, showcasing its ability to select minimal mode sets while maintaining high state and mode reconstruction accuracy.
    \item We analyze the interpretability of our \gls{nn} decoder by computing modes of the reconstructed data and provide physical insights into selected modes based on their order and characteristics.
\end{itemize}

Several recent works have also combined \gls{nn} expressivity with physically meaningful \gls{pod} bases for interpretable nonlinear decoders. For example,~\cite{dar2023Artificial,barnett2023Neuralnetworkaugmented} employ small linear \gls{pod} basis augmented with \gls{nn} correction terms, preserving interpretability while harnessing nonlinear expressivity. The authors of~\cite{somasekharan2025Kolmogorov} extend this approach by applying \gls{nn} corrections to both encoder and decoder, each weighted with trainable coefficients for improved fitting. Alternatively, some works augment linear reduction with nonlinear manifolds derived from probabilistic embeddings~\cite{guo2026Nonlinear}, user-defined reproducing kernel Hilbert spaces~\cite{diaz2025Kernel}, or Gaussian processes~\cite{aresdeparga2026Nonlinear}. Although these hybrid approaches maintain physical interpretability and sparsity by selecting few modes, they rely on energy-based heuristics rather than simultaneously identifying essential modes from a candidate set and learning the nonlinear manifold.

Another related work~\cite{csala2026Decomposed} recently introduced Decomposed Sparse Modal Optimization (DESMO), which leverages the Sparse Identification of Nonlinear Dynamics (SINDy) framework~\cite{brunton2016discovering,champion2019data,fukami2021sparse} to discover interpretable governing equations by promoting sparsity in candidate function libraries. DESMO applies SINDy to \gls{pod} or \gls{ae}-learned bases by constructing a nonlinear candidate library from modes using predefined basis functions and employing \(\ell_1\)-regularization to select the fewest active terms. While DESMO promotes sparsity in functional representations of a given modal basis, our approach identifies which modes themselves should be retained and simultaneously learns an expressive \gls{nn} manifold tailored to those selected modes.

The remainder of this paper is organized as follows. Section~\ref{sec:background} reviews background on encoder-decoder concepts. Section~\ref{sec:sparsemodesnet} presents the SparseModesNet framework, detailing LassoNet integration for mode selection and the \gls{nn} architecture for nonlinear mapping. Section~\ref{sec:numerics} demonstrates SparseModesNet's performance on benchmark problems and turbulent channel flow. Finally, Section~\ref{sec:conclusion} summarizes our findings and discusses future directions.
\section{Background}\label{sec:background}

This section covers the necessary background for this work. In~\Cref{sec:bg:encoder-decoder}, we explain the concepts of encoder-decoder pairs in the context of dynamical systems. In~\Cref{sec:bg:lassonet}, we introduce LassoNet, a \gls*{nn} framework for sparse feature selection, which we adapt for mode selection in this work.

\subsection{Data Encoder and Decoder on Manifolds}\label{sec:bg:encoder-decoder}

\subsubsection{Encoder-Decoder for Dimensionality Reduction}

Let a data matrix be \(\Xmat = [\xvec_1, \ldots, \xvec_n] \in \R^{d \times n}\) 
whose columns \(\xvec_1, \ldots, \xvec_n \in \R^d\) are \(n\) high-dimensional
data vectors. The dimension \(d\) is typically large, corresponding, for
instance, to the degrees of freedom of a discretized field of a \gls{pde}, the resolution of an image, or the number of recorded measurements. 

Dimensionality reduction seeks a lower-dimensional representation of each data vector by mapping it to reduced coordinates in a latent space \(\R^r\), where \(r \ll d\). This mapping is realized by an \emph{encoder} \(\Ecal : \R^d \to \R^r\) that compresses \(\xvec \) to a latent variable \(\zvec = \Ecal(\xvec) \in \R^r\), together with a \emph{decoder} \(\Dcal : \R^r \to \R^d\) that reconstructs an approximation of the original vector, \(\xvec \approx (\Dcal \circ \Ecal)(\xvec)\). The quality of this approximation is measured by the \emph{reconstruction error}
\begin{equation}\label{eqn:recon_error}
    \mathsf{RE}(\Xmat; \Ecal, \Dcal)
    = \frac{1}{n}\sum_{i=1}^n \| \xvec_i - (\Dcal \circ \Ecal)(\xvec_i) \|_2^2,
\end{equation}
which quantifies the discrepancy between the data points and their
reconstructions.

The simplest encoder-decoder pair is linear. Let \( \Xmat \approx \Umat_s \Sigmamat_s \Vmat_s^\top \) be the rank-\(s\) \gls*{svd} of the snapshot matrix, where \(s \leq \min{\{d,n\}}\), the matrices \(\Umat_s\in\R^{d\times s}\) and \(\Vmat_s\in\R^{n\times s}\) contain the left and right singular vectors, and \(\Sigmamat_s\in\R^{s\times s}\) is the diagonal matrix of singular values \( \sigma_1 \geq \cdots \geq \sigma_s > 0 \). The columns of \(\Umat_s\) are the \gls*{pod} modes, which form an orthonormal basis capturing the most energetic features of the data~\cite{berkooz1993Proper,holmes2012turbulence}. A linear encoder-decoder pair selects a subset of \( r < s \) modes, indexed by \( \Ical_r \subseteq [s] = \{1, \ldots, s\} \) with cardinality \( |\Ical_r| = r \), and defines \( \Ecal(\xvec) = \Umat_r^\top \xvec \) and \( \Dcal(\zvec) = \Umat_r \zvec \), where \(\Umat_r\in\R^{d\times r}\) contains the selected modes.

A common selection strategy is the \emph{energy heuristic}, which retains the leading \( r \) modes corresponding to the largest singular values. By the Eckart-Young theorem~\cite{eckart1936approximation,mirsky1960symmetric}, this choice minimizes the reconstruction error among all rank-\(r\) approximations:
\begin{equation}\label{eqn:lin_recon_error}
    \mathsf{RE}(\xvec;\Umat_r) = \frac{1}{n}\sum_{i=1}^n \| \xvec_i - \Umat_r \zvec_i \|_2^2 = \frac{1}{n}\sum_{i=r+1}^{\min \{d,n\}} \sigma_i^2.
\end{equation}
The limitation of linear encoder-decoder pairs is that reducing reconstruction error requires increasing \( r \), which becomes prohibitively expensive for certain data types. For example, transport-dominated or turbulent physics data with slowly decaying Kolmogorov \(n\)-widths~\cite{peherstorfer2022Breaking}. This motivates the linear encoder and nonlinear decoder pair, which we introduce in the next section.

\subsubsection{Linear Encoder with Nonlinear Decoder}

Data with slowly decaying Kolmogorov \(n\)-widths are known to be challenging, especially for linear encoder-decoder pairs, since reducing reconstruction error requires retaining prohibitively many modes. To address this limitation, practitioners have proposed retaining the linear encoder while augmenting the decoder with a nonlinear correction term:
\begin{displaymath}
    \zvec = \Ecal(\xvec) = \Umat_r^\top\xvec, \qquad \Dcal(\zvec) = \Umat_r \zvec + h(\zvec),
\end{displaymath}
where \(h: \R^r \to \R^d\) is a nonlinear function. Here, the \( r \) modes forming \( \Umat_r \) need not be the leading modes; they can be any subset of the candidate modes \( \Umat_s \). Since the nonlinear correction term \( h \) should remain low-dimensional for efficient evaluation, a natural choice for its input is the reduced representation \( \zvec \). This decoder structure with polynomial nonlinear mappings was introduced by~\cite{qiao2012Explicit} and has since been adopted in various model reduction contexts~\cite{xia2014Explicit,jain2017Quadratic,rutzmoser2017Generalization,diez2021Nonlinear,geelen2023Operator,schwerdtner2024Greedy}.

As shown in~\cite[eq.\,14]{geelen2023Operator}, the optimal nonlinear correction term \( h \) (in the linear least-squares sense) maps the reduced representation \( \zvec \) to the orthogonal complement of the subspace spanned by \(\Umat_r\). That is, \( h \) captures dynamics not represented by the selected modes \( \Umat_r \) alone, thereby enhancing expressivity beyond what the linear term provides. 
However, an arbitrary \( h \) may not satisfy this orthogonality condition, introducing redundancy with the linear term and potentially degrading reconstruction accuracy. Yet designing \( h \) to satisfy orthogonality a priori is nontrivial. To enforce orthogonality explicitly, many existing works~\cite{jain2017Quadratic,rutzmoser2017Generalization,diez2021Nonlinear,barnett2022Quadratica,geelen2023Operator,geelen2023Learning,geelen2024Learning,schwerdtner2024Greedy} define the decoder as
\begin{equation}\label{eqn:orth_decoder}
    \Dcal(\zvec) = \Umat_r \zvec + \Wmat h(\zvec),
\end{equation}
where \( h: \R^r \to \R^p \) and \( \Wmat\in\R^{d\times p} \) satisfies \( \Umat^\top_r\Wmat = \mathbf{0} \) for a mapping dimension \( p \). This formulation augments the linear reconstruction with information in its orthogonal complement through the composition of the nonlinearity \( h \) and rotation \( \Wmat \).

A possible choice is \( \Wmat = \overline{\Umat}_q \), where \( \overline{\Umat}_q \in \R^{d \times q} \) consists of \( q \) modes from \( \Umat_s \) that are orthogonal to \( \Umat_r \), i.e., \( \overline{\Umat}_q \subseteq \{\Umat_j : j \in [s] \setminus \Ical_r\} \). This formulation, adopted in~\cite{barnett2023Neuralnetworkaugmented,geelen2023Learning,geelen2024Learning,diaz2025Kernel}, provides a natural interpretation, where \( h \) maps the reduced coordinates \( \zvec \) to coefficients of modes orthogonal to those used in the linear reconstruction. However, this restricts the correction term to the span of modes from the predefined candidate set \( \Umat_s \), which may be insufficient in number. Rather, in this work, we will learn \( \Wmat \) from data.

\subsubsection{Linear Encoder with Neural Network Decoder}

A key limitation of the nonlinear decoder in~\eqref{eqn:orth_decoder} is that the functional form of \( h \) is typically fixed a priori (e.g., quadratic polynomials), which restricts expressivity. To overcome this, we parameterize the nonlinear correction term as a \gls*{nn}, following~\cite{barnett2023Neuralnetworkaugmented,somasekharan2025Kolmogorov}:
\begin{equation}\label{eqn:nn_decoder}
    \zvec = \Ecal(\xvec) = \Umat_r^\top\xvec, \qquad \Dcal(\zvec) = \Umat_r\zvec + \Wmat\hnn(\zvec),
\end{equation}
where \( \hnn: \R^r \to \R^p \) is the \gls*{nn} and \( \Wmat \in \R^{d \times p} \) satisfies \( \Umat_r^\top \Wmat = \mathbf{0} \) but is not restricted to columns from \( \Umat_s \). The \gls*{nn} learns the nonlinear mapping and weight \( \Wmat \) from data, while the linear term projects onto selected modes.

This formulation requires determining both (i) which modes to include in \( \Umat_r \) and (ii) how to train \( \hnn \) and \( \Wmat \) for the nonlinear correction---while ensuring orthogonality between the two terms. Rather than treating mode selection and \gls*{nn} training as separate problems, we seek a unified framework that jointly optimizes mode selection and the nonlinear mapping to minimize reconstruction error. In the following section, we introduce a feature selection method that enables this joint optimization for \gls*{nn}-based decoders.

\subsection{LassoNet for Feature Selection}\label{sec:bg:lassonet}

LassoNet~\cite{lemhadri2021Lassonet} is a \gls*{nn} framework for global feature selection that extends feature sparsity from linear models (e.g., Lasso regression~\cite{hastie2015statistical}) to deep \glspl*{nn}. While traditional Lasso employs an \(\ell_1\)-penalty to enforce sparsity on linear weights, LassoNet integrates feature selection into a \gls*{nn} via a linear skip (or residual) connection inspired by ResNets (residual \glspl{nn})~\cite{he2015Deep}.

\paragraph{Network Architecture} 
Consider an input feature vector \(\zhat \in \R^s\). Denote a feed-forward \gls{nn} \(\gnn(\zhat; \Wcal, \betavec): \R^s \to \R^s \) with weight parameters \(\Wcal = \{\Wcal^{(1)}, \ldots, \Wcal^{(L)}\} \) and bias parameters \(\betavec = \{\betavec^{(1)}, \ldots, \betavec^{(L)}\} \) for \(L\) layers, which can be expressed as a composition of layer-wise transformations: \( \gnn(\zhat; \Wcal, \betavec) = g^{(L)} \circ \cdots \circ g^{(1)}(\zhat) \), 
where \(g^{(k)}(\cdot) = \sigma_k(\Wcal^{(k)} \cdot + \betavec^{(k)})\) represents the \(k\)-th layer with weight matrix \(\Wcal^{(k)}\), bias vector \(\betavec^{(k)}\), and activation function \(\sigma_k(\cdot)\) applied element-wise. The LassoNet architecture augments this feed-forward \gls{nn} with a direct linear skip connection:
\begin{equation}\label{eqn:lassonet_architecture}
    \mathsf{LassoNet}(\zhat; \omegavec, \Wcal, \betavec) = \omegavec \odot \zhat + \gnn(\zhat; \Wcal, \betavec),
\end{equation}
where \(\omegavec \in \R^s\) are skip-layer weights, \(\odot \) denotes element-wise product, and each input feature \(\hat{z}_j\) connects to the corresponding output via skip weight \(\omega_j\). We denote by \(\Wcal_j^{(1)} \in \R^{p_1}\) the \(j\)-th column of the first layer weight matrix \(\Wcal^{(1)} \in \mathbb{R}^{p_1 \times s}\), representing all connections from input feature \(j\) to the \(p_1\) units in the first hidden layer.

\paragraph{Objective Function with Hierarchical Constraint}
The novelty of LassoNet lies in its training objective that enforces structured sparsity through
\begin{equation}\label{eqn:lassonet}
    \min_{\boldsymbol{\omega}, \mathcal{W}, \betavec} \mathcal{L}(\omegavec, \Wcal, \betavec) + \lambda \| \omegavec \|_1, \quad \text{subject to} \quad \|\Wcal_j^{(1)}\|_\infty \leq M|\omega_j| \quad \forall j,
\end{equation}
where \( \Lcal(\omegavec, \Wcal, \betavec) \) is the empirical loss (e.g., reconstruction error \( \mathsf{RE}(\xvec;\Ecal,\Dcal) \) defined in~\cref{eqn:recon_error}), \( \lambda \geq 0 \) controls sparsity via the \( \ell_1 \)-penalty, and \(M > 0\) governs the strength of the hierarchical constraint linking skip-layer and first hidden-layer weights.

Proposed in~\cite{lemhadri2021Lassonet}, this hierarchical constraint \( \| \Wcal_j^{(1)}\|_\infty \leq M|\omega_j|\) is key for feature selection. That is, when the \(\ell_1\)-penalty drives a skip-layer weight \(\omega_j\) to zero, this constraint forces all associated first hidden layer weights \( \Wcal_j^{(1)} \) to zero as well, completely eliminating feature \(j\) from the network. 

\paragraph{Training via Continuation Strategy}
LassoNet employs proximal gradient descent with a new hierarchical proximal operator that simultaneously enforces the \(\ell_1\)-penalty and hierarchical constraint (see~\cite{lemhadri2021Lassonet} for optimization details). 
Critically, rather than solving~\cref{eqn:lassonet} for a single fixed \(\lambda \), LassoNet implements a \emph{continuation strategy}~\cite{hale2007fixed,hale2008fixed} that solves a sequence of problems with increasing \(\lambda \) values \(\lambda_1 < \lambda_2 < \cdots < \lambda_K\). Starting from \(\lambda_1 \approx 0\) (yielding a dense model), each subsequent problem at \(\lambda_{k+1}\) is initialized from the solution at \(\lambda_k\), following the solution path as \(\lambda \) increases toward \(\lambda_K\). This path-following approach is computationally efficient---each warm-started subproblem converges faster than solving from scratch---and produces a sequence of models (i.e., regularization path~\cite{park2007L1,friedman2010regularization}) with progressively increasing sparsity. 
For mode selection, this continuation strategy provides a systematic framework: as \(\lambda \) increases along the path, modes are sequentially removed from the \gls*{nn} decoder inputs according to their importance for reconstruction. The sparsest model (fewest modes) that maintains acceptable reconstruction accuracy can then be selected, yielding an expressive nonlinear manifold with sparse, physically interpretable input features and reduced computational cost automatically.

\section{SparseModesNet: Sparse Mode Selecting Neural Network}\label{sec:sparsemodesnet}

This section presents SparseModesNet, our framework for selecting optimal \gls{pod} modes during nonlinear decoder training. We first formulate the mode selection problem in~\Cref{sec:sparsemodesnet:math}. In~\Cref{sec:sparsemodesnet:lassonet}, we show how to restructure the nonlinear decoder into a ResNet-style architecture incorporating feature sparsity~\cite{lemhadri2021Lassonet}. \Cref{sec:sparsemodesnet:training} describes the training procedure, and~\Cref{sec:sparsemodesnet:implementation} discusses implementation details for data normalization and \gls{nn} architecture.

\subsection{Mathematical Formulation for Mode Selection}\label{sec:sparsemodesnet:math}

The goal is to select optimal \gls{pod} modes by minimizing reconstruction error while training the \gls{nn} decoder. To select \( r \) modes from \( s \) candidate modes, let \(\Smat_r\in\R^{s\times r}\) be a column selector matrix~\cite[Sec.\,7.2]{boyd2018introduction} where each column has a single entry of one at the row corresponding to a selected left singular vector in \(\Umat_s\), such that \(\Umat_r = \Umat_s \,\Smat_r \).

Our \gls{nn} learns the matrix \(\Smat_r\Smat_r^\top \), which is a diagonal matrix with ones at entries corresponding to selected \(r\) left singular vectors. Define \( \hat\zvec_i = \Umat_s^\top\xvec_i \in \R^{s} \) and a binary vector \( \wvec\in{\{0,1\}}^s \) with \(r\) ones at entries corresponding to selected left singular vectors, such that \( \diag(\wvec) = \Smat_r\Smat_r^\top \in \R^{s \times s} \). The reconstruction error is then
\begin{equation}
    \mathsf{RE}(\xvec;\wvec) = \frac{1}{n}\sum_{i=1}^n \bigl \lVert \xvec_i - \Umat_s(\wvec \odot \hat\zvec_i) - 
    \Wmat\hnn(\wvec \odot \hat\zvec_i) \bigr \rVert_2^2 .
\end{equation}
Note the \gls{nn} mapping is now \( \hnn: \R^{s} \to \R^{p} \) since the input \( \hat\zvec_i \) is the full projection onto all \(s\) \gls{pod} modes, not just the selected \( r \) modes. However, \(\hat\zvec_i \) has zeros at rows corresponding to zero elements of \( \wvec \), ultimately selecting \(r\) modes from \(s\) candidates.

The set of nonzero entries in \(\wvec \) gives the indices of selected modes, i.e., \( \Ical_r = \{i \in [s] : w_i = \wvec(i) \neq 0\} \). We identify \( \Ical_r \) by training the \gls{nn} to minimize reconstruction error. To do so, we reformulate the nonlinear decoder as a ResNet-style architecture incorporating feature sparsity through a linear skip connection, following the LassoNet framework~\cite{lemhadri2021Lassonet}, as discussed in the next section.

\subsection{LassoNet Formulation of the Nonlinear Decoder}\label{sec:sparsemodesnet:lassonet}

To restructure the nonlinear decoder into a ResNet-style architecture, we first relax the problem to learn a continuous skip weight vector \( \omegavec \in \R^s \), allowing more flexible mode selection rather than learning a restrictive binary vector \( \wvec \). We then minimize the reconstruction error over the training data while enforcing two critical constraints.

\begin{figure}[htb!]
    \centering
    \includegraphics[width=0.82\textwidth]{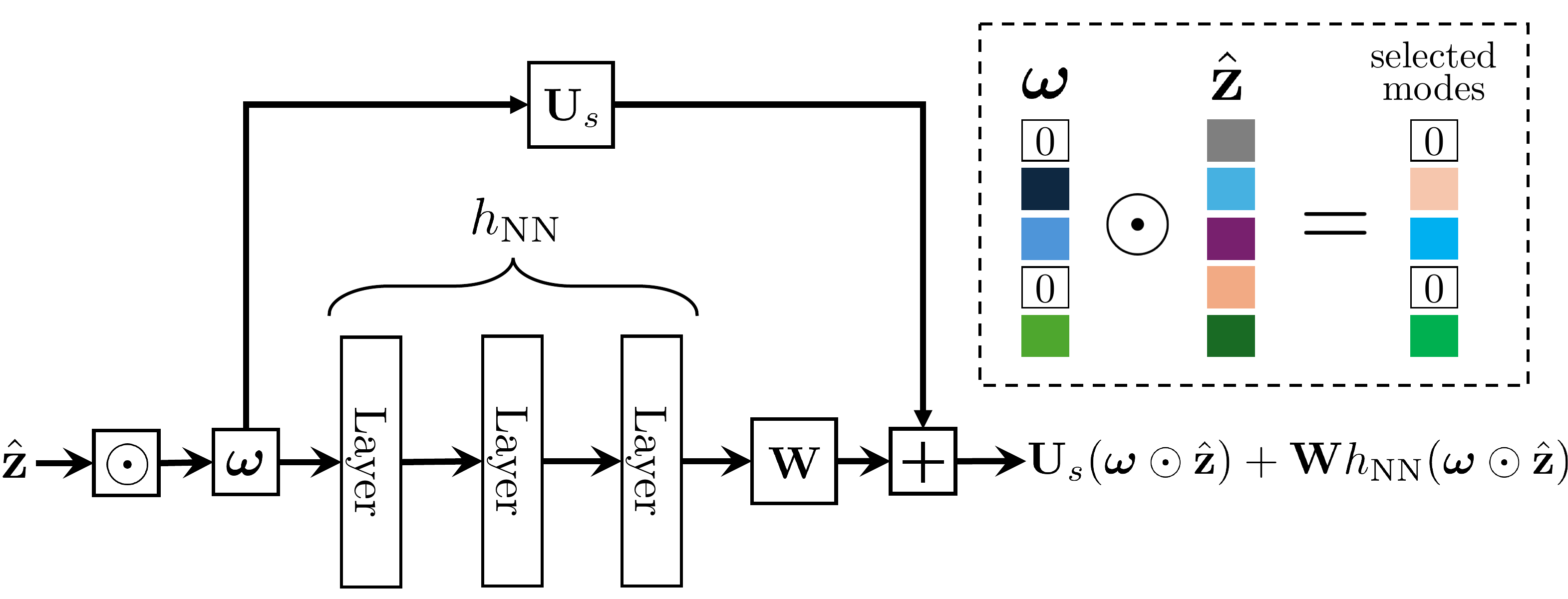}
    \caption{The SparseModesNet architecture with a linear skip connection \((\Umat_s(\omegavec \odot \hat{\zvec}))\) and a neural network branch \(( \Wmat\hnn(\omegavec \odot \hat{\zvec}) )\) forming a residual network. The top right illustrates how the mode selector zeros out unselected modes via element-wise multiplication (colored cells are nonzero).}\label{fig:sparse-modes-net}
\end{figure}

First, following LassoNet~\eqref{eqn:lassonet}, we impose an \( \ell_1 \)-penalty \( \lambda \| \omegavec \|_1 \) on the skip weights to promote sparsity. This is paired with the hierarchical constraint \( \| \Wcal_j^{(1)}\|_\infty \leq M|\omega_j| \) on the first-layer \gls{nn} weights, ensuring that if a mode is unselected (\( \omega_j = 0 \)), all associated first-layer weights \(\Wcal_j^{(1)}\) are also zero, preventing unselected modes from influencing the nonlinear mapping. Second, we impose an \emph{orthogonality constraint} \( [\Umat_s \diag(\1_{\{\omega_j\neq0\}}(\omegavec))]\tran\Wmat = 0 \) while learning \( \Wmat \) from data to ensure the linear and nonlinear decoder components do not interfere, maintaining clear separation of their contributions as discussed in~\Cref{sec:bg:encoder-decoder}. Here, \( \1_{\{\omega_j\neq 0\}}(\omegavec) \) is an indicator function returning \(1\) if \( \omega_j \neq 0 \) and \(0\) otherwise. 
Hence, the following optimization enables joint mode selection and \gls{nn} training:
\begin{equation}\label{eqn:lassonet-opt}
    \begin{gathered}
        \min_{\omegavec,\hnn, \Wmat}\frac{1}{n}\sum_{i=1}^n \bigl \lVert \xvec_i - \Umat_s (\omegavec \odot \hat{\zvec}_i) - \Wmat\hnn(\omegavec \odot \hat{\zvec}_i) \bigr \rVert_2^2 + \lambda \| \omegavec \|_1 \\
        \text{subject to} ~~ 
        \| \Wcal_j^{(1)}\|_\infty \leq M|\omega_j|~~
        \text{and}~~
        {[\Umat_s\diag(\1_{\{\omega_j\neq 0\}}(\omegavec))]}^\top\Wmat= 0, ~\forall j \in [s],
    \end{gathered}
\end{equation}
where \( \Wcal^{(1)} \), \( M \), and \(\lambda \) are equivalent to those in the LassoNet optimization~\eqref{eqn:lassonet}. 
To draw a clear connection to LassoNet, recognize that this decoder closely resembles the LassoNet architecture~\eqref{eqn:lassonet_architecture} with a linear skip connection and \gls{nn} branch:
\begin{displaymath}
    \Dcal(\zhat) = \underbrace{\Umat_s(\omegavec \odot \hat{\zvec})}_{\text{linear skip connection}} + ~\quad \underbrace{\Wmat\hnn(\omegavec \odot \hat{\zvec})}_{\text{neural network}}.
\end{displaymath}
This \gls{nn} architecture and mode selection mechanism is summarized in~\Cref{fig:sparse-modes-net}.

\subsubsection{Design Choices and Their Theoretical Justification}\label{sec:sparsemodesnet:design}

SparseModesNet differs from the original LassoNet and simple \( \ell_1 \)-regularization in two key respects, each essential for guaranteeing monotonic mode elimination (i.e., eliminated modes are never reactivated) when training via the continuation strategy.

\paragraph{Input masking} Unlike the original LassoNet~\eqref{eqn:lassonet_architecture}, which inputs the full feature vector \( \hat{\zvec} \) to the \gls{nn}, we employ \( \omegavec \odot \hat{\zvec} \) as the \gls{nn} input such that the unselected modes are effectively zero. Here, \(\omegavec \) remains a trainable parameter optimized via~\eqref{eqn:lassonet-opt}, while \(\omegavec \odot \hat{\zvec}\) serves as the parametrized input to \(\hnn(\cdot) \).

\paragraph{Hierarchical constraint} An alternative approach may be to remove the hierarchical constraint and keep only the \( \ell_1 \)-penalty, given the latent variable to both the linear and nonlinear terms of the decoder are already masked with \( \omegavec \). 

The input masking yields \emph{gradient severance}. The gradient of the loss with respect to the first-layer weights satisfies \( \partial \Lcal / \partial \Wcal_j^{(1)} = \mathbf{0} \) whenever \( \omega_j = 0 \). 
The hierarchical constraint \( \| \Wcal_j^{(1)}\|_\infty \leq M|\omega_j| \) complements this by enforcing \( \Wcal_j^{(1)} = \mathbf{0} \) whenever \( \omega_j = 0 \). 
Together, these properties completely sever mode \(j\)'s computational path through the \gls{nn}. No gradient signal can flow to update the zeroed-out weights, preventing the reactivation of an eliminated mode. 
Hence, combining gradient severance with the hierarchical constraint guarantees \emph{monotonic mode elimination}. That is, under mild regularity conditions on the solution path, modes eliminated at regularization level \( \lambda_k \) remain eliminated for all subsequent \( \lambda_\ell > \lambda_k \). This monotonicity ensures that SparseModesNet produces a stable, nested sequence of active modes, enabling reliable mode selection. Without both design choices, the nonzero weights may oscillate and spuriously reactivate previously eliminated modes as \(\lambda \) increases, forcing the network to repeatedly relearn mode suppression. 

In~\Cref{sec:sup:ablation}, empirical validations of our design choices are shown through an ablation study comparing SparseModesNet against variants without input masking and without the hierarchical constraint, demonstrating how our formulation achieves lower reconstruction error and more reliable mode selection. 
However, the monotonic mode elimination alone does not explain the better selection of modes. Specifically, we lack the mathematical machinery to explain how our method is less affected by the spectral bias~\cite{rahaman2019spectral} as we see in our empirical validations; while possible future directions would be to explore neural tangent kernels (NTKs)~\cite{jacot2018Neural}.
Hence, we postpone the formal statement and proof of monotonic mode elimination, including the precise regularity assumptions to~\Cref{sec:sup:sparsemodesnet:theory}. 

\subsection{Training the SparseModesNet}\label{sec:sparsemodesnet:training}

SparseModesNet training consists of two sequential phases, the first employs the continuation strategy from LassoNet~\cite{lemhadri2021Lassonet} to select modes, and the second retrains the decoder with the selected modes.

\paragraph{Phase 1: Mode Selection}
We solve~\eqref{eqn:lassonet-opt} via continuation, incrementally increasing \( \lambda \leftarrow (1 + \epsilon)\lambda \) with path multiplier \( \epsilon > 0 \) to promote sparsity while warm-starting each subproblem with the previous solution. At each gradient descent iteration, we apply a proximal gradient step: (i) compute gradients via backpropagation and update all parameters, then (ii) apply the hierarchical proximal operator from~\cite[Alg.\,2]{lemhadri2021Lassonet} to enforce the constraint \( \|\Wcal_j^{(1)}\|_\infty \leq M|\omega_j| \). To maintain orthogonality between the linear and nonlinear components, we project \( \Wmat \) onto the orthogonal complement of selected modes via \( \Wmat \gets \Wmat - \Umat_s\diag(\1_{\{\omega_j\neq0\}}(\omegavec))\Umat_s^\top\Wmat \) after each parameter update. This phase terminates when the number of active modes reaches the target \( r \), yielding the optimal mode selector \( \omegavec \).

\paragraph{Phase 2: Decoder Retraining}
With modes selected, we fix \( \zvec_i = \Umat_r^\top\xvec_i \in \R^r \) where \( \Umat_r = \Umat_s\Smat_r \in \R^{d\times r} \) contains the \( r \) selected \gls{pod} modes (see~\Cref{sec:sparsemodesnet:math}), and retrain the decoder from scratch without the \( \ell_1 \)-penalty or hierarchical constraint:
\begin{displaymath}
    \min_{\hnn,\Wmat} \frac{1}{n}\sum_{i=1}^n \bigl \lVert \xvec_i - \Umat_r \zvec_i - \Wmat\hnn(\zvec_i) \bigr \rVert_2^2 \quad \text{subject to} \quad \Umat_r^\top\Wmat= 0.
\end{displaymath}
Orthogonality is enforced at each iteration via \( \Wmat \gets \Wmat - \Umat_r(\Umat_r^\top\Wmat) \). Retraining without the penalty allows more effective learning by removing the downward bias on parameters~\cite{lemhadri2021Lassonet}. Upon convergence with \( \hnn \) fixed, we obtain the optimal \( \Wmat \) by solving the following linear least-squares problem only once:
\begin{equation}\label{eqn:weight-final-least-squares}
    \Wmat^\star = \arg\min_{\Wmat} \bigl\lVert \Xmat - \Umat_r\Umat_r^\top\Xmat - \Wmat\hnn(\Umat_r^\top\Xmat) \bigr\rVert_F^2 + \gamma \| \Wmat \|_F^2,
\end{equation}
where \( \hnn(\Umat_r^\top\Xmat) \in \R^{p\times n} \) applies \( \hnn \) column-wise and \(\gamma \) is the \(\ell_2\)-penalty parameter. This least-squares solution automatically satisfies the orthogonality constraint~\cite{geelen2023Operator}.

The complete training procedure is detailed in~\Cref{alg:lassonet-training,alg:decoder-training}. Hyperparameters (\( r, M, \lambda_0, \epsilon \)) are selected following recommendations from~\cite{lemhadri2021Lassonet} with problem-specific tuning during experimentation.

\begin{algorithm}[ht]
    \caption{SparseModesNet Phase 1: Mode Selection}\label{alg:lassonet-training}
    \begin{algorithmic}[1]
        \REQUIRE{Training data \(\Xmat, \hat{\Zmat} \), basis \(\Umat_s \), neural network \( \hnn \), target modes \( r \), initial \( \lambda_0 \), hierarchy multiplier \( M \), epochs \( B \), path multiplier \( \epsilon > 0 \)}
        \STATE{Initialize \( \hnn \) with parameters \( (\omegavec, \Wcal, \betavec, \Wmat) \) and set \( \lambda = \lambda_0 \), \( k = s \)}
        \WHILE{\( k > r \)}
            \FOR{\( b = 1,\ldots,B \)}
                \STATE{Perform proximal gradient step: compute gradients, update parameters, apply hierarchical proximal operator to \( (\omegavec, \Wcal^{(1)}) \)}
                \STATE{Project to orthogonal complement: \( \Wmat \gets \Wmat - \Umat_s\diag(\1_{\{\omega_j\neq0\}}(\omegavec))\Umat_s^\top\Wmat \)}
            \ENDFOR{}
            \STATE{Increment \( \lambda \gets (1 + \epsilon)\lambda \) and update \( k \gets \lvert \{i\in[s] : \omega_i \neq 0\} \rvert \)}
        \ENDWHILE{}
        \RETURN{mode selector \( \omegavec \), decoder parameters \( ( \Wcal, \betavec, \Wmat) \)}
    \end{algorithmic}
\end{algorithm}

\begin{algorithm}[ht]
    \caption{SparseModesNet Phase 2: Decoder Retraining}\label{alg:decoder-training}
    \begin{algorithmic}[1]
        \REQUIRE{Training data \(\Xmat \), selected basis \(\Umat_r \), neural network \( \hnn \), epochs \( B \)}
        \STATE{Initialize \( \hnn \) with parameters \( (\Wcal, \betavec, \Wmat) \)}
        \FOR{\( b = 1,\ldots,B \)}
            \STATE{Compute gradients via backpropagation and update parameters}
            \STATE{Project to orthogonal complement: \( \Wmat \gets \Wmat - \Umat_r(\Umat_r^\top\Wmat) \)}
        \ENDFOR{}
        \STATE{Solve least-squares problem~\eqref{eqn:weight-final-least-squares} for optimal \( \Wmat^\star \)}
        \RETURN{decoder parameters \( (\Wcal, \betavec, \Wmat^\star) \)}
    \end{algorithmic}
\end{algorithm}

\subsection{Additional Implementation Details}\label{sec:sparsemodesnet:implementation}

\paragraph{Data Normalization} 
We center the data around their temporal mean and apply min-max normalization to scale data to \( [-1,1] \), allowing training across different data scales. Centering and normalization are applied before computing \gls{pod} modes. During inference, outputs are denormalized and decentered to the original scale.

\paragraph{Stopping Criteria} 
Training continues until the number of active modes (nonzero entries in \( \omegavec \)) reaches the target \( r \). However, if the active mode count remains unchanged for a number of iterations (e.g., 50), we terminate training and select the \( r \) highest weighted modes in \( \omegavec \), ensuring convergence when the model has stabilized and allowing a fully automated mode selection even when \( r \) is unknown a priori.

\paragraph{Neural Network Architecture} 
For \( \hnn \), we employ a hybrid architecture consisting of a single fully connected layer followed by a deep polynomial \gls{nn} (\( \Pi \)-Net)~\cite{chrysos2022Deep}. The fully connected layer transforms input modes into a representation suitable for the \( \Pi \)-Net to capture polynomial nonlinearities and serves as the gate layer where the hierarchical sparsity constraint is applied during training. The \( \Pi \)-Net then models high-order polynomial functions, enabling rich representation of complex relationships. While this layer can be extended to multiple fully connected layers for more intricate problems, a single layer proved sufficient for the datasets considered.

We tested with other architectures, including standard multi-layer perceptrons, convolutional NN, and U-Nets, but found the hybrid fully connected-\( \Pi \)-Net architecture provided superior reconstruction accuracy and mode selection for our problems. Further details on the hybrid \( \Pi \)-Net architecture are provided in~\Cref{sec:sup:architecture}.

\paragraph{Computational Cost}
The basis \( \Umat_s \) is computed once upfront via a truncated \gls{svd}, after which the dominant cost is the \gls{nn} training over many epochs in both phases. Relative to training a standard decoder, SparseModesNet adds only two lightweight operations per iteration: the hierarchical proximal operator, which is a closed-form thresholding step, and the orthogonal-complement projection \( \Wmat \gets \Wmat - \Umat_s\diag(\1_{\{\omega_j\neq0\}}(\omegavec))\Umat_s^\top\Wmat \), an \( \mathcal{O}(p s d) \) matrix product. Both are negligible compared to backpropagation through \( \hnn \), so the per-epoch overhead of automated mode selection is marginal, and the entire cost is incurred offline during training. 

\section{Numerical Experiments}\label{sec:numerics}

In this section, we demonstrate SparseModesNet on three problems with high Kolmogorov \(n\)-width necessitating nonlinear decoders: (i) the linear transport equation, (ii) \gls{kse}, and (iii) turbulent channel flow. The turbulent channel flow simulations were performed on NLR's Kestrel high-performance computing cluster with H100 Nvidia GPUs. The SparseModesNet implementation and all experiments are publicly available\footnote{Code: \href{https://github.com/smallpondtom/sparsemodesnet}{https://github.com/smallpondtom/sparsemodesnet}}. Essential hyperparameters and nonlinear mapping dimensions are reported in~\Cref{tab:parameters_summary}.

\subsection{Experimental Setup}
In our experiments, we consider and compare several decoder structures across all experiments:
\begin{enumerate}
    \item A linear decoder selecting the leading \( r \) modes based on the energy heuristic.
    \item The Greedy Quadratic Manifold approach~\cite{schwerdtner2024Greedy}, which uses quadratic manifolds and greedy mode selection to minimize reconstruction error. We also include a Greedy Cubic Manifold extension (not in the original work) using cubic mappings for fair comparison with our \(\Pi_3\)-Net. Both use the authors' publicly available code\footnote{GreedyQM Code: \href{https://github.com/Algopaul/greedy\_quadratic\_manifolds}{https://github.com/Algopaul/greedy\_quadratic\_manifolds}}. For more details on greedy methods, see~\Cref{appsec:greedy_qm}.
    \item SparseModesNet with hybrid 2\textsuperscript{nd}-order (\(\Pi_2\)-Net) and 3\textsuperscript{rd}-order (\(\Pi_3\)-Net) deep polynomial networks as nonlinear decoders (architecture details are included in~\Cref{sec:sup:architecture}). For comparison, we also present results training these architectures with the leading \( r \) modes instead of SparseModesNet-selected modes, using identical hyperparameters to highlight mode selection importance.
\end{enumerate}

For each method, we report relative reconstruction errors:
\begin{equation}\label{eqn:rel-recon-error}
    \text{relative reconstruction error} = \frac{\| \Xmat - (\Dcal \circ \Ecal)(\Xmat) \|_F}{\| \Xmat \|_F},
\end{equation}
where \( \Xmat \in \R^{d \times n} \) contains all snapshots, and \( \Ecal \) and \( \Dcal \) are the encoder and decoder mappings applied column-wise to \( \Xmat \). 
Beyond reconstruction error, we assess decoder interpretability through \emph{mode fitting} as done in~\cite{solera-rico2024VVariational,csala2026Decomposed}. For mode fitting, we compute \gls{pod} modes from the reconstructed data from the decoder and compare them to the true modes from original data \(\Xmat \). While \emph{data fitting} evaluates pointwise accuracy via reconstruction errors, mode fitting reveals whether the decoder preserves underlying physical structures encoded in dominant spatial patterns. A decoder may achieve low reconstruction error while failing to capture the physically meaningful modes that govern system dynamics.

\subsection{Linear Transport Equation}\label{sec:numerics:pulse}

\subsubsection{Problem Setup}
Following~\cite{geelen2023Operator,schwerdtner2024Greedy}, we consider the one-dimensional linear transport equation
\begin{equation}
    \frac{\partial}{\partial t}x(\xi,t) + c\frac{\partial}{\partial\xi}x(\xi,t) = 0, 
    \quad \xi \in \R,
\end{equation} 
with spatial coordinate \( \xi \), time \( t \in [0, T] \), and constant advection velocity \( c \). The exact solution is \( x(\xi, t) = x_0(\xi - ct) \), where \( x_0 \) is a Gaussian bump initial condition:
\begin{displaymath}
    x_0 = x(\xi, 0) = \frac{1}{\sqrt{0.0005\pi}}\exp\left( -\frac{{(\xi - 0.1)}^2}{0.0005} \right).
\end{displaymath}
We define the spatial domain \( \xi \in [0, 1] \) with \( 1024 \) grid points and time domain \( t \in [0, 0.15] \) with \( 10^3 \) instances, yielding data dimension \( d = 1024 \) and \( n = 1000 \) snapshots. The advection velocity is set to \( c = 5 \). For mode selection, we use the leading \( s = 100 \) modes as candidates and select \( r = 15 \) modes. 

\subsubsection{Results}

\begin{figure}[htbp!]
    \centering
    \includegraphics[width=\textwidth]{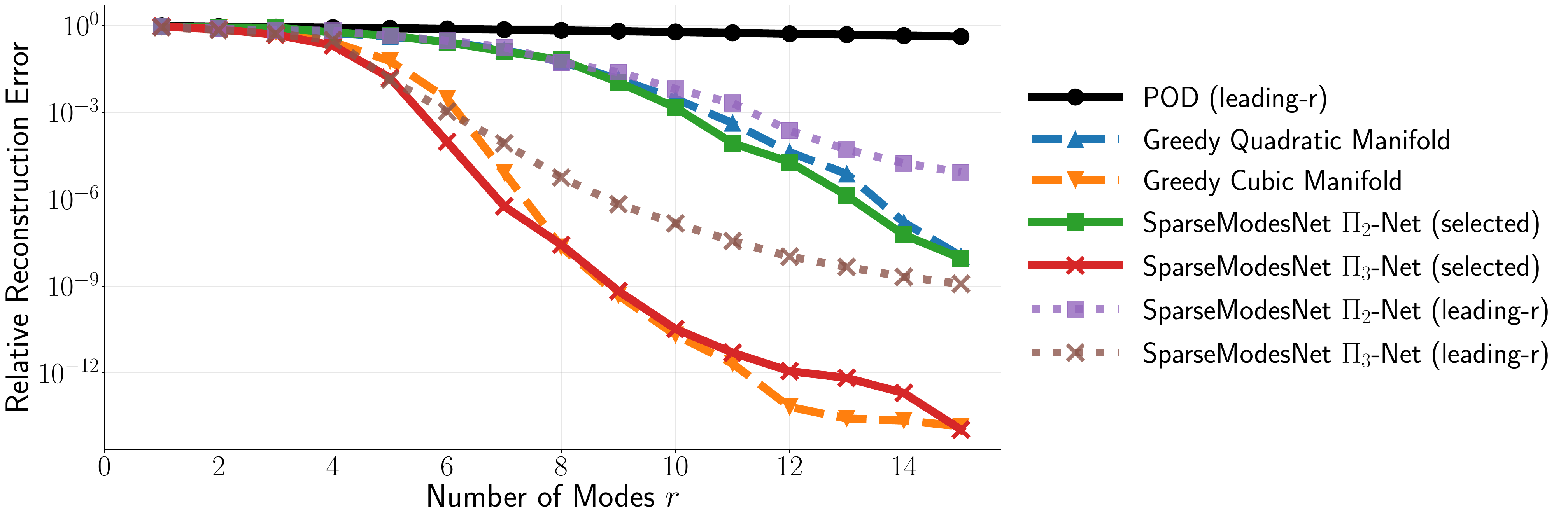} 
    \vspace{-2em}
    \caption{Relative reconstruction errors for the linear transport with different decoder structures.}\label{fig:pulse_recon_errors}
\end{figure}

\Cref{fig:pulse_recon_errors} presents relative reconstruction errors versus mode count for various decoder architectures. Standard \gls{pod} with leading-\( r \) modes (black) exhibits a persistent error plateau at \( 10^{-1} \), demonstrating that linear combinations of energetic modes cannot capture the nonlinear solution manifold structure.

SparseModesNet's \( \Pi_3 \)-Net achieves relative reconstruction error of \( 10^{-14} \) at \( r=15 \) compared to \( 10^{-8} \) for \( \Pi_2 \)-Net, amounting to an \(O(10^6)\) improvement. Comparing selected (solid) versus leading-r (dotted) modes shows learned mode selection is critical: \( \Pi_3 \)-Net with selected modes outperforms leading modes by \(O(10^5)\) at \( r=15 \). Greedy quadratic and cubic manifold approaches show similar performance to \(\Pi_2\)-Net and \(\Pi_3\)-Net, respectively, confirming polynomial decoders are effective and our method is competitive with state-of-the-art greedy methods. However, while greedy cubic manifold requires output dimension \( p = r(r+1)(r+2)/6 = 680 \) for \( r=15 \) (due to 680 total cubic monomial combinations), our \(\Pi_3\)-Net maintains fixed \( p = 400 \), achieving \textbf{40\% fewer} output dimensions (see~\Cref{tab:parameters_summary}).

\begin{figure}[t!]
    \centering
    \includegraphics[width=\textwidth]{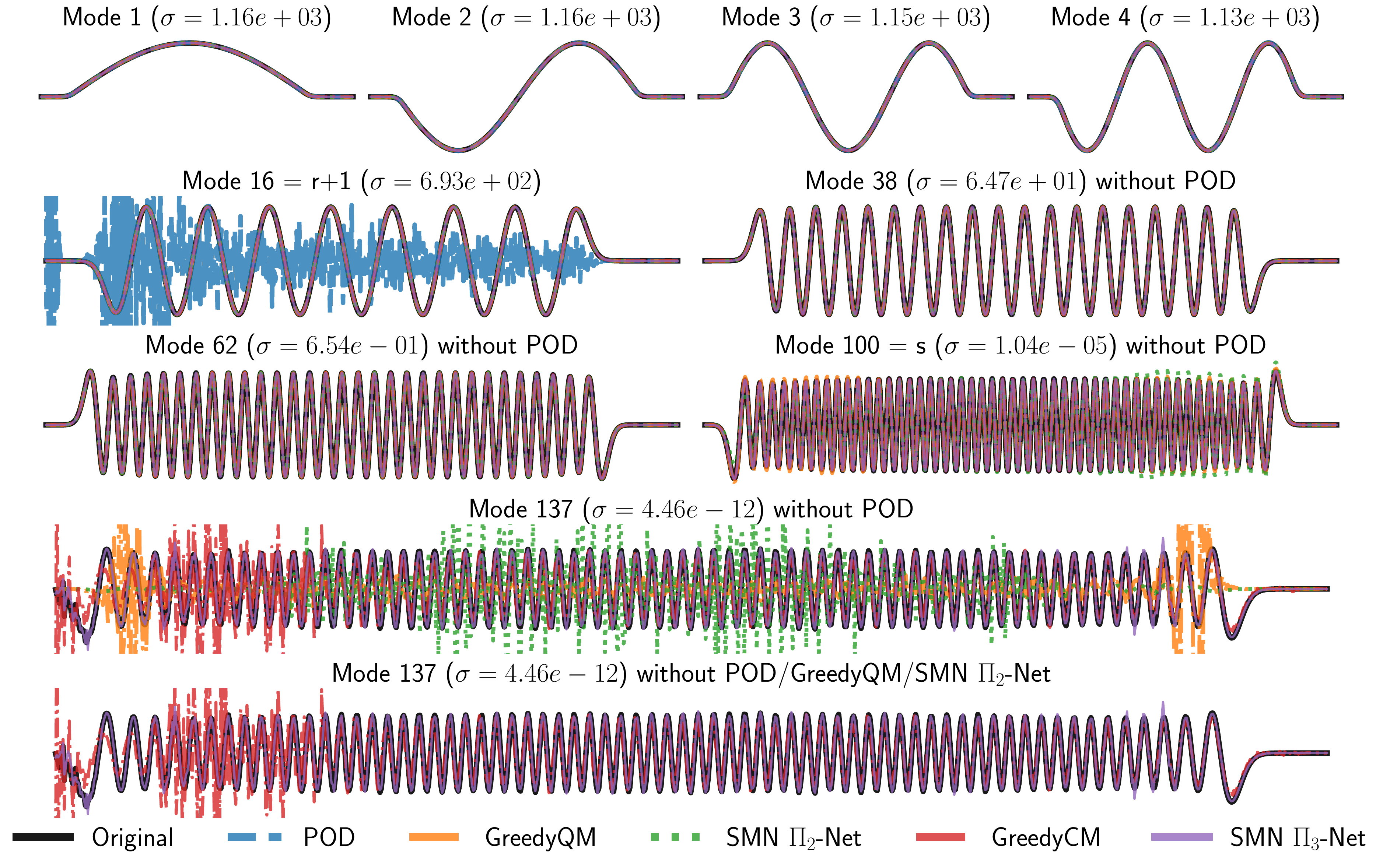}
    \vspace{-2em}
    \caption{Mode-fitting by different decoders for the linear transport, spanning modes within the leading-\(r\), within seen \(s\) candidates, and extrapolation beyond \(s\) to assess decoder interpretability. ``SMN'' indicates SparseModesNet. Once the method fails, it is excluded in higher modes.}\label{fig:pulse_modes}
\end{figure}

\Cref{fig:pulse_modes} compares POD modes of the reconstructed data for mode fitting across three regimes: leading-\(r\) (1--4), seen training candidates (16, 38, 62, 100), and extrapolation beyond training (mode 137). This linear transport equation exhibits structured wave patterns across the entire energy spectrum. SparseModesNet with \(\Pi_3\)-Net accurately reconstructs modes across all regimes, including mode 137 beyond the \(s=100\) training candidates, confirming the decoder captures underlying advection physics rather than memorizing data. Standard POD fails beyond leading-\(r\) modes, while greedy approaches and \(\Pi_2\)-Net show increasing extrapolation errors.For further qualitative assessment, we provide state reconstruction analysis in~\Cref{sec:sup:pulse:reconstructions}.

\Cref{tab:selected_modes} reveals distinct mode selection patterns. Analyzing the final selected modes are crucial as they provide further insight of the physical meaning of the selected modes rather than merely the successful sparsification of modes. The \( \Pi_3 \)-Net selects consecutive odd modes \( [1, 3, 5, 7, 9, \ldots, 19] \) initially, followed by higher-frequency modes \( [29, 24, \ldots, 42] \), suggesting low-frequency modes capture primary wave structure while selected high-frequency modes represent fine-scale features. The \( \Pi_2 \)-Net shows similar preference for odd modes with different high-frequency selection, indicating cubic architecture enables different mode utilization. In contrast, greedy methods exhibit less structured patterns, with Greedy Cubic Manifold notably including mode 92 early in greedy selection. Nevertheless, even though the SparseModesNet-selected modes are ordered by decreasing \( \omega \) magnitude rather than greedy selection order, it is notable that they share similar selection patterns. In~\Cref{sec:sup:pulse:mode-selection}, we further visualize the automated mode selection process during training.

These results show how SparseModesNet outperforms linear approaches and competes with state-of-the-art greedy methods. The combination of mode selection and cubic nonlinearity achieves machine-precision reconstruction with only 15 modes, providing computational efficiency for dimensionality reduction.

\begin{table}[htbp!]
    \centering
    \caption{Selected modes for each decoder in the linear transport equation, ordered by greedy selection order (GreedyQM/CM) or decreasing \( \omega \) magnitude (SparseModesNet).}\label{tab:selected_modes}
    \resizebox{0.8\columnwidth}{!}{%
    \begin{tabular}{c c}
        \toprule
        Decoder & Selected Modes \\
        \midrule
        GreedyQM  &              [1, 3, 5, 7, 4, 12, 10, 20, 28, 38, 15, 48, 23, 68, 75]\\
        GreedyCM  &              [2, 1, 3, 92, 4, 8, 5, 7, 29, 48, 69, 42, 70, 86, 18]\\
        SparseModesNet (\(\Pi_2\)-Net) &  [1, 3, 5, 7, 9, 13, 17, 33, 11, 31, 27, 29, 26, 47, 24] \\
        SparseModesNet (\(\Pi_3\)-Net) &  [1, 3, 5, 9, 13, 15, 17, 19, 29, 24, 22, 26, 18, 31, 42] \\
        \bottomrule
    \end{tabular} 
    }
\end{table}

\subsection{Kuramoto-Sivashinsky Equation}\label{sec:numerics:kse}

\subsubsection{Problem Setup}
The Kuramoto-Sivashinsky equation~\cite{kuramoto1978diffusion,sivashinksy1977nonlinear} is a one-dimensional nonlinear partial differential equation:
\begin{equation}
    \frac{\partial}{\partial t}x(\xi,t) + \frac{\partial^2}{\partial\xi^2}x(\xi,t) + \frac{\partial^4}{\partial\xi^4}x(\xi,t) + x(\xi,t)\frac{\partial}{\partial\xi}x(\xi,t) = 0.
\end{equation}
We use spatiotemporal domain \( [0, 32\pi] \times [0, 100] \) discretized with \( 1024 \) gridspaces and \( 2500 \) time instances, yielding \( d = 1024 \) and \( n = 2500 \). Due to the chaotic nature of this equation, a larger number of modes is required for accurate flow field reconstruction. 
As in the linear transport experiment, we use \( s = 100 \) candidate modes and select \( r = 15 \) modes. We assess performance through relative reconstruction error~\eqref{eqn:rel-recon-error} and compare mode fitting across the same decoder structures from the previous example.

\subsubsection{Results}

\begin{figure}[t!]
    \centering
    \includegraphics[width=\textwidth]{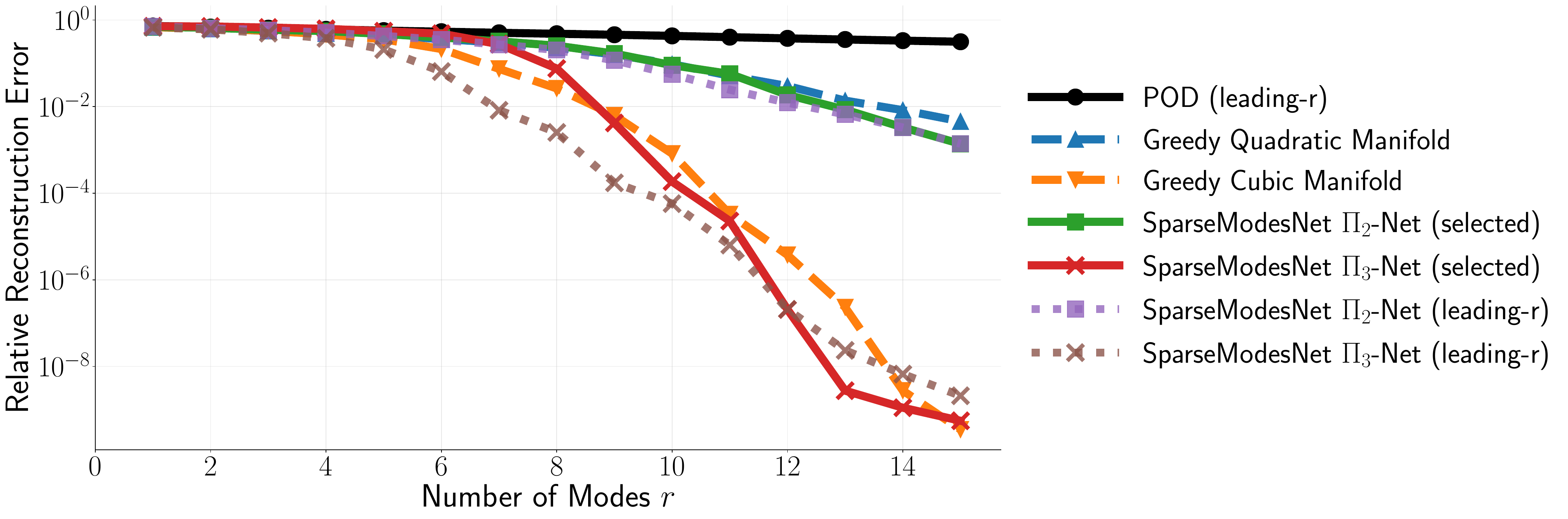} 
    \vspace{-2em}
    \caption{Relative reconstruction errors for \gls{kse} using different decoder structures.}\label{fig:kse_recon_errors}
\end{figure}

\Cref{fig:kse_recon_errors} presents relative reconstruction errors for \gls{kse}. While standard \gls{pod} with leading-\( r \) modes shows poor performance near \( 10^{-1} \) to \( 10^0 \), the performance gap between selected and leading modes is smaller than for linear transport. SparseModesNet with \( \Pi_3 \)-Net achieves approximately \( 10^{-9} \) relative error at \( r=15 \) for both selection strategies, contrasting the \(O(10^5)\) difference in the linear transport case. This suggests the \gls{kse} solution manifold is less sensitive to mode selection due to chaotic, diffusion-dominated dynamics. The \( \Pi_2 \)-Net achieves approximately \( 10^{-3} \) error at \( r=15 \), showing higher-order polynomial nonlinearity remains important. Greedy manifold approaches exhibit performance comparable to corresponding \( \Pi \)-Net architectures.

This reduced sensitivity to mode selection stems from two factors. First, chaotic \gls{kse} dynamics distribute energy more broadly across frequencies than coherent wave structures in linear transport, reducing the distinction between optimal and suboptimal mode choices. Second, joint training of the \gls{nn} decoder with mode selection weights allows the decoder to adapt its nonlinear mapping to work effectively with selected modes, extracting relevant information even when mode choice is energetically suboptimal. This suggests nonlinear mapping quality may be as important as mode selection for turbulent systems, highlighting the value of expressive \gls{nn} decoders.

\begin{figure}[t!]
    \centering
    \includegraphics[width=\textwidth]{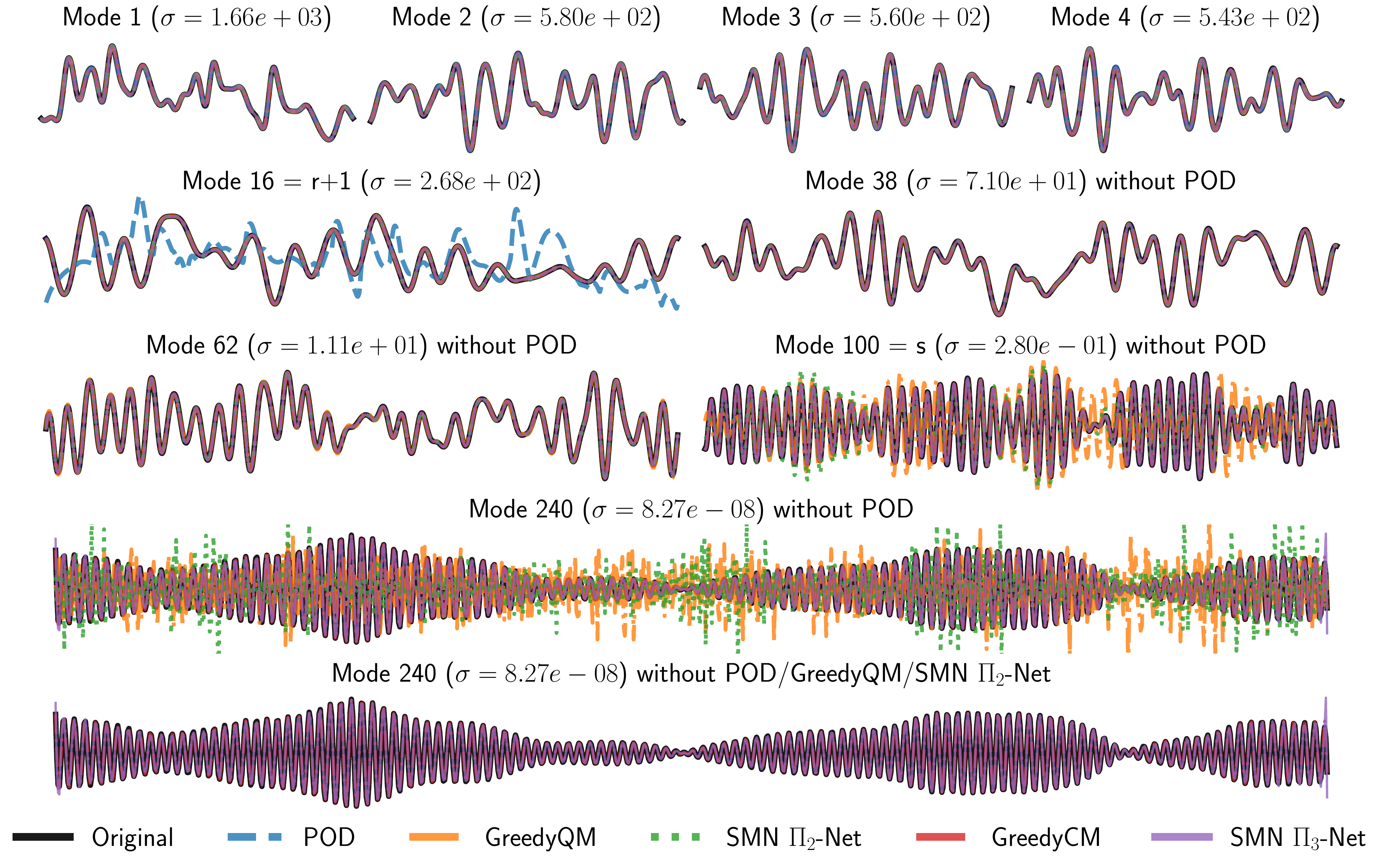}
    \vspace{-2em}
    \caption{Mode fitting by different decoders for \gls{kse}, spanning modes within leading-\(r\), within seen \(s\) candidates, and extrapolation beyond \(s\) to assess decoder interpretability. ``SMN'' indicates SparseModesNet. Once the method fails, it is excluded in higher modes.}\label{fig:kse_modes}
\end{figure}

\Cref{fig:kse_modes} compares reconstructed POD modes across three regimes: leading-\(r\) (1--4), seen training candidates (16, 38, 62, 100), and extrapolation (mode 240). Unlike linear transport, \gls{kse} shows high-energy modes (1--4) with chaotic, unstructured patterns while low-energy modes (100, 240) display well-defined periodic oscillations, reflecting how energy concentrates in large-scale chaotic motions while coherent structures persist at lower energies. SparseModesNet with \(\Pi_3\)-Net accurately reconstructs modes across all regimes, including mode 240 beyond the \(s=100\) training set, confirming it captures underlying chaotic dynamics. Greedy cubic manifold shows comparable performance consistent with \Cref{fig:kse_recon_errors}, while \(\Pi_2\)-Net and greedy quadratic manifold exhibit larger extrapolation errors. Standard \gls{pod} fails beyond leading-\(r\) modes. Successful reconstruction of structured low-energy modes demonstrates SparseModesNet maintains interpretability when capturing energetically subdominant features. For state reconstruction, we provide additional analysis in~\Cref{sec:sup:kse:reconstructions}.

\Cref{tab:kse_selected_modes} shows mode selection patterns differing from linear transport. The \( \Pi_3 \)-Net selects predominantly higher-frequency modes, with modes 60 and 56 most important, followed by gradual transition to lower frequencies \([18, 30, 19, \ldots, 3, 2]\), contrasting with the low-frequency odd mode dominance in linear transport. This higher-frequency preference reflects chaotic dynamics where energy cascades across scales and fine-scale structures are crucial. The \( \Pi_2 \)-Net shows mixed pattern from \([11, 6, 3, 16, \ldots] \) to \([\ldots, 10, 7, 2, 23]\), while greedy methods favor sequential low-frequency modes. That SparseModesNet achieves comparable accuracy with both selected and leading modes despite dramatically different indices underscores the \gls{nn} decoder's ability to learn effective nonlinear mappings compensating for non-optimal mode selection---valuable for turbulent systems where optimal mode identification is challenging. In~\Cref{sec:sup:kse:mode-selection}, we further visualize the automated mode selection process during training.

\begin{table}[t!]
    \centering
    \caption{Selected modes for each decoder in \gls{kse}, ordered by greedy selection order (GreedyQM/CM) or decreasing \( \omega \) magnitude (SparseModesNet).}\label{tab:kse_selected_modes}
    \resizebox{0.8\columnwidth}{!}{%
    \begin{tabular}{c c}
        \toprule
        Decoder & Selected Modes \\
        \midrule
        GreedyQM  &              [1, 2, 4, 3, 5, 6, 7, 10, 9, 8, 16, 13, 12, 14, 11]\\
        GreedyCM  &              [1, 2, 3, 5, 4, 7, 9, 10, 6, 13, 14, 12, 82, 92, 81]\\
        SparseModesNet \(\Pi_2\)-Net &  [11, 6, 3, 16, 5, 4, 8, 13, 1, 12, 18, 10, 7, 2, 23]\\
        SparseModesNet \(\Pi_3\)-Net &  [60, 56, 18, 30, 19, 14, 12, 10, 13,  9,  8,  5,  6,  3,  2]\\
        \bottomrule
    \end{tabular}
    }
\end{table}

These results demonstrate that while SparseModesNet maintains strong performance for chaotic \gls{kse}, achieving approximately \( 10^{-9} \) reconstruction error with 15 modes, learned sparse mode selection benefits are less pronounced than for transport-dominated problems. This result also suggests the possibility that linear encoding itself may be insufficient for \gls{kse}. Comparable performance of selected and leading modes suggests that for diffusion-dominated turbulent systems, nonlinear decoder mapping quality may be more critical than precise mode selection, favoring expressive \gls{nn} decoders. This result also confirms the energy heuristic is not optimal and alternative mode combinations could improve reconstruction.

\subsection{Turbulent Channel Flow}\label{sec:numerics:channel}

\begin{figure}[b!]
    \centering
    {\small\textbf{(a) Streamwise Velocity Component, \( u_1 \)}} \\[-0.1em]
    \includegraphics[width=0.495\textwidth]{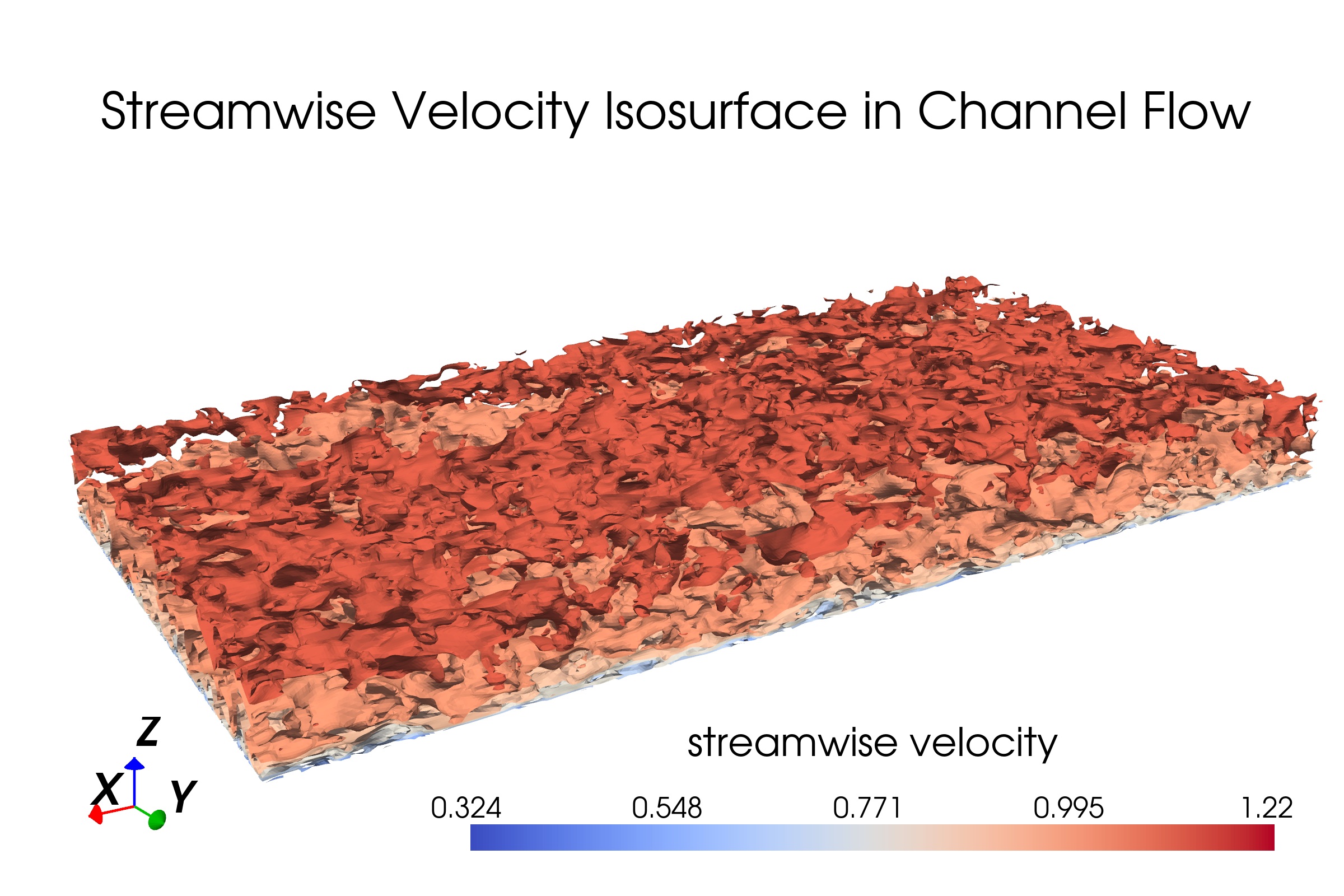} 
    \includegraphics[width=0.495\textwidth]{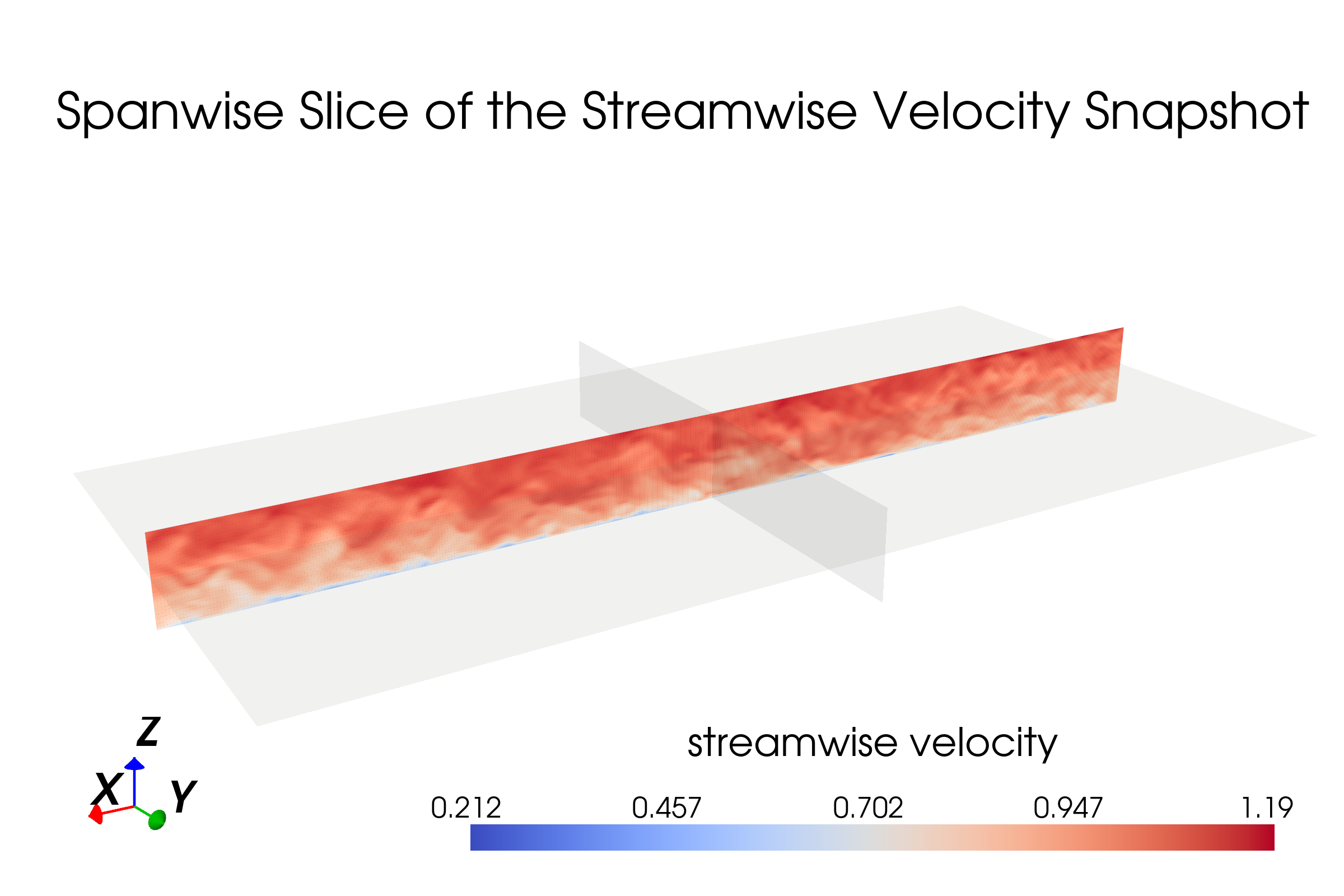} \\[0.2em]
    {\small\textbf{(b) Wall-Normal Velocity Component, \( u_3 \)}} \\[-0.1em]
    \includegraphics[width=0.495\textwidth]{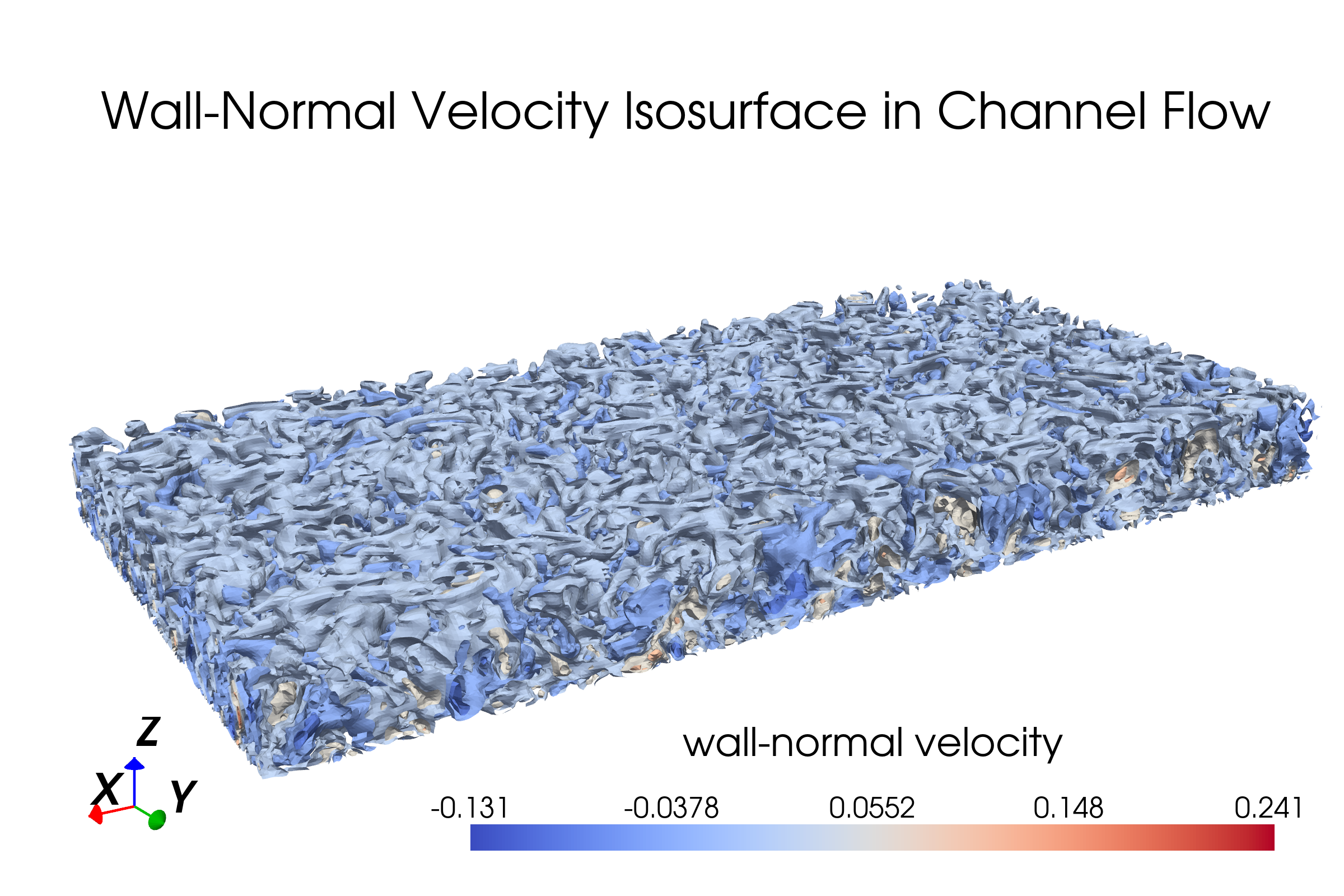} 
    \includegraphics[width=0.495\textwidth]{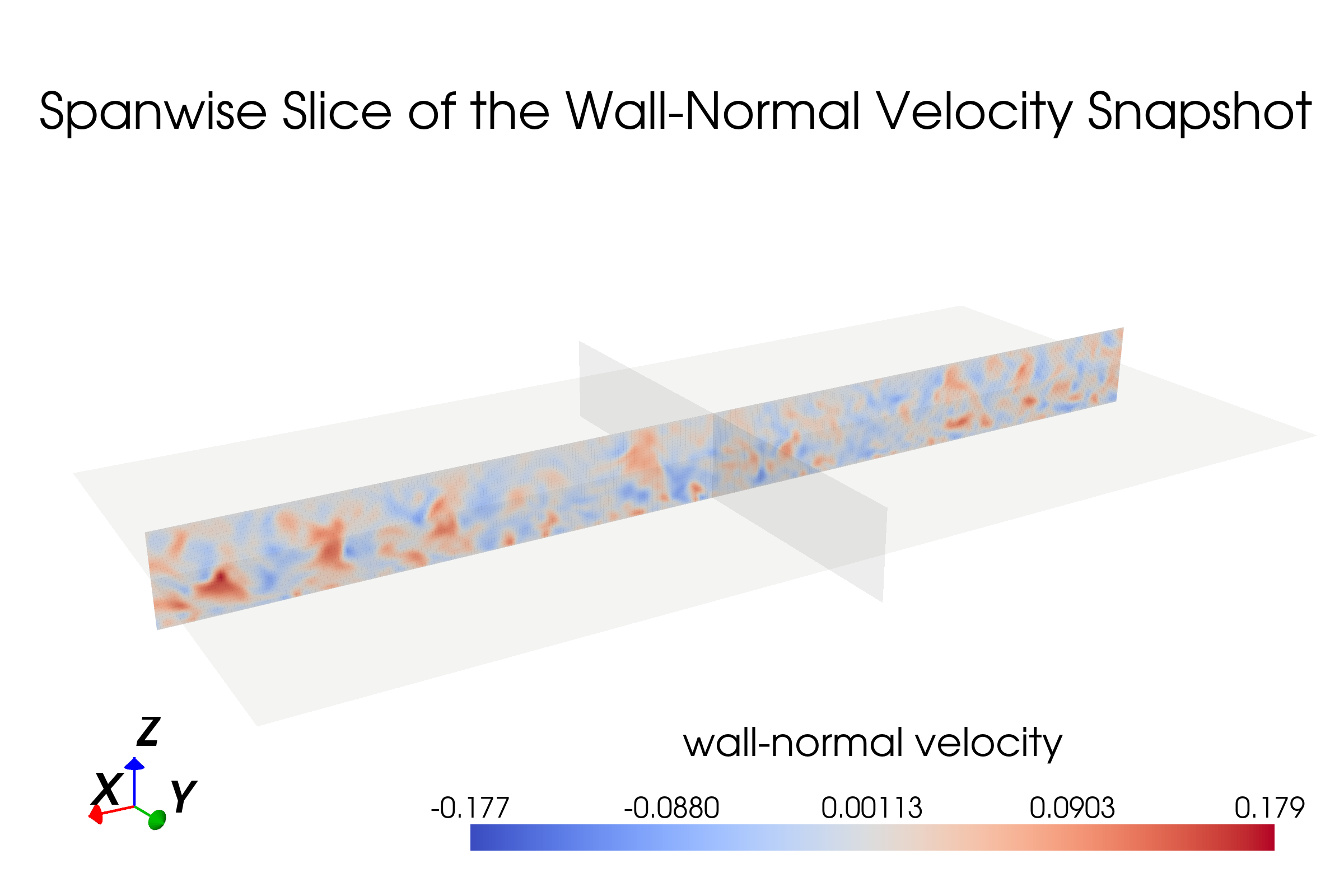} 
    \caption{(\textbf{Left}) Isosurface and (\textbf{right}) spanwise sliced flow field at a time instant for (a) streamwise and (b) wall-normal velocity components.}\label{fig:channel-surface-flow}
\end{figure}

\subsubsection{Problem Setup}
Following~\cite{schwerdtner2024Greedy,schwerdtner2025Online,koike2026streaming}, we consider turbulent channel flow data simulated using the AMR-Wind solver~\cite{sharma2024ExaWind,kuhn2025AMRWind}. Turbulent channel flow is common in engineering applications such as turbomachinery and vehicles~\cite{bredberg2000wall}. AMR-Wind solves the incompressible \gls{ns} equations in \gls{les} formulation. Using \( (\,\ot{\cdot}\,) \) to denote spatial filtering, the governing equations in Cartesian coordinates with Einstein summation convention are
\begin{align}
    \frac{\partial \ot{u_j}}{\partial \xi_j} & = 0,\label{eqn:ns-les-cont}\\
    \frac{\partial \ot{u_i}}{\partial t} +
    \frac{\partial \ot{u_i} \ot{u_j}}{\partial \xi_j} &=
    - \frac{1}{\rho}{\frac{\partial \ot{p}}{\partial \xi_i}}
    - {\frac{\partial \tau_{ij}}{\partial \xi_j} }
    + \nu \frac{\partial^2 \ot{u_i}}{\partial \xi_j \partial \xi_j}
    + {F_{i}},\label{eqn:ns-les-mom}
\end{align}
where \( \xi_i \) denotes spatial coordinate in direction \( i = \{1,2,3\} \), \( \ot{u_i} \) is the spatially-filtered velocity field, \( \ot{p} \) is pressure, \( \nu \) is kinematic viscosity, \( \rho \) is fluid density, and \( \tau_{ij} \) is the subgrid-scale stress tensor modeling unresolved turbulent scales: \( \tau_{ij} = \ot{u_i u_j} - \ot{u_i}\ot{u_j} \). The body force \( F_i \) maintains the flow, where the three coordinate directions represent streamwise, spanwise, and wall-normal components, respectively.

We simulate turbulent channel flow at friction Reynolds number \( Re_{\tau} = 5200 \) in domain \( 12\pi \delta \times 6\delta \times 1\delta \), where \( \delta \) is the channel half-width. The domain is discretized on a finite-volume grid of size \( 384 \times 192 \times 32 \). We collect \( n = 1000 \) snapshots after the flow reaches statistical stationarity and slice the data at the spanwise centerline to obtain two-dimensional velocity fields of size \( d = 384 \times 32 = 12288 \). We individually analyze streamwise (\( u_1 \)) and wall-normal (\( u_3 \)) velocity components, evaluating SparseModesNet on a larger-scale turbulent example where wall-normal velocity is more advection-dominated than streamwise velocity.

\begin{figure}[t!]
    \centering
    \includegraphics[width=0.85\textwidth]{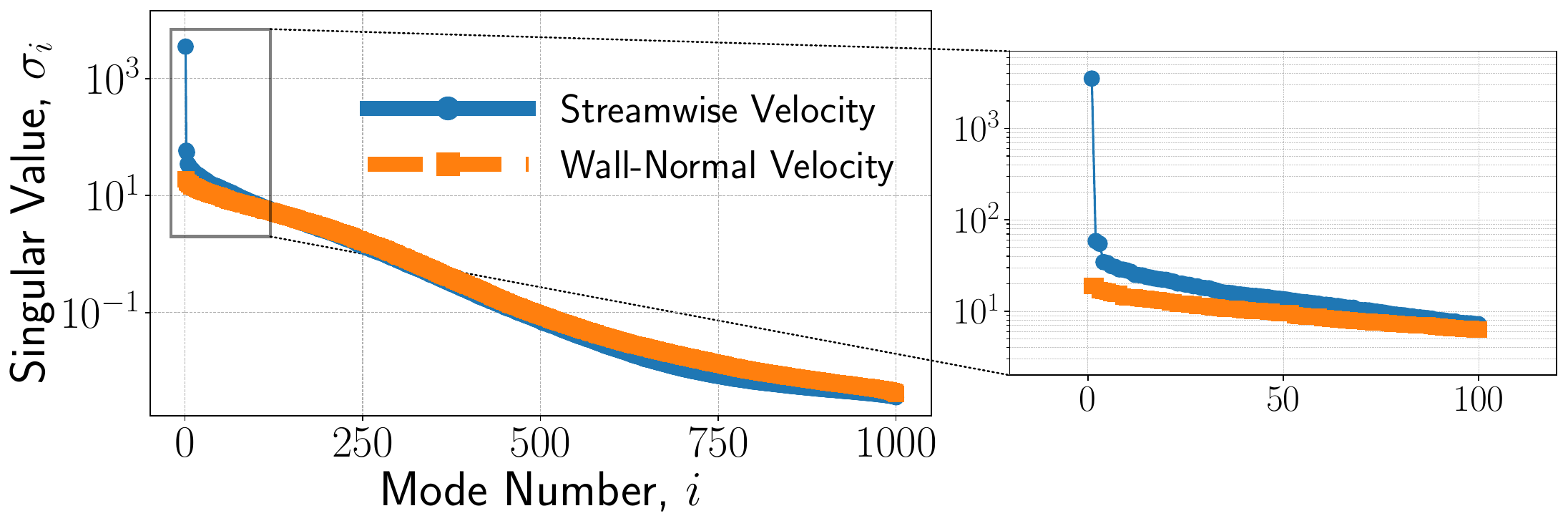} 
    \vspace{-1em}
    \caption{(\textbf{Left}) Spectral decay of \gls{pod} modes for turbulent channel flow for both velocity components and (\textbf{right}) zoomed view of leading 100 modes.}\label{fig:channel-spectral-decay}
\end{figure}

\Cref{fig:channel-surface-flow} visualizes the turbulent channel flow using isosurfaces and spanwise slices for both velocity components at a time instant. Intricate turbulent structures in both components highlight the challenges in reconstructing such high-dimensional, nonlinear flow fields. \Cref{fig:channel-spectral-decay} shows the spectral decay of \gls{pod} modes for both velocity components, including a zoomed view of the leading 100 modes. The slow energy decay across modes indicates many modes are necessary to capture flow energy for linear reconstruction methods, emphasizing the problem's difficulty. 
For both velocity components, the data matrix is \( \Xmat \in \R^{12288 \times 1000}\) with \( s = 100 \) candidate and \( r = 15 \) selected modes. As in previous examples, we evaluate relative reconstruction error~\eqref{eqn:rel-recon-error} and mode fitting for standard POD with leading-\( r \) modes, SparseModesNet with \( \Pi_3 \)-Net, and Greedy Quadratic and Cubic Manifold approaches.

\subsubsection{Results: Streamwise and Wall-Normal Velocity Components}

\begin{figure}[t!]
    \centering
    \includegraphics[width=0.49\textwidth]{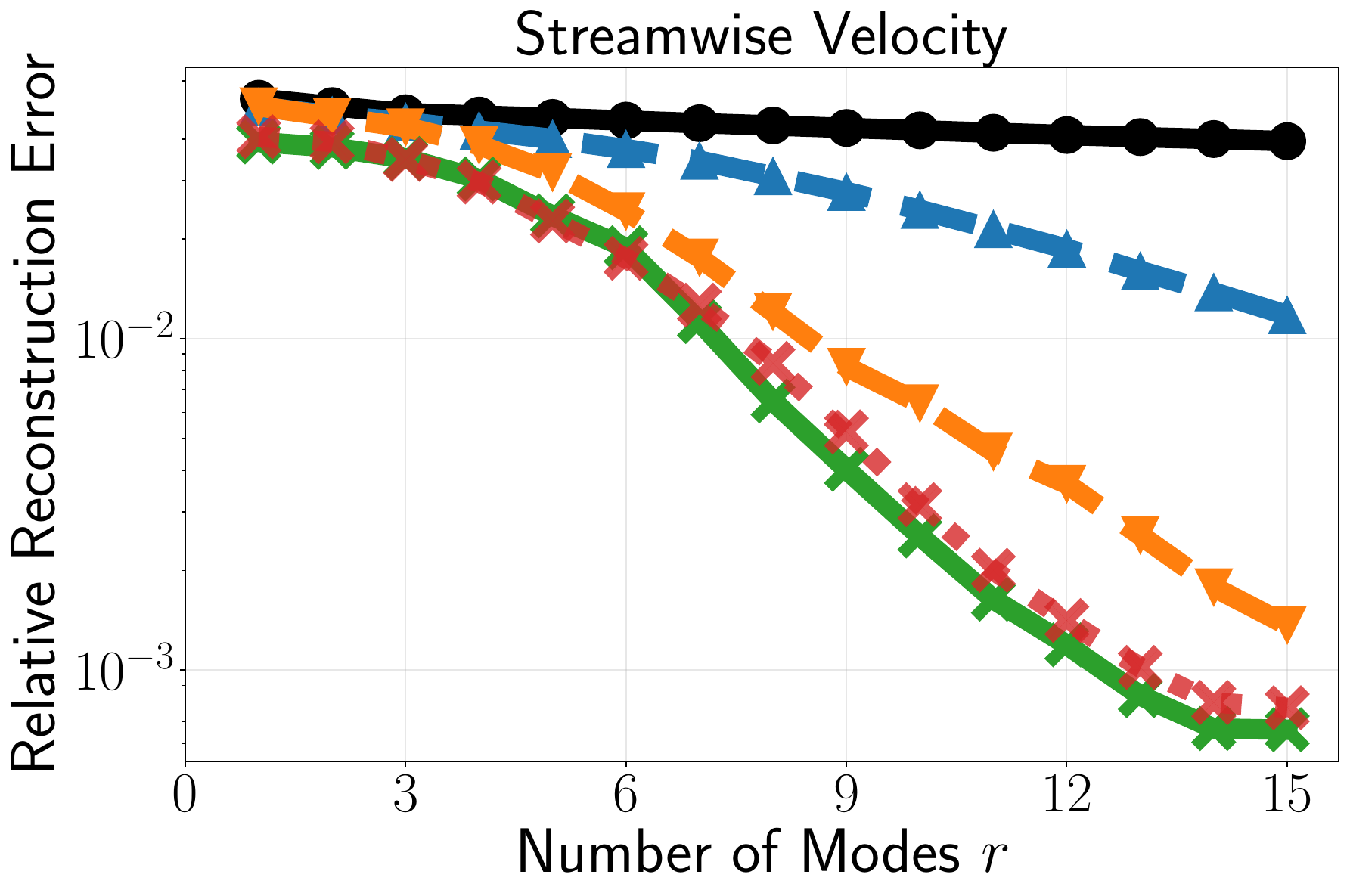} 
    \includegraphics[width=0.49\textwidth]{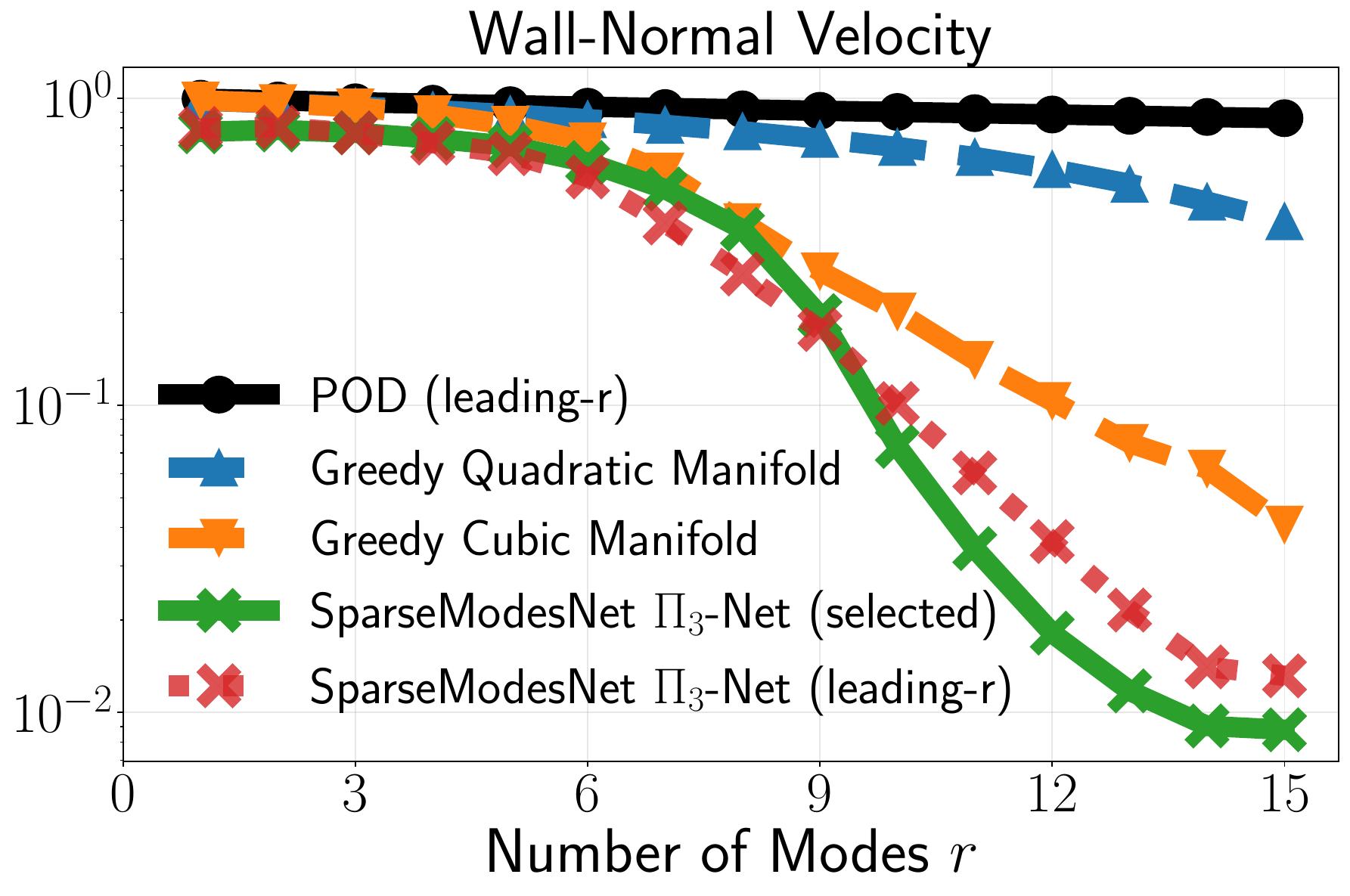} 
    \caption{Relative reconstruction errors for (\textbf{left}) streamwise velocity and (\textbf{right}) wall-normal velocity of the turbulent channel flow using different decoders.}\label{fig:channel_recon_errors}
\end{figure}

The top plot of \Cref{fig:channel_recon_errors} presents relative reconstruction errors for the streamwise velocity component. SparseModesNet with \( \Pi_3 \)-Net achieves superior performance compared to all baselines. Standard POD with leading-\( r \) modes plateaus at approximately \( 4 \times 10^{-2} \), confirming linear methods are inadequate for this high-Reynolds-number flow. Greedy Quadratic Manifold achieves error slightly above \( 1 \times 10^{-2} \) at \( r=15 \), while Greedy Cubic Manifold improves to approximately \( 1.4 \times 10^{-3} \). SparseModesNet with \( \Pi_3 \)-Net achieves approximately \( 6.6 \times 10^{-4} \) at \( r=15 \), representing \textbf{51\% reduction} versus Greedy Cubic Manifold and over \textbf{94\% reduction} versus Greedy Quadratic Manifold. Similar to \gls{kse}, the performance gap between selected and leading modes is minimal (\( 7.7 \times 10^{-4} \) at \( r=15 \)), suggesting that for large-scale turbulent systems with less advection dominance, the \gls{nn} decoder's ability to learn effective nonlinear mappings is as critical as precise mode selection.

The bottom plot of \Cref{fig:channel_recon_errors} presents reconstruction errors for the wall-normal velocity component, revealing similar trends. The wall-normal velocity exhibits more intricate, smaller-scale turbulent structures (see~\Cref{fig:channel-surface-flow}). Despite this complexity, SparseModesNet achieves approximately \( 8.7 \times 10^{-3} \) at \( r=15 \), outperforming Greedy Cubic Manifold by \textbf{78\%} and Greedy Quadratic Manifold by \textbf{97\%}. For wall-normal velocity, SparseModesNet with mode selection achieves \textbf{33\% error reduction} versus leading \( r \) modes, indicating that for the more advection-dominated component, careful mode selection provides additional benefits beyond \gls{nn} decoder expressivity.

\begin{figure}[htbp!]
    \centering
    \includegraphics[width=0.85\textwidth]{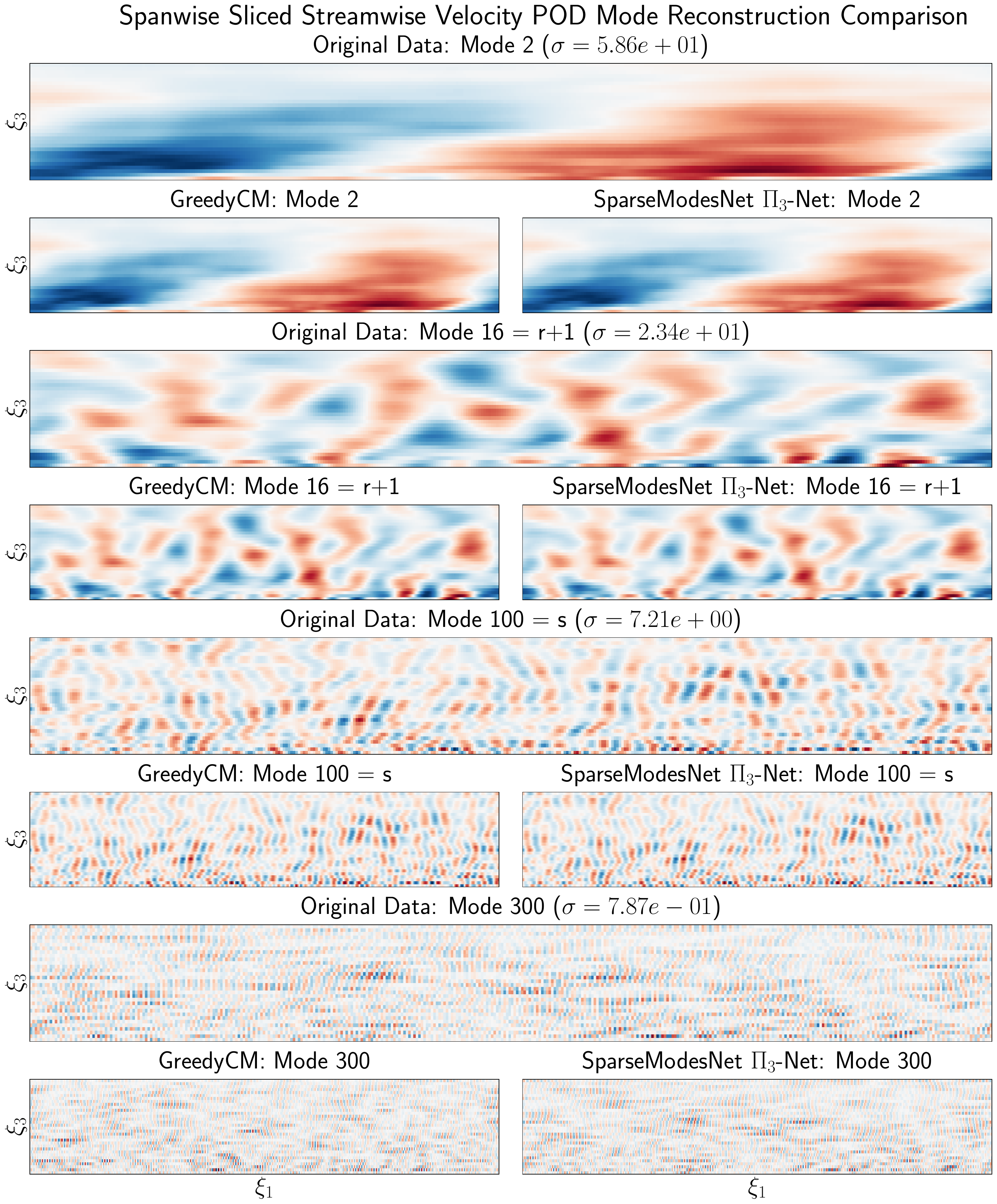} 
    \caption{Mode fitting by different decoders for streamwise velocity, spanning modes within leading-\(r\), within seen \(s\) candidates, and extrapolation beyond \(s\) to assess interpretability.}\label{fig:channel_u_modes}
\end{figure}

\begin{figure}[htbp!]
    \centering
    \includegraphics[width=0.85\textwidth]{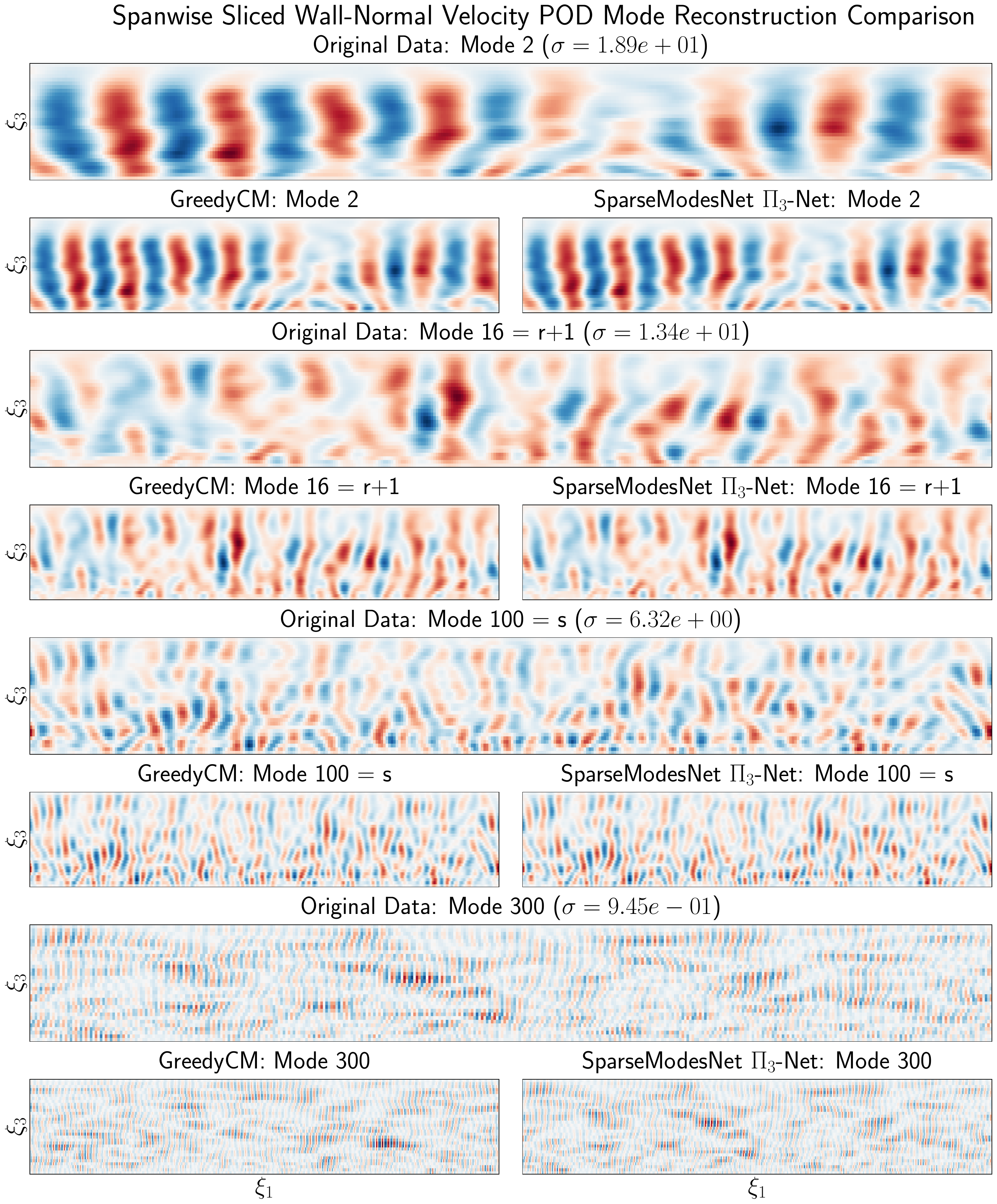}
    \caption{Mode fitting by different decoders for wall-normal velocity, spanning modes within leading-\(r\), within seen \(s\) candidates, and extrapolation beyond \(s\) to assess interpretability.}\label{fig:channel_w_modes}
\end{figure}

\Cref{fig:channel_u_modes,fig:channel_w_modes} compare reconstructed \gls{pod} modes for both velocity components across mode 2, mode 16 (r+1), mode 100 (s), and mode 300 (extrapolation). Both components exhibit progression from large-scale coherent structures in leading modes to fine-scale periodic patterns in lower-energy modes, characteristic of wall-bounded turbulence where energy cascades across scales. SparseModesNet with \(\Pi_3\)-Net accurately reconstructs modes across all regimes for both components, including mode 300 beyond the \(s=100\) training set, demonstrating the decoder captures underlying turbulent physics. Greedy cubic manifold exhibits similar reconstruction quality. For additional qualitative assessments, we provide comparisons of flow field reconstructions in~\Cref{sec:sup:channel:reconstructions}.

\Cref{tab:channel_selected_modes} shows distinct mode selection patterns between velocity components, providing physical insight into turbulent channel flow structure. For streamwise velocity, SparseModesNet selects mixed distribution \( [1, 16, 18, 41, \ldots, 3, 29, 13, 44, 20, 6] \), beginning with the mean flow (mode 1) followed by low-to-intermediate frequency modes with some higher-frequency modes, reflecting characteristic large-scale coherent structures and elongated streaks. Wall-normal velocity selects predominantly higher-frequency modes \( [1, 51, 45, 55, \ldots, 60, 21, 12, 20] \), with six of the top seven non-mean modes in the 45--60 range. This higher-frequency preference corresponds physically to wall-normal velocity's role in turbulent transport perpendicular to the wall, characterized by smaller-scale ejection and sweep events dominating turbulent momentum transfer~\cite{wallace1972Wall,robinson1991Coherent}. Greedy methods select more sequential low-to-intermediate frequency modes, with Greedy Cubic Manifold occasionally selecting very high-frequency modes (e.g., modes 97, 98 for streamwise; mode 100 for wall-normal), potentially indicating sensitivity to initialization or local optima.

These results demonstrate SparseModesNet achieves state-of-the-art performance for dimensionality reduction of high-Reynolds-number turbulent channel flow, providing superior quantitative accuracy and qualitative reconstruction fidelity across velocity components. Achieving approximately \( 6.6 \times 10^{-4} \) or \( 8.7 \times 10^{-3} \) relative error with only 15 modes—over \textbf{51\% and 78\% error reduction} versus existing nonlinear approaches—represents significant advance in computational efficiency. Critically, SparseModesNet achieves this while using lower-dimensional nonlinear mapping (\( p = 400 \)) compared to generic cubic mapping (\( p = 680 \)) across all experiments (see~\Cref{tab:parameters_summary}), reducing inference costs. The comparable performance of selected versus leading modes suggests that for large-scale turbulent systems, investing computational effort in training high-quality nonlinear decoders may be more beneficial than exhaustive mode selection optimization, though learned mode selection can provide improved accuracy and valuable physical insights into multi-scale turbulent structure.

\begin{table}[htbp!]
    \centering
    \caption{Selected modes for each decoder in turbulent channel flow for (a) streamwise and (b) wall-normal velocities, ordered by greedy selection order (GreedyQM/CM) or decreasing \( \omega \) magnitude (SparseModesNet).}\label{tab:channel_selected_modes}
    {\small \textbf{(a) Streamwise Velocity Component, \( u_1 \)}}\\
    \resizebox{0.8\columnwidth}{!}{%
    \begin{tabular}{c c}
        \toprule
        Decoder & Selected Modes \\
        \midrule
        GreedyQM  & [1, 2, 5, 6, 9, 11, 8, 7, 3, 21, 25, 26, 19, 48, 54]\\
        GreedyCM  & [1, 2, 3, 4, 12, 13, 5, 9, 19, 27, 72, 78, 97, 98, 73]\\
        SparseModesNet \(\Pi_3\)-Net &  [1, 16, 18, 41, 17, 33, 12, 10, 32, 3, 29, 13, 44, 20, 6]\\
        \bottomrule
    \end{tabular}
    }
    \\[1em]
    {\small \textbf{(b) Wall-Normal Velocity Component, \( u_3 \)}}\\
    \resizebox{0.8\columnwidth}{!}{%
    \begin{tabular}{c c}
        \toprule
        Decoder & Selected Modes \\
        \midrule
        GreedyQM  & [1, 2, 9, 3, 8, 12, 5, 6, 30, 15, 27, 28, 4, 13, 7]\\
        GreedyCM  & [1, 2, 3, 6, 5, 4, 8, 18, 7, 9, 29, 56, 32, 65, 100]\\
        SparseModesNet \(\Pi_3\)-Net & [1, 51, 45, 55, 52, 36, 19, 46, 50, 47, 43, 60, 21, 12, 20]\\
        \bottomrule
    \end{tabular}
    }
\end{table}

\begin{table}[htbp!]
    \centering
    \caption{Summary of SparseModesNet hyperparameters and nonlinear mapping dimensions used for each numerical experiment.\  \( \lambda_0 \) is the initial \( \ell_1 \) penalty, \( \epsilon \) is the \( \ell_1 \) penalty's increment factor, \( M \) is the hierarchy constraint scaling parameter, and \( \gamma \) is the \(\ell_2\) penalty parameter for \( \Wmat^\star \) computation.}\label{tab:parameters_summary}
    \resizebox{\columnwidth}{!}{%
    \begin{tabular}{ccccccccc}
        \toprule
        & \multicolumn{4}{c}{Hyperparameters} & \multicolumn{4}{c}{Nonlinear Mapping \( h: \R^r \to \R^p \)} \\
        & \( \lambda_0 \) & \( \epsilon \) & \( M \) & \( \gamma \) & Quadratic & Cubic  & \( \Pi_2 \)-Net  & \( \Pi_3 \)-Net \\
        \midrule
        Linear Transport & \( 3.0 \) & \( 0.0005 \) & 12 & 1e--15 & \(\R^{15} \to \R^{120} \) & \(  \R^{15} \to \R^{680} \) & \(  \R^{15} \to \R^{225} \) & \(  \R^{15} \to \R^{400} \) \\
        \gls{kse} & \( 3.0 \) & \( 0.01 \) & 12 & 1e--15 & \(\R^{15}\to\R^{120}\) & \(\R^{15}\to\R^{680}\) & \(\R^{15}\to\R^{300}\) & \(\R^{15}\to\R^{680}\) \\
        Channel Flow \(u_1,u_3\) & \( 10 \) & \( 0.1 \) & 12 & 1e--15 & \(\R^{15}\to\R^{120}\) & \(\R^{15}\to\R^{680}\) & \longdash[4] & \(\R^{15}\to\R^{624}\) \\
        % Channel Flow \(u_3\) & \( 10 \) & \( 0.1 \) & 12 & \(\R^{15}\to\R^{120}\) & \(\R^{15}\to\R^{680}\) & \longdash[4] & \(\R^{15}\to\R^{624}\) \\
        \bottomrule
    \end{tabular}
    }
\end{table}

\section{Conclusion}\label{sec:conclusion}

In this work, we introduced SparseModesNet, a novel \gls*{nn} architecture that automates the selection of sparse, interpretable \gls*{pod} modes for nonlinear manifold learning and dimensionality reduction. By integrating LassoNet's hierarchical \(\ell_1\)-regularization strategy with a nonlinear manifold decoder, SparseModesNet simultaneously learns low-dimensional representations while enforcing sparsity in the selected modes. This approach identifies a small subset of the most relevant modes from a larger candidate set while minimizing reconstruction error.

Our numerical experiments across three problems with increasing complexity, the linear transport equation, \gls*{kse}, and turbulent channel flow, demonstrated that SparseModesNet outperforms traditional energy-based mode selection and state-of-the-art greedy algorithms in reconstruction accuracy. The SparseModesNet with \(\Pi_3\)-Net architecture achieves 50--90\% reduction in reconstruction error compared to greedy manifold methods for the turbulent channel flow while selecting different modes, confirming that energy content is not an optimal criterion for mode selection in turbulent systems. SparseModesNet accurately reconstructs not only the candidate \gls*{pod} modes within the training set but also extrapolates to unseen modes beyond the candidate set. This extrapolation confirms the interpretability of the learned \gls*{nn} decoder through its ability to capture underlying physics via physically meaningful \gls*{pod} modes, rather than merely memorizing training data. Results highlight the potential of SparseModesNet to facilitate more accurate, interpretable, and efficient reduced-order models across diverse scientific and engineering applications. 

Future work could pursue several promising directions. First, integrating SparseModesNet with operator learning methods such as Operator Inference (OpInf)~\cite{peherstorfer2016Datadriven,kramer2024Learningb,parish2026NNOpInf,codega2026Machine} could enable joint learning of encoder-decoder pairs with \glspl*{rom}. Second, incorporating physical constraints directly into the mode selection and manifold learning process through physics-informed neural networks~\cite{raissi2019Physicsinformeda} could further enhance interpretability and generalization. Finally, we can extend SparseModesNet to other modal bases, such as wavelet bases~\cite{tilki2026wavelet}, to assess its versatility.
\section*{Acknowledgements}
The authors were supported by the Department of Energy Office of Science Advanced Scientific Computing Research, DOE Award DE-SC0024721. 
This work was authored in part by National Laboratory of the Rockies (NLR) for the U.S. Department of Energy (DOE), operated under Contract No. DE-AC36-08GO28308. The views expressed in the article do not necessarily represent the views of the DOE or the U.S. Government. The U.S. Government retains and the publisher, by accepting the article for publication, acknowledges that the U.S. Government retains a nonexclusive, paid-up, irrevocable, worldwide license to publish or reproduce the published form of this work, or allow others to do so, for U.S. Government purposes. A portion of the research was performed using computational resources sponsored by the Department of Energy's Office of Critical Minerals and Energy Innovation and located at the National Laboratory of the Rockies.

%==================================%
% Appendix
%==================================%
\appendix
\section{Greedy Quadratic Manifold Construction}\label{appsec:greedy_qm}

We summarize the greedy quadratic manifold (Greedy-QM) approach of~\cite{schwerdtner2024Greedy}, which serves as a main comparison method in our numerical experiments. 
A quadratic manifold decoder augments linear \gls{pod} with a quadratic correction:
\begin{displaymath}
    \Dcal(\zvec) = \Umat_r \zvec + \Wmat h_2(\zvec),
\end{displaymath}
where \( \Umat_r \in \R^{d \times r} \) contains \( r \) orthonormal modes, \( \Wmat \in \R^{d \times {p_2}} \) is a weight matrix, and \( h_2: \R^r \to \R^{p_2} \) with \( p_2 = r(r+1)/2 \) is the quadratic feature map
\begin{equation}\label{eqn:quad_feature}
    h_2(\zvec) = \begin{bmatrix} z_1 z_1 & z_1 z_2 & \cdots & z_1 z_r & z_2 z_2 & \cdots & z_r z_r \end{bmatrix}^\top,
\end{equation}
for \( \zvec = [z_1, \ldots, z_r]^\top \). The encoder is linear: \( \Ecal(\xvec) = \Umat_r^\top \xvec \).

Let the \(j\)-th column of \(\Umat\) be \(\uvec_j\). A key insight of~\cite{schwerdtner2024Greedy} is that selecting \( \Umat_r \) as the leading \( r \) \gls{pod} modes (as in~\cite{geelen2023Operator,jain2017Quadratic}) can be suboptimal: the resulting latent coordinates \( \zvec = \Umat_r^\top \xvec \) may discard information critical for the quadratic correction. Instead, the greedy method selects \( r \) modes \( \uvec_{j_1}, \ldots, \uvec_{j_r} \) from a candidate pool of the first \( s > r \) \gls{pod} modes, where the indices \( j_1, \ldots, j_r \in [s] \) need not be the leading \( r \). 
In the greedy selection algorithm, given basis \( \Umat_{i-1} = [\uvec_{j_1}, \ldots, \uvec_{j_{i-1}}] \) with selected index set \( \Ical_{i-1} = \{j_1,\ldots,j_{i-1}\} \) at iteration \( i \), the next mode is chosen by
\begin{equation}\label{eq:greedy_selection}
    j_i = \argmin_{j \in [s]\setminus\Ical_{i-1}} \min_{\Wmat' \in \R^{d \times p_2}} \left\| \Xmat - \Pmat_{[\Umat_{i-1},\uvec_j]} \Xmat - \Wmat' h_2\bigl([\Umat_{i-1},\uvec_j]^\top \Xmat\bigr) \right\|_F^2 + \gamma \|\Wmat'\|_F^2,
\end{equation}
where \( \Pmat_{\Umat} \) denotes the orthogonal projector onto the column space of \( \Umat \) and \( \gamma > 0 \) is a \( \ell_2 \)-regularization parameter. After \( r \) iterations, \( \Umat_r = [\Umat(:,j_1), \ldots, \Umat(:,j_r)] \) is formed and \( \Wmat \) is computed via linear least squares. The method exploits the precomputed SVD to reformulate~\eqref{eq:greedy_selection} so that the computational cost scales with the number of snapshots \( n \) rather than the state dimension \( d \); see~\cite[Section~3.2]{schwerdtner2024Greedy}.

\paragraph{Extension to cubic manifolds}
The framework extends to a greedy cubic manifold (Greedy-CM) approach with cubic corrections by replacing~\eqref{eqn:quad_feature} with a cubic feature map \( h_3: \R^r \to \R^{p_3} \) containing all cubic monomials:
\begin{displaymath}
    h_3(\zvec) = \begin{bmatrix} z_1^3 & z_1^2 z_2 & \cdots & z_1^2 z_r & z_1 z_2^2 & z_1 z_2 z_3 & \cdots & z_r^3 \end{bmatrix}^\top,
\end{displaymath}
where \( p_3 = r(r+1)(r+2)/6 \). Note that as \( r \) increases, the feature dimension \( p \) and number of parameters in \( \Wmat \) in the greedy manifold approach grow rapidly, whereas SparseModesNet's use of \gls{nn} decoder allows to select more modes with smaller feature dimension \( p \) by training the \gls{nn} in lieu of fixing the mapping form a priori.

%==================================%
% Bibliography
%==================================%
\bibliographystyle{siamplain}
\bibliography{references}

%==================================%
% ARXIV SUPPLEMENT (INCLUDE ALL PAGES)
%==================================%
% THIS IS FOR ARXIV SUBMISSION ONLY
% COMMENT OUT FOR FORMAL SIAM SUBMISSION
\clearpage
\ifarXiv
    

\section*{Supplementary material}

# SUPPLEMENTARY MATERIALS: SPARSE POD MODE SELECTION AND MANIFOLD DIMENSIONALITY REDUCTION WITH NEURAL NETWORKS*

TOMOKI KOIKE†, PRAKASH MOHAN‡, MARC T. HENRY DE FRAHAN‡, ELIZABETH QIAN†, AND JULIE BESSAC‡

**SM1. Theoretical Analysis of SparseModesNet's Design Choice.** We establish theoretical foundations for the design choices in SparseModesNet.

**SM1.1. Assumptions & Constraints.** Throughout this section, we impose the following standing assumptions and constraints.

*Assumption* SM1.1. The training data $\{\mathbf{x}_i\}_{i=1}^n \subset \mathbb{R}^d$ as well as projected data $\{\hat{\mathbf{z}}_i\}_{i=1}^n \subset \mathbb{R}^s$ satisfy $\hat{\mathbf{z}}_i \neq \mathbf{0}$ for all $i \in [n]$.

*Assumption* SM1.2. The neural network (NN) $h_{\text{NN}} : \mathbb{R}^s \to \mathbb{R}^p$ is continuously differentiable, composed of several affine transformations and smooth activations (e.g., tanh). Its Jacobian is denoted $\mathbf{J}_{h_{\text{NN}}}(\boldsymbol{\zeta}) = \partial h_{\text{NN}}/\partial \boldsymbol{\zeta} \in \mathbb{R}^{p\times s}$.

*Constraint* SM1.3. The first-layer weights $\mathcal{W}^{(1)} \in \mathbb{R}^{p_1\times s}$ and skip weights $\boldsymbol{\omega} \in \mathbb{R}^s$ satisfy $\|\mathcal{W}_j^{(1)}\|_\infty \leq M|\omega_j|$ for all $j \in [s]$, where $M > 0$ is the hierarchy coefficient and $\mathcal{W}_j^{(1)} \in \mathbb{R}^{p_1}$ denotes the $j$-th column of $\mathcal{W}^{(1)}$ [SM4].

**SM1.2. Notation.** The standard basis vector in $\mathbb{R}^s$ with 1 in position $j$ is denoted $\mathbf{e}_j$. For $\boldsymbol{\omega} \in \mathbb{R}^s$, we define $\mathsf{diag}(\boldsymbol{\omega}) \in \mathbb{R}^{s\times s}$ as the diagonal matrix with entries $[\mathsf{diag}(\boldsymbol{\omega})]_{jj} = \omega_j$. Note that the elementwise product $\boldsymbol{\omega} \odot \hat{\mathbf{z}}_i$ equals $\mathsf{diag}(\boldsymbol{\omega})\hat{\mathbf{z}}_i$. We denote $\Theta$ to be the collection of NN parameters.

We distinguish two types of superscripts throughout this section:

- For a regularization parameter $\lambda > 0$, the pair $(\boldsymbol{\omega}^*(\lambda), \Theta^*(\lambda))$ denotes a *stationary point* of the regularized objective (SM1.2), i.e., a point satisfying the first-order optimality (KKT) conditions for the constrained problem.
- Given current values $(\omega_j, \mathcal{W}_j^{(1)})$, the notation $(\omega_j^+, \mathcal{W}_j^{(1)+})$ denotes the output of a single proximal gradient iteration.

For a regularization parameter $\lambda > 0$, the *active set* at a stationary point is

$$\mathcal{A}(\lambda) = \{j \in [s] : \omega_j^*(\lambda) \neq 0\}.$$

**SM1.3. The SparseModesNet Objective.** Define the mean-squared error loss functional $\mathcal{L} : \mathbb{R}^s \times \Theta \to \mathbb{R}$ by

$$\mathcal{L}(\boldsymbol{\omega}, \Theta) = \frac{1}{n}\sum_{i=1}^n \left\|\mathbf{x}_i - \mathbf{U}_s \mathsf{diag}(\boldsymbol{\omega})\hat{\mathbf{z}}_i - \mathbf{W} h_{\text{NN}}(\mathsf{diag}(\boldsymbol{\omega})\hat{\mathbf{z}}_i; \Theta)\right\|_2^2. \tag{SM1.1}$$

The regularized objective subject to the Hierarchical Constraint SM1.3 is then

$$\min_{\boldsymbol{\omega},\Theta} \mathcal{J}_\lambda(\boldsymbol{\omega}, \Theta) := \mathcal{L}(\boldsymbol{\omega}, \Theta) + \lambda\|\boldsymbol{\omega}\|_1. \tag{SM1.2}$$

*Submitted to the editors June 29, 2026.

**Funding:** The authors were supported by the Department of Energy Office of Science Advanced Scientific Computing Research, DOE Award DE-SC0024721.

†School of Aerospace Engineering, Georgia Institute of Technology, Atlanta, GA ({tkoike, eqian}@gatech.edu)

‡Computational Science Center, National Laboratory of the Rockies (NLR), Golden, CO ({julie.bessac}@nlr.gov)

**SM1.4. Gradient Structure Analysis.** We first establish the precise form of the gradient and contrast it with the standard LassoNet architecture.

DEFINITION SM1.4 (Masked vs. unmasked architectures).

- *The* masked architecture *(SparseModesNet) uses the NN input that masks* $\hat{\mathbf{z}}_i$ *with the mode selector weight* $\boldsymbol{\omega}$*, i.e.,* $\boldsymbol{\zeta}_i(\boldsymbol{\omega}) = \mathsf{diag}(\boldsymbol{\omega})\hat{\mathbf{z}}_i \in \mathbb{R}^s$*.*
- *The* unmasked architecture *(standard LassoNet, cf.* **??***) uses the NN input* $\boldsymbol{\zeta}_i = \hat{\mathbf{z}}_i \in \mathbb{R}^s$*, independent of* $\boldsymbol{\omega}$*.*

DEFINITION SM1.5. *Define the residual vector* $\mathbf{r}_i(\boldsymbol{\omega}, \Theta) \in \mathbb{R}^d$ *by*

$$\mathbf{r}_i(\boldsymbol{\omega}, \Theta) = \mathbf{x}_i - \mathbf{U}_s \mathsf{diag}(\boldsymbol{\omega})\hat{\mathbf{z}}_i - \mathbf{W} h_{\mathrm{NN}}(\mathsf{diag}(\boldsymbol{\omega})\hat{\mathbf{z}}_i; \Theta). \tag{SM1.3}$$

*Then, the loss functional admits the representation*

$$\mathcal{L}(\boldsymbol{\omega}, \Theta) = \frac{1}{n}\sum_{i=1}^{n} \|\mathbf{r}_i(\boldsymbol{\omega}, \Theta)\|_2^2$$

*and is continuously differentiable in* $(\boldsymbol{\omega}, \Theta)$*.*

The loss representation is immediate from Definition SM1.5 and its continuous differentiability follows from Assumption SM1.2 and the chain rule, noting that $\mathsf{diag}(\boldsymbol{\omega})\hat{\mathbf{z}}_i$ is linear in $\boldsymbol{\omega}$ and $h_{\mathrm{NN}}$ is $C^1$ by assumption.

LEMMA SM1.6. *Under Assumptions SM1.1 and SM1.2, the gradient of the loss functional with respect to* $\omega_j$ *for* $j \in [s]$ *is*

$$\frac{\partial \mathcal{L}}{\partial \omega_j} = -\frac{2}{n}\sum_{i=1}^{n} \hat{z}_{ij} \cdot \mathbf{r}_i^\top \left[\mathbf{u}_j + \mathbf{W}\mathbf{J}_{h_{\mathrm{NN}}}(\mathsf{diag}(\boldsymbol{\omega})\hat{\mathbf{z}}_i)\mathbf{e}_j\right], \tag{SM1.4}$$

*where* $\mathbf{u}_j \in \mathbb{R}^d$ *is the* $j$*-th column of* $\mathbf{U}_s$*,* $\hat{z}_{ij} = [\hat{\mathbf{z}}_i]_j$ *is the* $j$*-th component of* $\hat{\mathbf{z}}_i$*, and* $\mathbf{J}_{h_{\mathrm{NN}}}(\cdot) \in \mathbb{R}^{p \times s}$ *is the Jacobian of* $h_{\mathrm{NN}}$*.*

*Proof.* Fix $j \in [s]$. Differentiating under the sum:

$$\frac{\partial \mathcal{L}}{\partial \omega_j} = \frac{1}{n}\sum_{i=1}^{n} \frac{\partial}{\partial \omega_j}\|\mathbf{r}_i\|_2^2 = \frac{2}{n}\sum_{i=1}^{n} \mathbf{r}_i^\top \frac{\partial \mathbf{r}_i}{\partial \omega_j}.$$

We compute $\partial \mathbf{r}_i / \partial \omega_j$ term by term.

**Term 1: Linear skip connection.** Since $[\mathsf{diag}(\boldsymbol{\omega})\hat{\mathbf{z}}_i]_k = \omega_k \hat{z}_{ik}$, we have $\partial[\mathsf{diag}(\boldsymbol{\omega})\hat{\mathbf{z}}_i]_k / \partial \omega_j = \delta_{jk}\hat{z}_{ik} = \hat{z}_{ij}\delta_{jk}$, giving $\partial[\mathsf{diag}(\boldsymbol{\omega})\hat{\mathbf{z}}_i]/\partial \omega_j = \hat{z}_{ij}\mathbf{e}_j$. Therefore,

$$\frac{\partial}{\partial \omega_j}\left[\mathbf{U}_s \mathsf{diag}(\boldsymbol{\omega})\hat{\mathbf{z}}_i\right] = \hat{z}_{ij}\mathbf{U}_s \mathbf{e}_j = \hat{z}_{ij}\mathbf{u}_j.$$

**Term 2: NN branch.** Define $\boldsymbol{\zeta}_i = \mathsf{diag}(\boldsymbol{\omega})\hat{\mathbf{z}}_i$. By the chain rule:

$$\frac{\partial}{\partial \omega_j}\left[\mathbf{W} h_{\mathrm{NN}}(\boldsymbol{\zeta}_i)\right] = \mathbf{W}\mathbf{J}_{h_{\mathrm{NN}}}(\boldsymbol{\zeta}_i)\frac{\partial \boldsymbol{\zeta}_i}{\partial \omega_j} = \hat{z}_{ij}\mathbf{W}\mathbf{J}_{h_{\mathrm{NN}}}(\boldsymbol{\zeta}_i)\mathbf{e}_j.$$

**Combining terms:** Since $\mathbf{r}_i = \mathbf{x}_i - \mathbf{U}_s \mathsf{diag}(\boldsymbol{\omega})\hat{\mathbf{z}}_i - \mathbf{W} h_{\mathrm{NN}}(\boldsymbol{\zeta}_i)$,

$$\frac{\partial \mathbf{r}_i}{\partial \omega_j} = -\hat{z}_{ij}\left[\mathbf{u}_j + \mathbf{W}\mathbf{J}_{h_{\mathrm{NN}}}(\mathsf{diag}(\boldsymbol{\omega})\hat{\mathbf{z}}_i)\mathbf{e}_j\right].$$

Substituting yields (SM1.4). ∎

LEMMA SM1.7. *In the unmasked architecture (Definition SM1.4) where the NN input is* $\hat{\mathbf{z}}_i$*, the gradient of the loss with respect to* $\omega_j$ *reduces to*

$$\frac{\partial \mathcal{L}}{\partial \omega_j} = -\frac{2}{n}\sum_{i=1}^{n} \hat{z}_{ij} \cdot \mathbf{r}_i^\top \mathbf{u}_j. \tag{SM1.5}$$

*Proof.* In the unmasked case, the NN term $\mathbf{W}h_{\mathrm{NN}}(\hat{\mathbf{z}}_i;\Theta)$ is independent of $\boldsymbol{\omega}$, so $\frac{\partial}{\partial \omega_j}[\mathbf{W}h_{\mathrm{NN}}(\hat{\mathbf{z}}_i;\Theta)] = \mathbf{0}$. The result follows from Term 1 of the proof of Lemma SM1.6. ∎

DEFINITION SM1.8. *Define the* skip gradient *and* network gradient *components:*

$$G_j^{\mathrm{skip}}(\boldsymbol{\omega},\Theta) = -\frac{2}{n}\sum_{i=1}^{n} \hat{z}_{ij} \cdot \mathbf{r}_i(\boldsymbol{\omega},\Theta)^\top \mathbf{u}_j, \tag{SM1.6}$$

$$G_j^{\mathrm{net}}(\boldsymbol{\omega},\Theta) = -\frac{2}{n}\sum_{i=1}^{n} \hat{z}_{ij} \cdot \mathbf{r}_i(\boldsymbol{\omega},\Theta)^\top \mathbf{W}\mathbf{J}_{h_{\mathrm{NN}}}(\mathsf{diag}(\boldsymbol{\omega})\hat{\mathbf{z}}_i)\mathbf{e}_j. \tag{SM1.7}$$

*Then for SparseModesNet:* $\partial\mathcal{L}/\partial\omega_j = G_j^{\mathrm{skip}} + G_j^{\mathrm{net}}$*, while for standard LassoNet:* $\partial\mathcal{L}/\partial\omega_j = G_j^{\mathrm{skip}}$*.*

*Remark* SM1.9. As training progresses and mode $j$ becomes less informative, the residuals $\mathbf{r}_i$ are increasingly explained by the remaining active modes or by those beyond the $s$ candidates, causing $\mathbf{r}_i$ to lie approximately orthogonal to $\mathbf{u}_j$, and hence drives $|G_j^{\mathrm{skip}}|$ toward zero. Consequently, the network gradient $G_j^{\mathrm{net}}$ in (SM1.7) can dominate the overall gradient in SparseModesNet, accelerating the shrinkage of $\omega_j$.

**SM1.5. Proof of Monotonic Mode Elimination.** We now prove the monotonic mode elimination property stated in Subsection 3.2.1 of the main text. We begin by establishing lemmas on the nullification of weights and gradients induced by the hierarchical constraint and input masking.

**SM1.5.1. Preliminary Lemmas.**

LEMMA SM1.10 (Constraint-induced weight nullification). *Under the Hierarchical Constraint SM1.3, if* $\omega_j = 0$ *for some* $j \in [s]$*, then* $\mathcal{W}_j^{(1)} = \mathbf{0} \in \mathbb{R}^{p_1}$*.*

*Proof.* The constraint $\|\mathcal{W}_j^{(1)}\|_\infty \le M|\omega_j| = 0$ implies $|[\mathcal{W}_j^{(1)}]_k| \le 0$ for all $k \in [p_1]$. Since $|\cdot| \ge 0$, we have $[\mathcal{W}_j^{(1)}]_k = 0$ for all $k$, i.e., $\mathcal{W}_j^{(1)} = \mathbf{0}$. ∎

LEMMA SM1.11 (Input nullification). *If* $\omega_j = 0$*, then the* $j$*-th component of the masked input vanishes:* $[\mathsf{diag}(\boldsymbol{\omega})\hat{\mathbf{z}}_i]_j = 0$ *for all* $i \in [n]$*.*

*Proof.* Direct computation: $[\mathsf{diag}(\boldsymbol{\omega})\hat{\mathbf{z}}_i]_j = \omega_j \hat{z}_{ij} = 0$. ∎

LEMMA SM1.12 (Gradient severance). *Suppose* $\omega_j = 0$ *for some* $j \in [s]$*. Under the masked architecture (Definition SM1.4), the gradient of the loss* $\mathcal{L}$ *with respect to the first-layer weights* $\mathcal{W}_j^{(1)}$ *satisfies*

$$\frac{\partial \mathcal{L}}{\partial \mathcal{W}_j^{(1)}} = \mathbf{0} \in \mathbb{R}^{p_1}.$$

*Proof.* Let the NN have the form $h_{\mathrm{NN}}(\boldsymbol{\zeta};\Theta) = {h_{\mathrm{NN}}}^{(L)} \circ \cdots \circ {h_{\mathrm{NN}}}^{(2)} \circ {h_{\mathrm{NN}}}^{(1)}(\boldsymbol{\zeta})$, where ${h_{\mathrm{NN}}}^{(1)}(\boldsymbol{\zeta}) = \sigma(\mathcal{W}^{(1)}\boldsymbol{\zeta} + \boldsymbol{\beta}^{(1)})$ with elementwise activation $\sigma$. Define the preactivation $\mathbf{a}^{(1)}(\boldsymbol{\zeta}) = \mathcal{W}^{(1)}\boldsymbol{\zeta} + \boldsymbol{\beta}^{(1)} \in \mathbb{R}^{p_1}$.

For the masked input $\boldsymbol{\zeta}_i = \mathsf{diag}(\boldsymbol{\omega})\hat{\mathbf{z}}_i$, the $k$-th component of the preactivation is

$$[\mathbf{a}^{(1)}(\boldsymbol{\zeta}_i)]_k = \sum_{m=1}^{s} [\mathcal{W}_m^{(1)}]_k \omega_m \hat{z}_{im} + [\boldsymbol{\beta}^{(1)}]_k.$$

Differentiating with respect to $[\mathcal{W}_j^{(1)}]_k$ (the $(k,j)$-entry of $\mathcal{W}^{(1)}$):

$$\frac{\partial [\mathbf{a}^{(1)}(\boldsymbol{\zeta}_i)]_k}{\partial [\mathcal{W}_j^{(1)}]_k} = \omega_j \hat{z}_{ij}.$$

By the chain rule:

$$\frac{\partial \mathcal{L}}{\partial \mathcal{W}_j^{(1)}} = \frac{2}{n}\sum_{i=1}^{n} \frac{\partial \mathcal{L}}{\partial \mathbf{a}^{(1)}(\boldsymbol{\zeta}_i)} \cdot \frac{\partial \mathbf{a}^{(1)}(\boldsymbol{\zeta}_i)}{\partial \mathcal{W}_j^{(1)}} = \frac{2}{n}\sum_{i=1}^{n} \omega_j \hat{z}_{ij} \cdot \frac{\partial \mathcal{L}}{\partial \mathbf{a}^{(1)}(\boldsymbol{\zeta}_i)}.$$

When $\omega_j = 0$, $\frac{\partial \mathcal{L}}{\partial \mathcal{W}_j^{(1)}} = \mathbf{0}$ regardless of the values of $\hat{z}_{ij}$ and $\partial \mathcal{L}/\partial \mathbf{a}^{(1)}(\boldsymbol{\zeta}_i)$. ∎

*Remark* SM1.13. Lemma SM1.12 shows that input masking causes the gradient with respect to $\mathcal{W}_j^{(1)}$ to vanish when $\omega_j = 0$. This severs mode $j$: no gradient signal can update $\mathcal{W}_j^{(1)}$. In the unmasked architecture, by contrast, $\partial \mathcal{L}/\partial \mathcal{W}_j^{(1)}$ is generically nonzero even when $\omega_j = 0$, since the NN receives the full input $\hat{\mathbf{z}}_i$. However, gradient severance alone only *freezes* $\mathcal{W}_j^{(1)}$ at its current value; it does not force $\mathcal{W}_j^{(1)} = \mathbf{0}$. The hierarchical constraint (Lemma SM1.10) is necessary to ensure that when $\omega_j = 0$, the constraint enforces $\mathcal{W}_j^{(1)} = \mathbf{0}$.

**SM1.5.2. The Hier-Prox Operator.** The optimization problem (SM1.2) is solved via the continuation strategy described in Subsection 2.2, using the hierarchical proximal (Hier-Prox) operator developed in [SM4].

Definition SM1.14. *In a proximal gradient iteration with step size $\alpha > 0$, given current values $(\omega_j, \mathcal{W}_j^{(1)})$, the* preproximal values *are*

$$\widetilde{\omega}_j = \omega_j - \alpha \frac{\partial \mathcal{L}}{\partial \omega_j}, \qquad \widetilde{\mathcal{W}}_j^{(1)} = \mathcal{W}_j^{(1)} - \alpha \frac{\partial \mathcal{L}}{\partial \mathcal{W}_j^{(1)}}. \tag{SM1.8}$$

Definition SM1.15 (Hier-Prox operator). *For mode $j \in [s]$, step size $\alpha > 0$, regularization parameter $\lambda > 0$, hierarchy coefficient $M > 0$, and preproximal values $(\widetilde{\omega}_j, \widetilde{\mathcal{W}}_j^{(1)}) \in \mathbb{R} \times \mathbb{R}^{p_1}$, the Hier-Prox operator is*

$$\mathrm{HierProx}_{\alpha\lambda,M}\big(\widetilde{\omega}_j, \widetilde{\mathcal{W}}_j^{(1)}\big) := \operatorname*{arg\,min}_{(\omega_j,\mathbf{w})\in\mathcal{C}} \frac{1}{2}(\omega_j - \widetilde{\omega}_j)^2 + \frac{1}{2}\|\mathbf{w} - \widetilde{\mathcal{W}}_j^{(1)}\|_2^2 + \alpha\lambda|\omega_j|, \tag{SM1.9}$$

*where the constraint set is $\mathcal{C} = \{(\omega_j, \mathbf{w}) \in \mathbb{R} \times \mathbb{R}^{p_1} : \|\mathbf{w}\|_\infty \le M|\omega_j|\}$.*

The constraint set $\mathcal{C}$ is a double cone with apex at the origin. The Hier-Prox operator produces the updated values $(\omega_j^+, \mathcal{W}_j^{(1)+}) = \mathrm{HierProx}_{\alpha\lambda,M}(\widetilde{\omega}_j, \widetilde{\mathcal{W}}_j^{(1)})$.

Proposition SM1.16 (Zero attractor condition). *The Hier-Prox operator satisfies $\mathrm{HierProx}_{\alpha\lambda,M}(\widetilde{\omega}_j, \widetilde{\mathcal{W}}_j^{(1)}) = (0, \mathbf{0})$ if and only if*

$$|\widetilde{\omega}_j| + M\|\widetilde{\mathcal{W}}_j^{(1)}\|_1 \le \alpha\lambda. \tag{SM1.10}$$

*Proof.* Define $f(\omega_j, \mathbf{w}) = \frac{1}{2}(\omega_j - \widetilde{\omega}_j)^2 + \frac{1}{2}\|\mathbf{w} - \widetilde{\mathcal{W}}_j^{(1)}\|_2^2 + \alpha\lambda|\omega_j|$. For brevity, write $\widetilde{\omega} = \widetilde{\omega}_j$ and $\widetilde{\mathbf{w}} = \widetilde{\mathcal{W}}_j^{(1)}$.

Since $\mathcal{C}$ is a cone, its tangent cone at the origin equals itself:

$$T_{\mathcal{C}}(0, \mathbf{0}) = \mathcal{C} = \big\{(d_\omega, \mathbf{d}) \in \mathbb{R} \times \mathbb{R}^{p_1} : \|\mathbf{d}\|_\infty \le M|d_\omega|\big\}.$$

For any direction $(d_\omega, \mathbf{d}) \in T_{\mathcal{C}}(0, \mathbf{0})$, the directional derivative of $f$ at $(0, \mathbf{0})$ is

$$f'((0, \mathbf{0}); (d_\omega, \mathbf{d})) = \lim_{t \to 0^+} \frac{f(td_\omega, t\mathbf{d}) - f(0, \mathbf{0})}{t}.$$

Expanding $f(td_\omega, t\mathbf{d})$:

$$\begin{aligned} f(td_\omega, t\mathbf{d}) &= \frac{1}{2}(td_\omega - \widetilde{\omega})^2 + \frac{1}{2}\|t\mathbf{d} - \widetilde{\mathbf{w}}\|_2^2 + \alpha\lambda|td_\omega| \\ &= \frac{1}{2}\widetilde{\omega}^2 + \frac{1}{2}\|\widetilde{\mathbf{w}}\|_2^2 - td_\omega\widetilde{\omega} - t\mathbf{d}^\top\widetilde{\mathbf{w}} + \alpha\lambda t|d_\omega| + O(t^2). \end{aligned}$$

Since $f(0, \mathbf{0}) = \frac{1}{2}\widetilde{\omega}^2 + \frac{1}{2}\|\widetilde{\mathbf{w}}\|_2^2$, we obtain

$$f'((0, \mathbf{0}); (d_\omega, \mathbf{d})) = -d_\omega\widetilde{\omega} - \mathbf{d}^\top\widetilde{\mathbf{w}} + \alpha\lambda|d_\omega|. \tag{SM1.11}$$

Following the first-order optimality conditions, since $f$ and a single cone of $\mathcal{C}$ with the apex at the origin are convex, $(0, \mathbf{0})$ is optimal if and only if $f'((0, \mathbf{0}); (d_\omega, \mathbf{d})) \ge 0$ for all $(d_\omega, \mathbf{d}) \in T_{\mathcal{C}}(0, \mathbf{0})$. We analyze three cases.

*Case 1:* $d_\omega = 0$. The constraint forces $\mathbf{d} = \mathbf{0}$, giving $f'((0, \mathbf{0}); (0, \mathbf{0})) = 0 \ge 0$.

*Case 2:* $d_\omega > 0$. The directional derivative (SM1.11) becomes $f'((0, \mathbf{0}); (d_\omega, \mathbf{d})) = d_\omega(\alpha\lambda - \widetilde{\omega}) - \mathbf{d}^\top\widetilde{\mathbf{w}}$. Maximizing $\mathbf{d}^\top\widetilde{\mathbf{w}}$ over $\|\mathbf{d}\|_\infty \le Md_\omega$ is achieved at $d_k = Md_\omega \cdot \mathsf{sign}(\widetilde{w}_k)$ for each coordinate $k$ where $\widetilde{w}_k = [\widetilde{\mathbf{w}}]_k$, yielding

$$\max_{\|\mathbf{d}\|_\infty \le Md_\omega} \mathbf{d}^\top\widetilde{\mathbf{w}} = \sum_{k=1}^{p_1} Md_\omega|\widetilde{w}_k| = Md_\omega\|\widetilde{\mathbf{w}}\|_1.$$

Thus, the infimum of the directional derivative over $d_\omega > 0$ and feasible $\mathbf{d}$ is

$$\inf_{d_\omega > 0,\, \|\mathbf{d}\|_\infty \le Md_\omega} f'((0, \mathbf{0}); (d_\omega, \mathbf{d})) = d_\omega\big(\alpha\lambda - \widetilde{\omega} - M\|\widetilde{\mathbf{w}}\|_1\big).$$

For this to be nonnegative for all $d_\omega > 0$, we require

$$\widetilde{\omega} + M\|\widetilde{\mathbf{w}}\|_1 \le \alpha\lambda. \tag{SM1.12}$$

*Case 3:* $d_\omega < 0$. Writing $d_\omega = -\delta$ with $\delta > 0$, a similar argument yields

$$-\widetilde{\omega} + M\|\widetilde{\mathbf{w}}\|_1 \le \alpha\lambda. \tag{SM1.13}$$

Combining (SM1.12) and (SM1.13) gives $|\widetilde{\omega}| + M\|\widetilde{\mathbf{w}}\|_1 \le \alpha\lambda$, which is (SM1.10). ∎

**SM1.5.3. The Jacobian Structure.** The following fact establishes how the network Jacobian depends on the first-layer weights.

*Fact* SM1.17. Consider a NN with first layer $h_{\mathrm{NN}}^{(1)}(\boldsymbol{\zeta}) = \sigma(\mathcal{W}^{(1)}\boldsymbol{\zeta} + \boldsymbol{\beta}^{(1)})$, where $\sigma$ is an elementwise activation with derivative $\sigma'$. Let $\mathbf{J}^{(L:2)}$ denote the Jacobian of the composition of layers 2 through $L$. Then the $j$-th column of the network Jacobian satisfies

$$\mathbf{J}_{h_{\mathrm{NN}}}(\boldsymbol{\zeta})\mathbf{e}_j = \mathbf{J}^{(L:2)} \cdot \mathsf{diag}\big(\sigma'(\mathcal{W}^{(1)}\boldsymbol{\zeta} + \boldsymbol{\beta}^{(1)})\big) \cdot \mathcal{W}_j^{(1)}. \tag{SM1.14}$$

In particular, if $\mathcal{W}_j^{(1)} = \mathbf{0}$, then $\mathbf{J}_{h_{\mathrm{NN}}}(\boldsymbol{\zeta})\mathbf{e}_j = \mathbf{0}$; and if $\mathcal{W}_j^{(1)} \ne \mathbf{0}$, then generically $\mathbf{J}_{h_{\mathrm{NN}}}(\boldsymbol{\zeta})\mathbf{e}_j \ne \mathbf{0}$.

By the chain rule, $\mathbf{J}_{h_{\mathrm{NN}}}(\boldsymbol{\zeta}) = \mathbf{J}^{(L:2)} \cdot \mathsf{diag}(\sigma'(\mathcal{W}^{(1)}\boldsymbol{\zeta} + \boldsymbol{\beta}^{(1)})) \cdot \mathcal{W}^{(1)}$. The $j$-th column is obtained by right-multiplying by $\mathbf{e}_j$, which extracts $\mathcal{W}_j^{(1)}$. This fact follows from the linearity of this expression in $\mathcal{W}_j^{(1)}$.

**SM1.5.4. Additional Assumptions for Monotonicity.** To establish monotonic mode elimination, we impose two regularity conditions on the solution path. These assumptions involve the stationary points $(\boldsymbol{\omega}^*(\lambda), \Theta^*(\lambda))$ of the regularized objective (SM1.2) at parameter $\lambda$.

*Assumption* SM1.18 (Lipschitz regularity of the skip gradient). There exists a constant $L > 0$ such that for all regularization parameters $\lambda, \lambda'$ along the path and all eliminated modes $j \in [s] \setminus \mathcal{A}(\lambda) \cap [s] \setminus \mathcal{A}(\lambda')$:

$$\left|G_j^{\mathrm{skip}}(\boldsymbol{\omega}^*(\lambda'), \Theta^*(\lambda')) - G_j^{\mathrm{skip}}(\boldsymbol{\omega}^*(\lambda), \Theta^*(\lambda))\right| \le L|\lambda' - \lambda|,$$

where $G_j^{\mathrm{skip}}$ is defined in (SM1.6) and $(\boldsymbol{\omega}^*(\lambda), \Theta^*(\lambda))$ denotes a stationary point of (SM1.2) at parameter $\lambda$.

*Assumption* SM1.19 (Strict complementarity with margin). There exists a margin $\delta \in (0, 1)$ such that at each stationary point $(\boldsymbol{\omega}^*(\lambda), \Theta^*(\lambda))$ of (SM1.2), for all eliminated modes $j \in [s] \setminus \mathcal{A}(\lambda)$:

$$|G_j^{\mathrm{skip}}(\boldsymbol{\omega}^*(\lambda), \Theta^*(\lambda))| \le (1 - \delta)\lambda.$$

Assumption SM1.18 requires the skip gradient for eliminated modes to vary Lipschitz-continuously along the solution path. For eliminated modes $j$, we obtain $\omega_j^*(\lambda) = 0$, so $G_j^{\mathrm{skip}}$ varies only through changes in the residual induced by the active parameters; this dependence is typically well-behaved. Assumption SM1.19 posits a strict gap between the gradient magnitude $|G_j^{\mathrm{skip}}|$ and the threshold $\lambda$. Recalling Remark SM1.9, $|G_j^{\mathrm{skip}}|$ is small given $G_j^{\mathrm{skip}}$ is a weighted sum of inner products $\mathbf{r}_i^\top \mathbf{u}_j$ (cf. (SM1.6)) with $\mathbf{r}_i$ and $\mathbf{u}_j$ being nearly orthogonal. The margin $\delta$ quantifies this gap as a fraction of $\lambda$, ensuring it grows with the regularization strength.

**SM1.5.5. Main Result.**

THEOREM SM1.20 (Monotonic mode elimination). *Let Assumptions SM1.1, SM1.2, SM1.18, and SM1.19 and Constraint SM1.3 hold. Consider the regularization path $\lambda_{k+1} = (1 + \epsilon)\lambda_k$ for $k = 0, 1, \ldots, K - 1$ with path multiplier $\epsilon > 0$. Under the SparseModesNet architecture (input masking + hierarchical constraint) optimizing objective (SM1.2), if mode $j$ is eliminated at $\lambda_k$, i.e., $\omega_j^*(\lambda_k) = 0$, then*

$$\omega_j^*(\lambda_\ell) = 0 \quad \textit{for all} \quad \ell \ge k,$$

*provided one of the following conditions holds:*

1. *$L \le 1$ and $\epsilon > 0$ is arbitrary, or*
2. *$L > 1$ and $0 < \epsilon \le \frac{\delta}{L-1}$,*

*where $L$ and $\delta$ are from Assumptions SM1.18 and SM1.19, respectively. Equivalently, the active sets are nested: $\lambda' > \lambda \implies \mathcal{A}(\lambda') \subseteq \mathcal{A}(\lambda)$.*

*Proof.* Suppose $\omega_j^*(\lambda_k) = 0$ for some mode $j$ and index $k$. We show $\omega_j^*(\lambda_{k+1}) = 0$; the full result follows by induction.

By Lemma SM1.10, the hierarchical constraint forces $\mathcal{W}_j^{(1)*}(\lambda_k) = \mathbf{0}$. At $\lambda_{k+1}$, warm-start initialization sets $(\omega_j, \mathcal{W}_j^{(1)}) = (0, \mathbf{0})$.

Consider a proximal gradient iteration with step size $\alpha > 0$. From Definition SM1.14, the preproximal values starting from $(0, \mathbf{0})$ are

$$\widetilde{\omega}_j = 0 - \alpha \frac{\partial \mathcal{L}}{\partial \omega_j}\Big|_{(\omega_j, \mathcal{W}_j^{(1)})=(0,\mathbf{0})}, \qquad \widetilde{\mathcal{W}}_j^{(1)} = \mathbf{0} - \alpha \frac{\partial \mathcal{L}}{\partial \mathcal{W}_j^{(1)}}\Big|_{(\omega_j, \mathcal{W}_j^{(1)})=(0,\mathbf{0})}.$$

By Lemma SM1.12, $\frac{\partial \mathcal{L}}{\partial \mathcal{W}_j^{(1)}}\big|_{\omega_j=0} = \mathbf{0}$, hence $\widetilde{\mathcal{W}}_j^{(1)} = \mathbf{0}$.

Then, from Proposition SM1.16, the zero attractor condition becomes

$$|\widetilde{\omega}_j| + M \cdot 0 = \alpha \left| \frac{\partial \mathcal{L}}{\partial \omega_j}\Big|_{(\omega_j, \mathcal{W}_j^{(1)})=(0,\mathbf{0})} \right| \leq \alpha \lambda_{k+1}.$$

Since $\omega_j = 0$ and $\mathcal{W}_j^{(1)} = \mathbf{0}$, by Fact SM1.17 we have $\mathbf{J}_{h_{\text{NN}}}(\boldsymbol{\zeta})\mathbf{e}_j = \mathbf{0}$, hence $G_j^{\text{net}} = 0$ by (SM1.7). The condition simplifies to

$$|G_j^{\text{skip}}(\boldsymbol{\omega}, \Theta)| \leq \lambda_{k+1}.$$

By Assumption SM1.19, at the stationary point for $\lambda_k$:

$$|G_j^{\text{skip}}(\boldsymbol{\omega}^*(\lambda_k), \Theta^*(\lambda_k))| \leq (1-\delta)\lambda_k.$$

By Assumption SM1.18, at the stationary point for $\lambda_{k+1}$:

$$\begin{aligned} |G_j^{\text{skip}}(\boldsymbol{\omega}^*(\lambda_{k+1}), \Theta^*(\lambda_{k+1}))| &\leq |G_j^{\text{skip}}(\boldsymbol{\omega}^*(\lambda_k), \Theta^*(\lambda_k))| + L|\lambda_{k+1} - \lambda_k| \\ &\leq (1-\delta)\lambda_k + L\epsilon\lambda_k = \lambda_k(1-\delta+L\epsilon). \end{aligned}$$

For (7) to hold, we need $\lambda_k(1-\delta+L\epsilon) \leq \lambda_{k+1} = (1+\epsilon)\lambda_k$, i.e.,

$$0 < 1-\delta+L\epsilon \leq 1+\epsilon \quad \iff \quad (L-1)\epsilon \leq \delta.$$

*Case* $L \leq 1$: For any $\epsilon > 0$, $(L-1)\epsilon \leq 0 < \delta$, so the inequality holds.

*Case* $L > 1$: The inequality holds if and only if $\epsilon \leq \delta/(L-1)$.

Under the stated conditions, the Hier-Prox operator returns $(\omega_j^+, \mathcal{W}_j^{(1)+}) = (0, \mathbf{0})$ at each iteration. The algorithm converges to a stationary point with $\omega_j^*(\lambda_{k+1}) = 0$. By Assumption SM1.19, this new stationary point satisfies $|G_j^{\text{skip}}| \leq (1-\delta)\lambda_{k+1}$, so the argument repeats for subsequent $\lambda_\ell$, and thus, completing the induction. ∎

*Remark* SM1.21 (Interpretation of the step size bound). When $L > 1$, the bound $\epsilon \leq \delta/(L-1)$ balances two effects:

- The margin $\delta$ measures how far below $\lambda_k$ the gradient $|G_j^{\text{skip}}|$ lies. Larger $\delta$ permits larger steps.
- The Lipschitz constant $L$ measures how rapidly $G_j^{\text{skip}}$ can change. Larger $L$ requires smaller steps.

When $L \leq 1$, the gradient changes more slowly than the threshold of the zero attractor condition grows, so any path multiplier $\epsilon$ suffices.

**SM1.6. Failure of Monotonicity Without Both Design Choices.** We now show that both input masking and the hierarchical constraint are essential and neither alone guarantees monotonic mode elimination.

### SM1.6.1. Failure Without Input Masking.

COROLLARY SM1.22. *In the unmasked architecture (Definition SM1.4) with the Hierarchical Constraint SM1.3 and Assumptions SM1.18 and SM1.19, there exist problem instances where a mode $j$, eliminated at $\lambda_k$, is reactivated at $\lambda_{k+1} = (1+\epsilon)\lambda_k$ for any $\epsilon > 0$.*

*Proof.* By Lemma SM1.10, warm-start initialization sets $(\omega_j, \mathcal{W}_j^{(1)}) = (0, \mathbf{0})$ at $\lambda_{k+1}$. The preproximal values after one gradient step are

$$\widetilde{\omega}_j = -\alpha \frac{\partial \mathcal{L}}{\partial \omega_j}\Big|_{(\omega_j, \mathcal{W}_j^{(1)})=(0,\mathbf{0})}, \qquad \widetilde{\mathcal{W}}_j^{(1)} = -\alpha \frac{\partial \mathcal{L}}{\partial \mathcal{W}_j^{(1)}}\Big|_{(\omega_j, \mathcal{W}_j^{(1)})=(0,\mathbf{0})}.$$

In the unmasked architecture, the NN input is $\hat{\mathbf{z}}_i$ rather than $\mathsf{diag}(\boldsymbol{\omega})\hat{\mathbf{z}}_i$. The gradient with respect to $\mathcal{W}_j^{(1)}$ becomes

$$\frac{\partial \mathcal{L}}{\partial \mathcal{W}_j^{(1)}}\Big|_{(\omega_j, \mathcal{W}_j^{(1)})=(0,\mathbf{0})} = \frac{2}{n}\sum_{i=1}^{n} \hat{z}_{ij} \cdot \frac{\partial \mathcal{L}}{\partial \mathbf{a}^{(1)}(\hat{\mathbf{z}}_i)},$$

which is generically *nonzero* since the NN observes the full input $\hat{\mathbf{z}}_i$. Hence $\widetilde{\mathcal{W}}_j^{(1)} \neq \mathbf{0}$.

By Proposition SM1.16, the Hier-Prox operator returns $(0, \mathbf{0})$ only if

$$\left|\frac{\partial \mathcal{L}}{\partial \omega_j}\right| + M \left\|\frac{\partial \mathcal{L}}{\partial \mathcal{W}_j^{(1)}}\right\|_1 \le \lambda_{k+1}.$$

By Assumption SM1.19, $|\partial \mathcal{L}/\partial \omega_j| \le (1-\delta)\lambda_k$ since $\partial \mathcal{L}/\partial \omega_j = G_j^{\text{skip}}$ from Definition SM1.8. However, there is no bound on $\|\partial \mathcal{L}/\partial \mathcal{W}_j^{(1)}\|_1$. Reactivation occurs when

$$M \left\|\frac{\partial \mathcal{L}}{\partial \mathcal{W}_j^{(1)}}\right\|_1 > (\delta + \epsilon)\lambda_k,$$

which can be satisfied for any $\epsilon > 0$ in problem instances where mode $j$ is informative through the network path. ∎

Without input masking, $\partial \mathcal{L}/\partial \mathcal{W}_j^{(1)} \neq \mathbf{0}$ even when $\omega_j = 0$. This produces a nonzero $\widetilde{\mathcal{W}}_j^{(1)}$, contributing the term $M\|\widetilde{\mathcal{W}}_j^{(1)}\|_1$ to the Hier-Prox threshold. If this term is large enough, the threshold is exceeded and the mode reactivates.

### SM1.6.2. Failure with $\ell_1$-Penalty Alone.

COROLLARY SM1.23. *Consider an architecture with $\ell_1$-penalty on $\boldsymbol{\omega}$ but without the Hierarchical Constraint SM1.3. Monotonic mode elimination is not guaranteed, even with input masking and Assumptions SM1.18 and SM1.19.*

*Proof.* Without the hierarchical constraint, the proximal update for $\omega_j$ is soft-thresholding: $\omega_j^+ = \mathcal{S}_{\alpha\lambda}(\widetilde{\omega}_j)$, where $\mathcal{S}_\tau(x) = \text{sign}(x)\max(|x| - \tau, 0)$ [SM5].

Consider the iteration at which $\omega_j$ first reaches zero while $\mathcal{W}_j^{(1)}$ may be nonzero. Suppose at iteration $t$, we have $\omega_j^{(t)} = 0$ but $\mathcal{W}_j^{(1)(t)} \neq \mathbf{0}$. By Fact SM1.17, $\mathcal{W}_j^{(1)} \neq \mathbf{0}$ implies $\mathbf{J}_{h_{\text{NN}}}(\boldsymbol{\zeta})\mathbf{e}_j \neq \mathbf{0}$, hence $G_j^{\text{net}} \neq 0$ by (SM1.7).

The preproximal value is

$$\widetilde{\omega}_j = -\alpha(G_j^{\text{skip}} + G_j^{\text{net}}).$$

Soft-thresholding gives $\omega_j^{(t+1)} = 0$ only if $|G_j^{\text{skip}} + G_j^{\text{net}}| \leq \lambda$ [SM5]. Even if $|G_j^{\text{skip}}| \leq (1-\delta)\lambda$ per Assumption SM1.19, the term $G_j^{\text{net}}$ can cause the sum to exceed $\lambda$, yielding

$$\omega_j^{(t+1)} = \mathcal{S}_{\alpha\lambda}(-\alpha(G_j^{\text{skip}} + G_j^{\text{net}})) \neq 0,$$

and mode $j$ is reactivated. ∎

The mechanism underlying Corollary SM1.23 is the dependence of $\mathbf{J}_{h_{\text{NN}}}(\boldsymbol{\zeta})\mathbf{e}_j$ on $\mathcal{W}_j^{(1)}$ established in Fact SM1.17. Even though input masking ensures the NN *output* does not depend on $\mathcal{W}_j^{(1)}$ when $\omega_j = 0$, the *sensitivity* $\mathbf{J}_{h_{\text{NN}}}\mathbf{e}_j$ does depend on $\mathcal{W}_j^{(1)}$. This sensitivity appears in $G_j^{\text{net}}$ and hence in $\partial\mathcal{L}/\partial\omega_j$.

In summary, neither condition alone suffices: without the hierarchical constraint, $\mathcal{W}_j^{(1)}$ is not forced to zero (Corollary SM1.23); without input masking, gradients are not severed (Corollary SM1.22). SparseModesNet employs both, achieving the monotonic mode elimination of Theorem SM1.20.

**SM2. Comparison Study of Different Mode Selection Strategies.** As discussed in Subsection 3.2.1, SparseModesNet differs from the original LassoNet framework in that the NN component receives only the selected modes ($\boldsymbol{\omega} \odot \mathbf{z}$) rather than the full candidate set ($\mathbf{z}$). Following the original LassoNet framework instead would yield the optimization problem

(SM2.1)
$$\min_{\boldsymbol{\omega}, h_{\text{NN}}, \mathbf{W}} \sum_{i=1}^{n} \left\| \mathbf{x}_i - \mathbf{U}_s(\boldsymbol{\omega} \odot \mathbf{z}_i) - \mathbf{W} h_{\text{NN}}(\mathbf{z}_i) \right\|_2^2 + \lambda \|\boldsymbol{\omega}\|_1$$
$$\text{subject to} \quad \|\mathcal{W}_j^{(1)}\|_\infty \leq M|\omega_j| \quad \text{and} \quad [\mathbf{U}_s \mathsf{diag}(\mathbb{1}_{\{\omega_j \neq 0\}}(\boldsymbol{\omega}))]^\top \mathbf{W} = 0, \ \forall j \in [s].$$

We also discussed how, when both components receive the same selected-mode input, the hierarchical sparsity constraint may be unnecessary and a simpler $\ell_1$ regularization on $\boldsymbol{\omega}$ might suffice, yielding the alternative problem:

$$\min_{\boldsymbol{\omega}, h_{\text{NN}}, \mathbf{W}} \sum_{i=1}^{n} \left\| \mathbf{x}_i - \mathbf{U}_s(\boldsymbol{\omega} \odot \mathbf{z}_i) - \mathbf{W} h_{\text{NN}}(\boldsymbol{\omega} \odot \mathbf{z}_i) \right\|_2^2 + \lambda \|\boldsymbol{\omega}\|_1 \tag{SM2.2}$$
$$\text{subject to} \quad [\mathbf{U}_s \mathsf{diag}(\mathbb{1}_{\{\omega_j \neq 0\}}(\boldsymbol{\omega}))]^\top \mathbf{W} = 0, \ \forall j \in [s].$$

To evaluate SparseModesNet against these alternatives, we compare three mode selection strategies:

- **SparseModesNet (Proposed)**: as in (3.2), the NN component receives only the selected modes ($\boldsymbol{\omega} \odot \mathbf{z}$) and the hierarchical sparsity constraint is enforced.
- **LassoNet with $h_{\text{NN}}(\mathbf{z})$ (unmasked)**: the NN component receives the full candidate set ($\mathbf{z}$) while the hierarchical sparsity constraint is retained, as in (SM2.1).
- **$\ell_1$-penalty only**: the NN component receives only the selected modes ($\boldsymbol{\omega} \odot \mathbf{z}$) but uses only $\ell_1$ regularization on $\boldsymbol{\omega}$, without the hierarchical sparsity constraint, as in (SM2.2).

We adopt the experimental setup of Subsection 4.2, training a nonlinear decoder for a $\Pi_2$-Net model while selecting modes that minimize the reconstruction error for the linear transport equation. To account for variability due to random initialization of NN weights and biases, we repeat the training 100 times per method and report the statistics of the final reconstruction error at reduced dimension $r = 15$, along

with the mode selection patterns. For a fair comparison, the stopping criteria and hyperparameters are kept consistent across all three methods, with mode selection stopped once the number of modes decreases to $r = 15$.

Figure SM1 presents the frequency of selecting each POD mode over the 100 runs per method, while Figure SM2 shows the corresponding box plots of reconstruction errors. SparseModesNet selects a more diverse range of modes, including lower-energy ones, whereas both the $\ell_1$-penalty only and LassoNet-style variants exhibit a strong bias toward higher-energy modes, commonly known as the *spectral bias* [SM6].

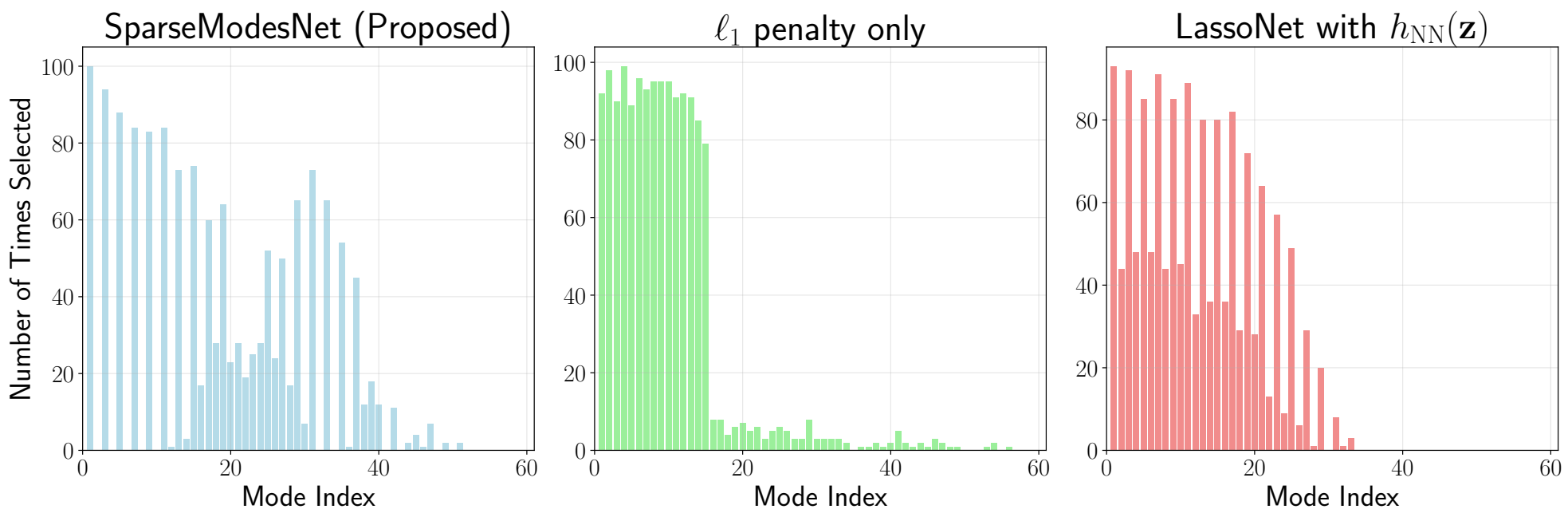


Fig. SM1. *Comparison of the three different mode selection strategies in terms of the frequency of selecting each POD mode over 100 independent training runs. The proposed SparseModesNet framework* ***(left)*** *selects a wider range of modes including low-energy modes, while the $\ell_1$-penalty only variant* ***(middle)*** *and LassoNet-style variant* ***(right)*** *exhibit a strong bias toward selecting higher-energy modes.*

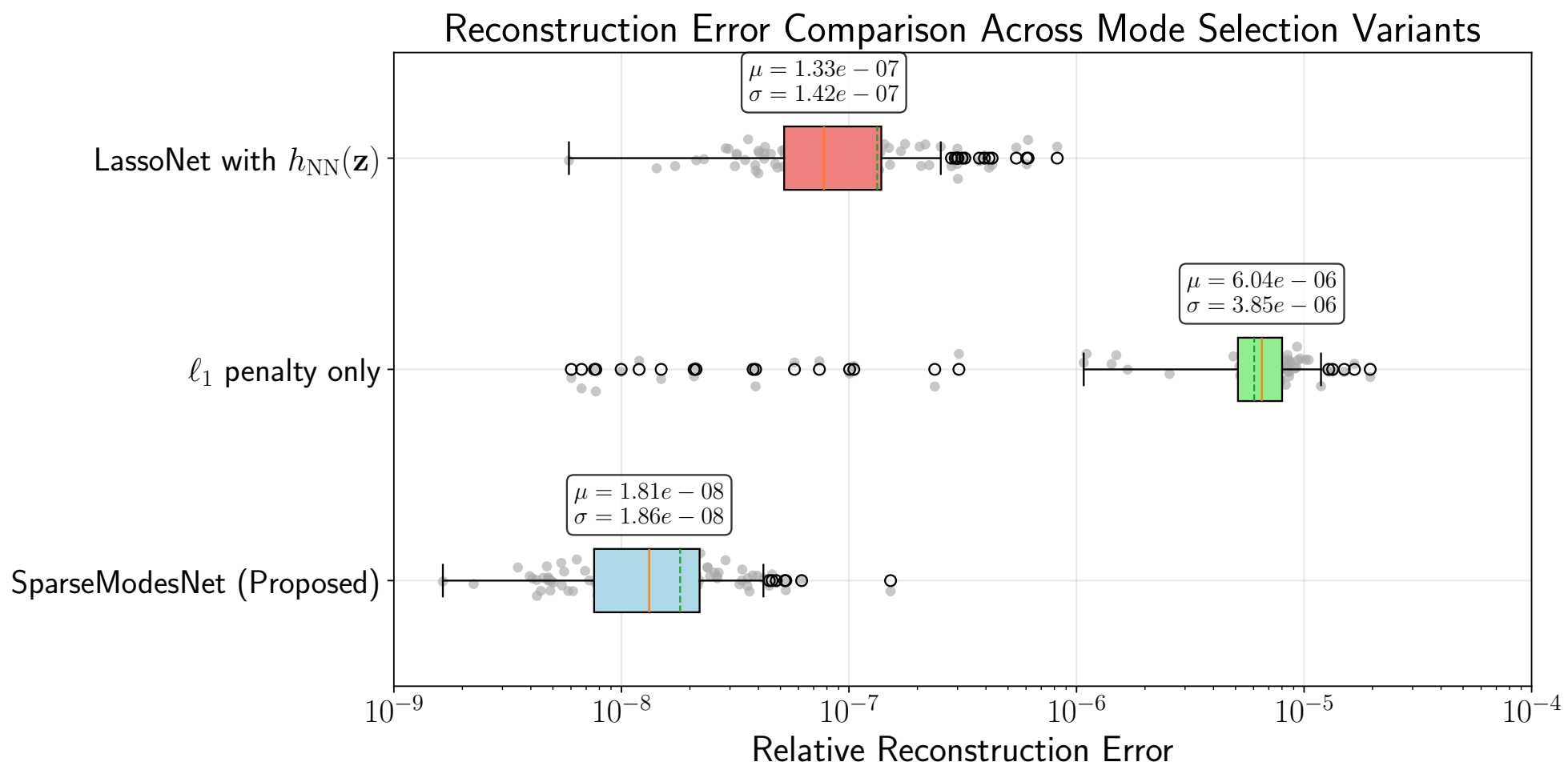


Fig. SM2. *Box plot of the reconstruction error for the three different mode selection strategies over 100 independent training runs. The proposed SparseModesNet framework (bottom) achieves significantly lower reconstruction error with substantially less variability compared to the other two variants.*

Quantitatively, SparseModesNet achieves the lowest mean reconstruction error ($\mu = 1.81 \times 10^{-8}$) with minimal variability ($\sigma = 1.86 \times 10^{-8}$). The LassoNet-style variant is moderate ($\mu = 1.33 \times 10^{-7}$, $\sigma = 1.42 \times 10^{-7}$), roughly a 7-fold increase over

SparseModesNet, while the $\ell_1$-penalty only variant is far worse ($\mu = 6.04 \times 10^{-6}$, $\sigma = 3.85 \times 10^{-6}$)—more than 300 times higher than the proposed method. This suggests that the hierarchical sparsity constraint, combined with feeding selected modes to both components, encourages the selection of modes that may not be the most energetic yet yield superior reconstruction.

The $\ell_1$-penalty only variant strongly favors the most energetic modes and rarely selects lower-energy ones. In the rare cases when it does, it attains reconstruction errors comparable to SparseModesNet, as seen in the low-error outliers in Figure SM2, but this behavior is inconsistent and infrequent. The LassoNet-style variant, which enforces the hierarchical sparsity constraint but feeds the full candidate set to the NN component, also biases toward higher-energy modes; this likely drives its higher reconstruction errors relative to SparseModesNet, though it still substantially outperforms the $\ell_1$-penalty only approach. Overall, SparseModesNet consistently identifies relevant modes that yield superior reconstruction compared to the alternatives.

Figure SM3 illustrates the mode elimination process of a single training instance over the entire regularization path, including intermediate epochs for each $\lambda$. SparseModesNet eliminates modes monotonically, whereas the other two variants spuriously reactivate previously eliminated modes—consistent with our monotonic mode elimination results in Subsection SM1.5. We also observe faster sparsification in SparseModesNet than in the LassoNet-style variant, matching the insight from Remark SM1.9 that the network gradient can accelerate the shrinkage of $\omega_j$ as it becomes less informative.

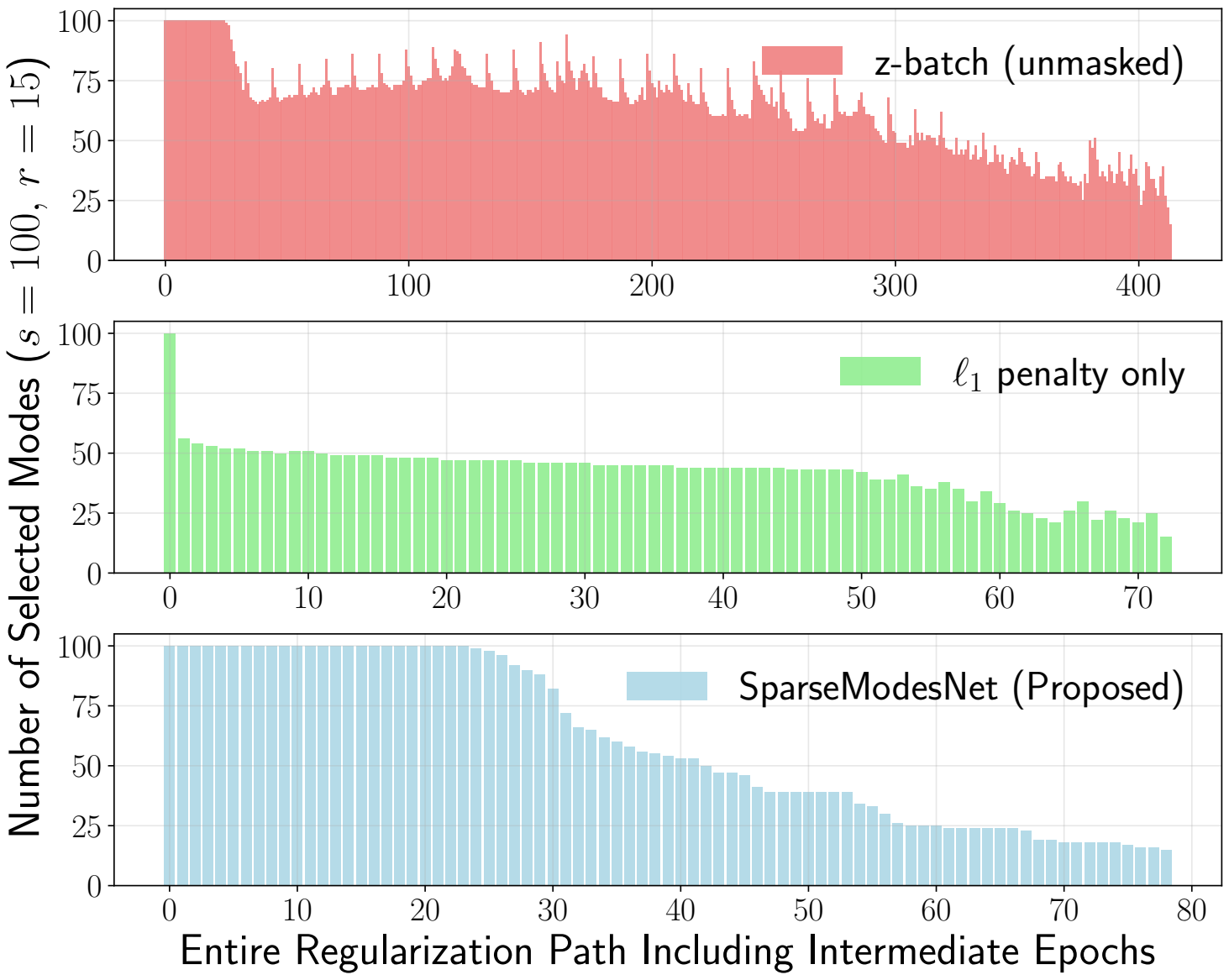


Fig. SM3. *Comparison of mode elimination for a single training instance throughout the entire regularization path including intermediate epochs for each $\lambda$ value. Our proposed SparseModesNet shows monotonic mode elimination while the other shows spurious reactivation of previously eliminated modes.*

**SM3. Hybrid Neural Network Architecture.** This section details the NN architecture used in SparseModesNet, consisting of a Deep Polynomial NNs known as Π-Nets [SM2], and their combination with a linear layer to form the hybrid architec-

ture used in our experiments.

**SM3.1. Deep Polynomial Neural Networks.** Deep Polynomial NNs, or Π-Nets [SM2], are function approximators whose output is a high-order polynomial of the input, in contrast to standard Deep Convolutional NNs (DCNNs), which compose linear operators with nonlinear activations. While any smooth function admits a polynomial approximation, a direct implementation suffers a combinatorial explosion in parameters, growing as $\mathcal{O}(o^N)$ for an $o$-dimensional input and an $N^{th}$-order polynomial. To overcome this, Π-Nets cast parameter estimation as a coupled tensor factorization: decomposing the high-order parameter tensors and enforcing factor sharing sharply reduces the number of learnable parameters and enables an efficient implementation through hierarchical neural network architectures.

A vector-valued function $g(\mathbf{y}) : \mathbb{R}^o \to \mathbb{R}^p$ (with $o$ an arbitrary input dimension) is modeled as an $N^{th}$-order multivariate polynomial of $\mathbf{y}$, whose vectorized output $\mathbf{v}$ is

$$\mathbf{v} = g(\mathbf{y}) = \boldsymbol{\beta}_\pi + \sum_{n=1}^{N} \left( \mathcal{W}_\pi^{(n)} \prod_{j=2}^{n+1} \times_j \mathbf{y} \right) \tag{SM3.1}$$

where $\boldsymbol{\beta}_\pi \in \mathbb{R}^p$ is a bias term and $\{\mathcal{W}_\pi^{(n)} \in \mathbb{R}^{p\times o\times\cdots\times o}\}_{n=1}^{N}$ are the learnable parameter tensors for each order $n$. We distinguish the first-layer weight $\mathcal{W}^{(1)}$ of our SparseModesNet architecture from the Π-Net layer tensors $\mathcal{W}_\pi^{(n)}$, $n = 1, \dots, N$.

The core innovation of Π-Nets is structuring these tensors through joint decompositions: all $\{\mathcal{W}_\pi^{(n)}\}_{n=1}^{N}$ are jointly factorized via a coupled CANDECOMP/PARAFAC (CP) decomposition [SM1, SM3], with factors for lower-order terms shared. Denote the matrix form of the CP decomposition of $\mathcal{W}_\pi^{(n)}$ by $\mathbf{W}_\pi^{(n)} \in \mathbb{R}^{p\times o^n}$, and the factor matrices by $\mathbf{C} \in \mathbb{R}^{p\times k}$ and $\mathbf{P}_n \in \mathbb{R}^{o\times k}$, $n = 1, \dots, N$, where $k$ is the CP rank. For example, a 3rd-order polynomial is

$$\begin{aligned} g(\mathbf{y}) &= \boldsymbol{\beta}_\pi + \mathbf{C}\mathbf{P}_1^\top \mathbf{y} + \mathbf{C}(\mathbf{P}_3 * \mathbf{P}_1)^\top (\mathbf{y} * \mathbf{y}) + \mathbf{C}(\mathbf{P}_2 * \mathbf{P}_1)^\top (\mathbf{y} * \mathbf{y}) \\ &\quad + \mathbf{C}(\mathbf{P}_3 * \mathbf{P}_2 * \mathbf{P}_1)^\top (\mathbf{y} * \mathbf{y} * \mathbf{y}) \end{aligned}$$

where $*$ denotes the Khatri-Rao or column-wise Kronecker product. This polynomial can then be generalized by the following recursive relation:

$$\mathbf{v}_n = (\mathbf{P}_n^\top \mathbf{y}) \odot \mathbf{v}_{n-1} + \mathbf{v}_{n-1}, \quad n = 2, \dots, N \tag{SM3.2}$$

with $\mathbf{v}_1 = \mathbf{C}(\mathbf{P}_1^\top \mathbf{y})$ and $\mathbf{v} = \mathbf{C}\mathbf{v}_N + \boldsymbol{\beta}_\pi$, where $\odot$ is the element-wise product and $\boldsymbol{\beta}_\pi$ an optional bias. This recursive form shows how a standard linear or convolutional layer becomes a polynomial one; see [SM2] for details. In practice, one defines the polynomial order $N$ and CP rank $k$, takes $\mathbf{C}$, $\mathbf{P}_n$, and $\boldsymbol{\beta}_\pi$ as learnable parameters, and computes the forward model via the recursion (SM3.2), enabling efficient training and inference.

**SM3.2. Hybrid Architecture: Fully Connected Layer + Π-Net.** Although the Π-Net is itself highly expressive through polynomial expansions, a hybrid approach can be beneficial when reconstructing data with both linear and nonlinear characteristics from a limited number of modes. We therefore combine a standard fully connected layer with a Π-Net, leveraging the strengths of both standard NN transformations and polynomial approximations. Figure SM4 shows the schematic.

The portion following the first layer realizes the recursion (SM3.2) for a 3[rd]-order polynomial, which takes the first-layer output as input in the form:

$$\mathbf{y} = \mathcal{W}^{(1)}(\boldsymbol{\omega} \odot \hat{\mathbf{z}}) + \boldsymbol{\beta}^{(1)} \in \mathbb{R}^{o}, \tag{SM3.3}$$

with an identity activation, first-layer weights $\mathcal{W}^{(1)} \in \mathbb{R}^{o\times s}$, and bias $\boldsymbol{\beta}^{(1)} \in \mathbb{R}^{o}$. This affine layer maps the input features/modes into a representation better suited for the Π-Net to capture polynomial nonlinearities, and also serves as the gate layer where the hierarchical sparsity constraint is applied during training.

Hybrid Architecture ($\Pi_3$-Net): Fully Connected Layer + 3$^{\text{rd}}$-Order Π-Net

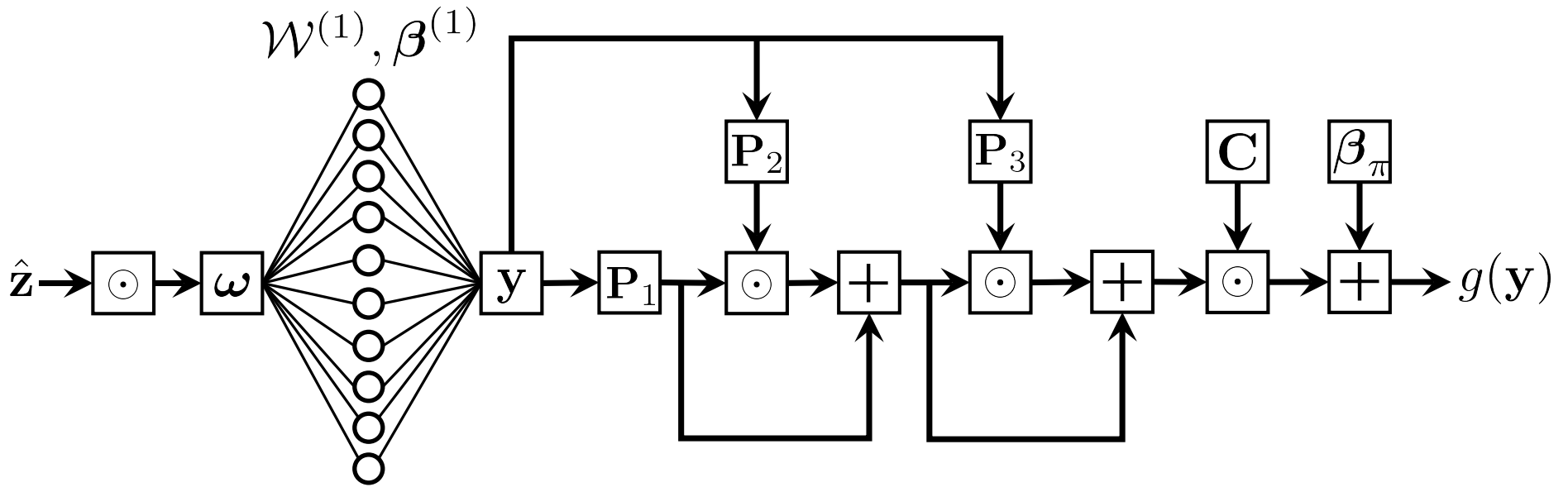


Fig. SM4. *Schematic of the hybrid architecture combining a fully connected layer with a* Π*-Net. The input first passes through a fully connected layer, which is then fed into the* Π*-Net to produce the final output. This architecture leverages the strengths of both linear transformations and polynomial approximations for enhanced expressivity.*

This choice is necessary because applying the hierarchical sparsity constraint directly to the Π-Net CP factors would be non-trivial and complicate training. Placing it on the fully connected layer lets us control mode selection while retaining the polynomial expressivity of the Π-Net.

We note that this shallow hybrid—a single fully connected layer with no nonlinear activation followed by a Π-Net—performed best in our experiments, outperforming deeper variants with additional fully connected layers before or after the Π-Net. This suggests that complicated architectures are unnecessary for reconstructing data from a limited number of modes, as the Π-Net already provides sufficient expressivity.

## SM4. Additional Numerical Results.

**SM4.1. Wave profile reconstructions of the linear transport equation.** For the linear transport equation example, we also conduct comparisons of the reconstructed wave profiles in Figure SM5. This analysis provides qualitative validation of the reconstruction accuracy at three different time instances. At all times, the POD reconstruction exhibits significant amplitude and phase errors, while both the Greedy Quadratic Manifold and SparseModesNet ($\Pi_3$-Net) maintain high fidelity.

**SM4.2. Sparse mode selection process for the linear transport equation.** Figure SM6 shows mode selection weight $\omega$ evolution during training for $\Pi_2$-Net. This plot is essential as it demonstrates the automated mode selection process, in which SparseModesNet identifies 15 important modes (solid lines converging to $10^{-2}$–$10^{-4}$) while pruning others toward zero (faded lines decaying to $10^{-6}$ or below). Convergence to $r = 15$ occurs within 6–8 $\lambda$ iterations (100 epochs each), demonstrating efficient sparse selection. The $O(10^2)$ magnitude spread among selected $\omega$ values reveals an importance hierarchy. Furthermore, this plot demonstrates the monotonic

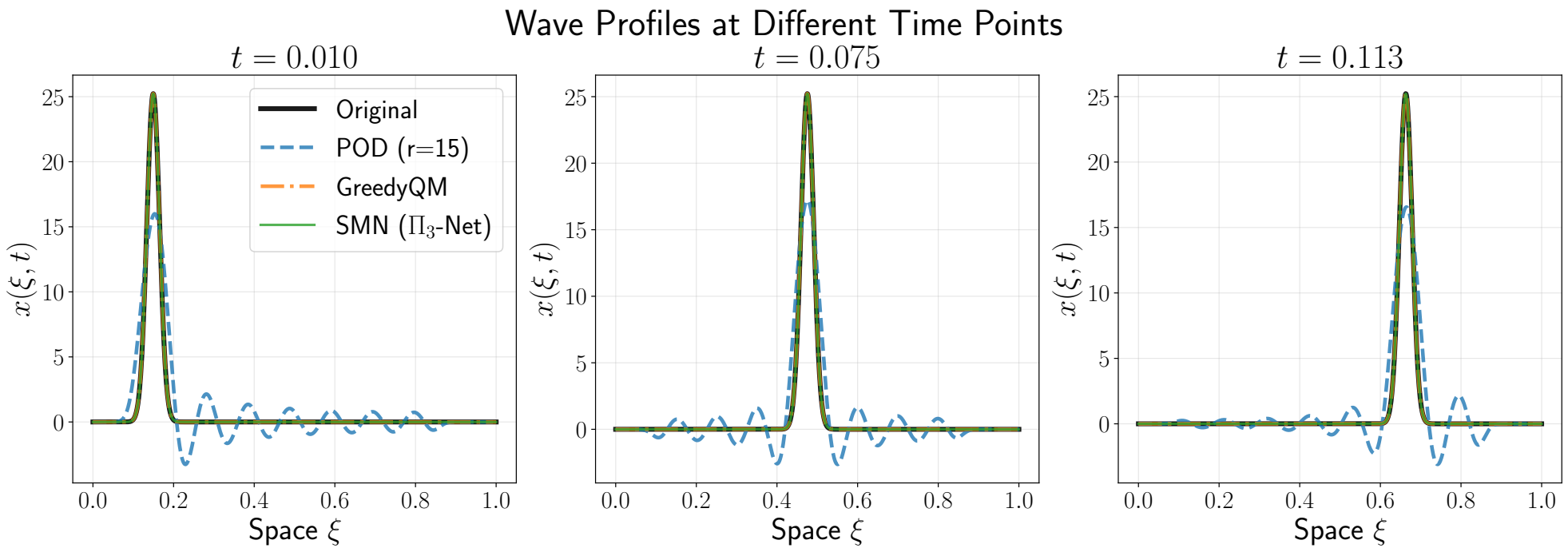


FIG. SM5. *Comparison of the wave profiles at different time instances for the exact solution and the reconstructed solutions for different decoder structures for the linear transport equation. "SMN" is a shorthand for SparseModesNet.*

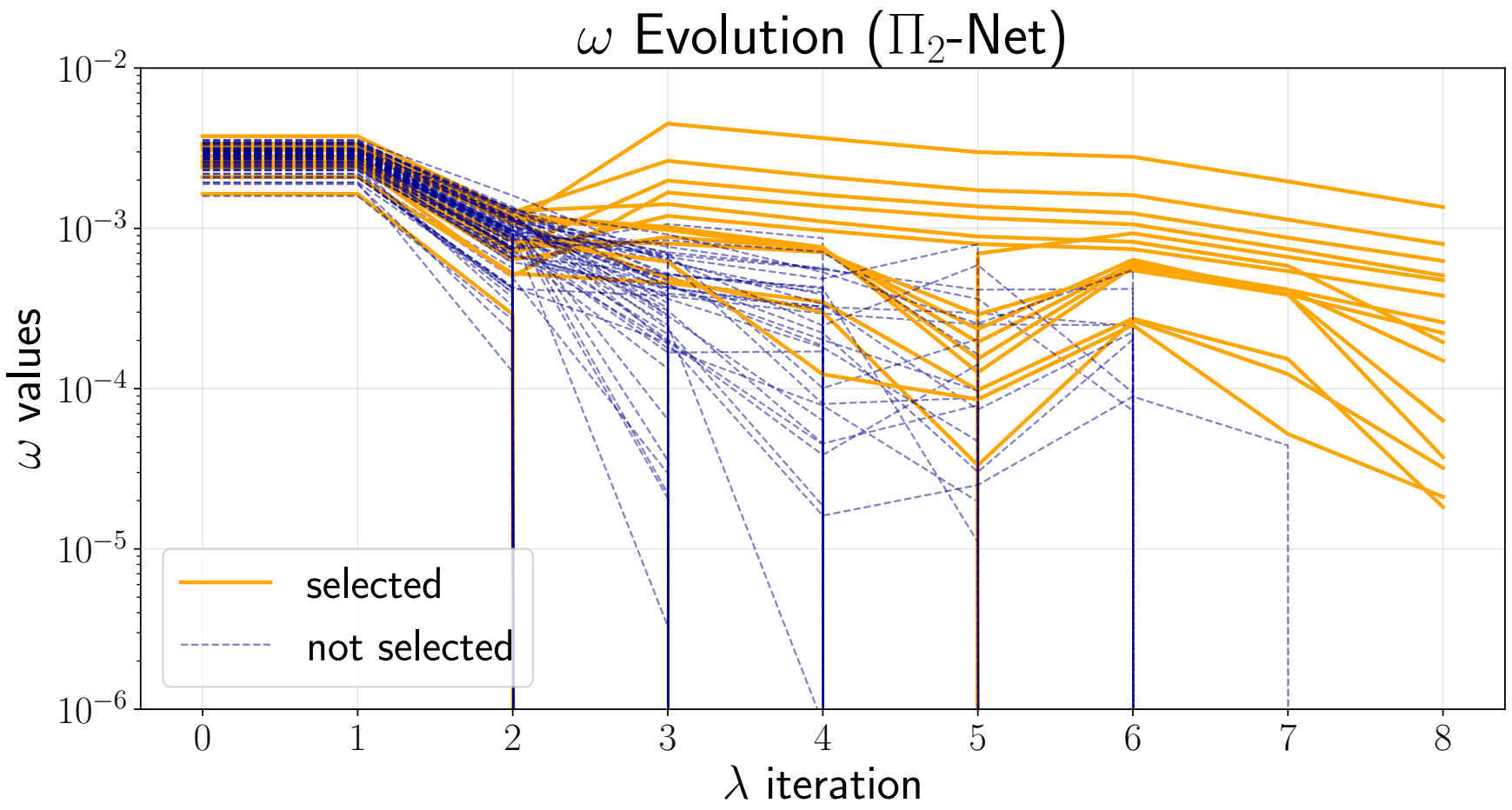


FIG. SM6. *Evolution of selected modes $\omega$ over sparsity parameter $\lambda$ updates for SparseModesNet with $\Pi_2$-Net for the linear transport, showcasing the automatic mode selection process.*

mode elimination as stated in Theorem SM1.20. The $\Pi_3$-Net exhibits qualitatively similar behavior and is omitted for brevity.

**SM4.3. Wave profile reconstructions of the KSE.** Similar to the linear transport equation, we compare wave profiles in Figure SM7, which demonstrate the reconstruction quality at three time instances spanning the turbulent evolution. Across all times, the POD reconstruction fails to capture the complex turbulent structures, exhibiting significant amplitude errors similar to the previous example. The Greedy Quadratic Manifold and SparseModesNet ($\Pi_3$-Net) maintain high fidelity to the exact solution.

**SM4.4. Sparse mode selection process for the KSE.** Figure SM8 shows mode selection weight evolution during training for $\Pi_3$-Net. Compared to linear transport, KSE requires longer training, with modes remaining constant at 16 for an extended period rather than 6–8 iterations. After 50 iterations, the mode with lowest weight in $\boldsymbol{\omega}$ was automatically removed to select the best 15 modes. This automatic truncation based on the weight magnitudes demonstrate the mode selection

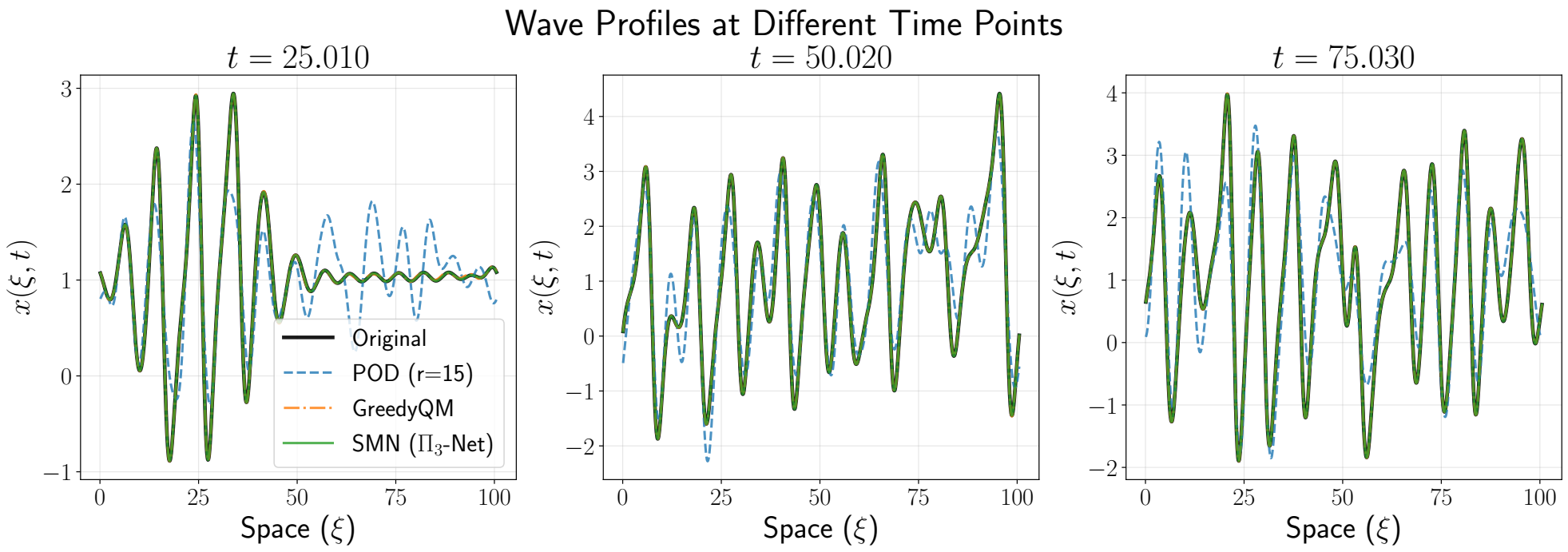


Fig. SM7. *Comparison of the wave profiles at different time instances for the exact solution and the reconstructed solutions for different decoder structures for the KSE. "SMN" is a shorthand for SparseModesNet.*

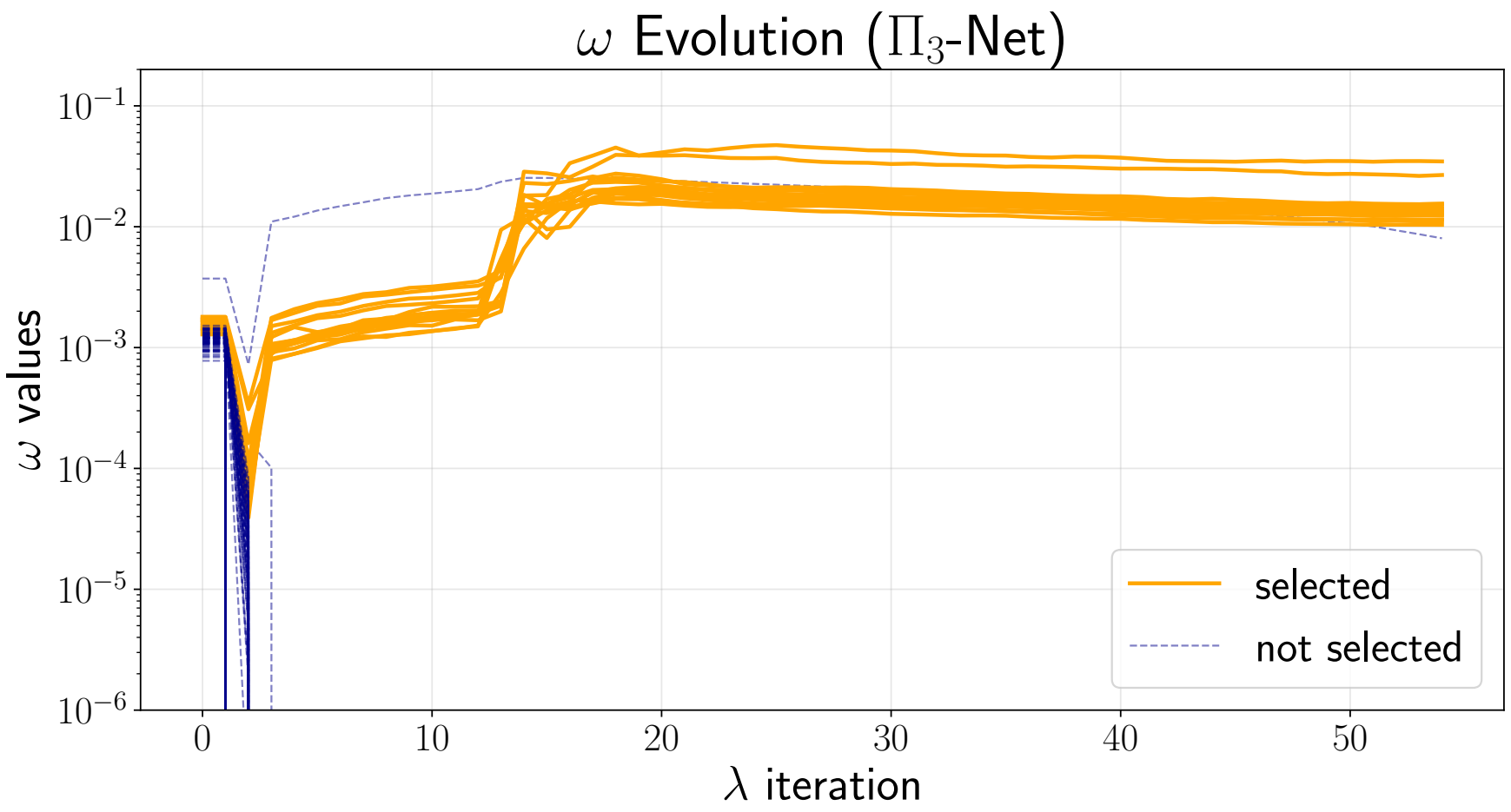


Fig. SM8. *Evolution of selected modes $\omega$ over sparsity parameter $\lambda$ updates for SparseModesNet $\Pi_3$-Net for KSE, showcasing the automatic mode selection with the continuation strategy.*

automation in SparseModesNet, which is especially beneficial when the target reduced rank-$r$ to achieve a specific target reconstruction error is unknown a priori. Selected modes (solid lines) converge to $10^{-2}$ to $10^{-1}$, while non-selected modes are pruned to $10^{-6}$ or below.

**SM4.5. Reconstructed Velocities for the Turbulent Channel Flow.** Figures SM9 and SM10 provide qualitative validation through spanwise-sliced flow field visualizations, comparing reconstructions against the Greedy Quadratic and Cubic Manifold methods, respectively. The top four panels show the original data and reconstructions from different methods, revealing that all nonlinear approaches capture the large-scale turbulent structures. However, the error fields in the bottom three panels expose differences in reconstruction fidelity. When compared to the Greedy Quadratic Manifold (Figure SM9), the error field shows absolute errors spanning $\pm 0.06$ throughout the domain, with particularly pronounced errors in high-shear regions near the bottom wall. Both the SparseModesNet with leading-$r$ modes and SparseModesNet with intelligent mode selection shows reduced errors. Similarly, Figure SM10 demon-

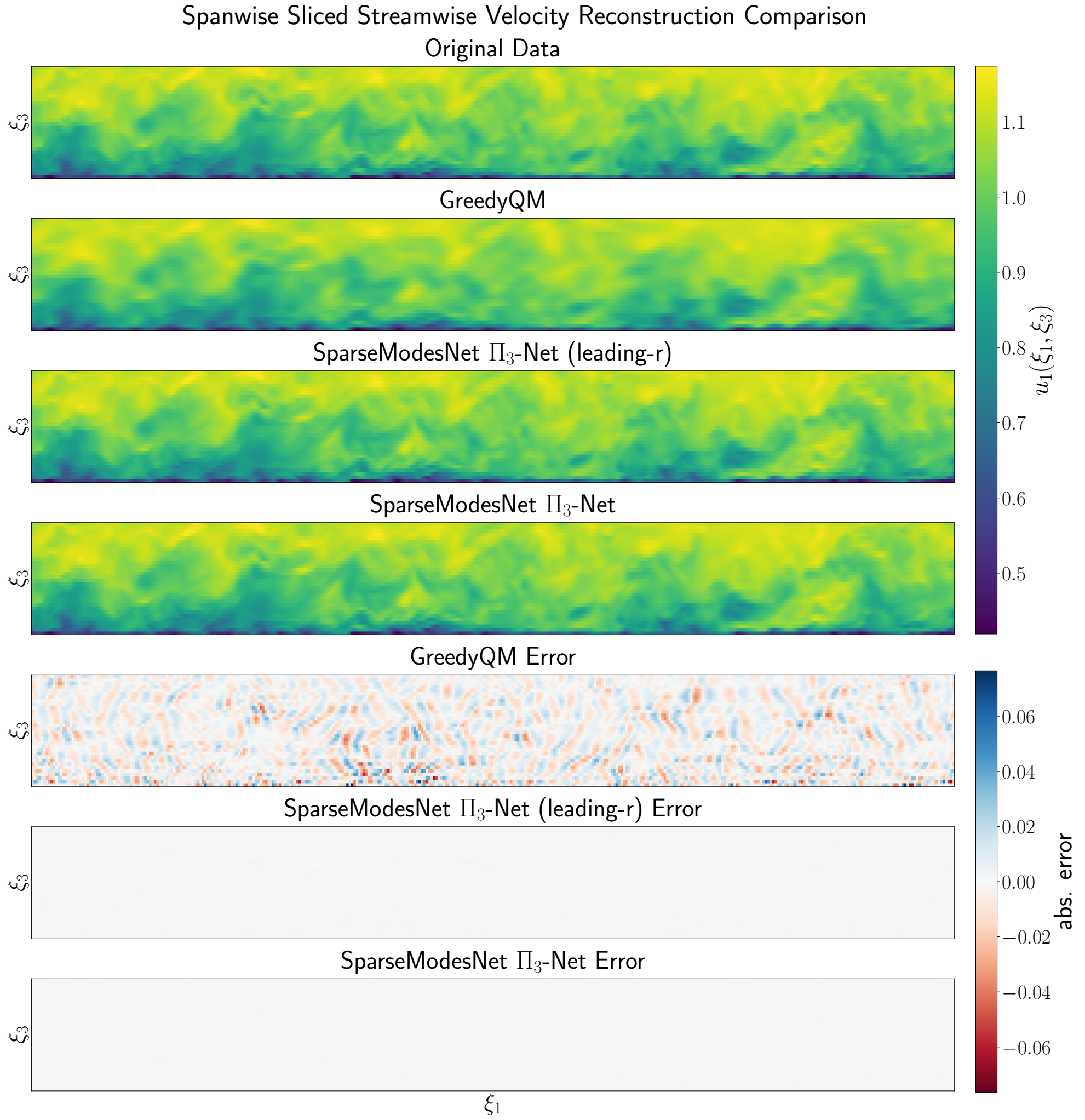


Fig. SM9. *Spanwise sliced streamwise velocity field reconstructions at a certain time $t$ using different decoder structures compared to the Greedy quadratic manifold (GreedyQM) method.*

strates that even the Greedy Cubic Manifold, which represents the state-of-the-art, produces visible errors with magnitudes up to $\pm 0.010$, whereas both SparseModesNet variants achieve reconstruction errors that are visually indistinguishable from zero. These qualitative comparisons demonstrate that SparseModesNet not only achieves better quantitative metrics but also produces accurate flow field reconstructions that preserve fine-scale turbulent structures throughout the domain.

In Figures SM11 and SM12 we see improved wall-normal velocity reconstructions compared to the streamwise component. Figure SM11 shows the Greedy Quadratic Manifold has errors up to $\pm 0.08$ and fails to capture fine-scale turbulent eddies and coherent structures. Figure SM12 shows the Greedy Cubic Manifold, while superior to its quadratic counterpart, still produces visible errors compared to SparseModesNet. SparseModesNet successfully preserves the complex multi-scale turbulent structures of wall-bounded turbulence at $Re_\tau = 5200$. This is particularly noteworthy for the

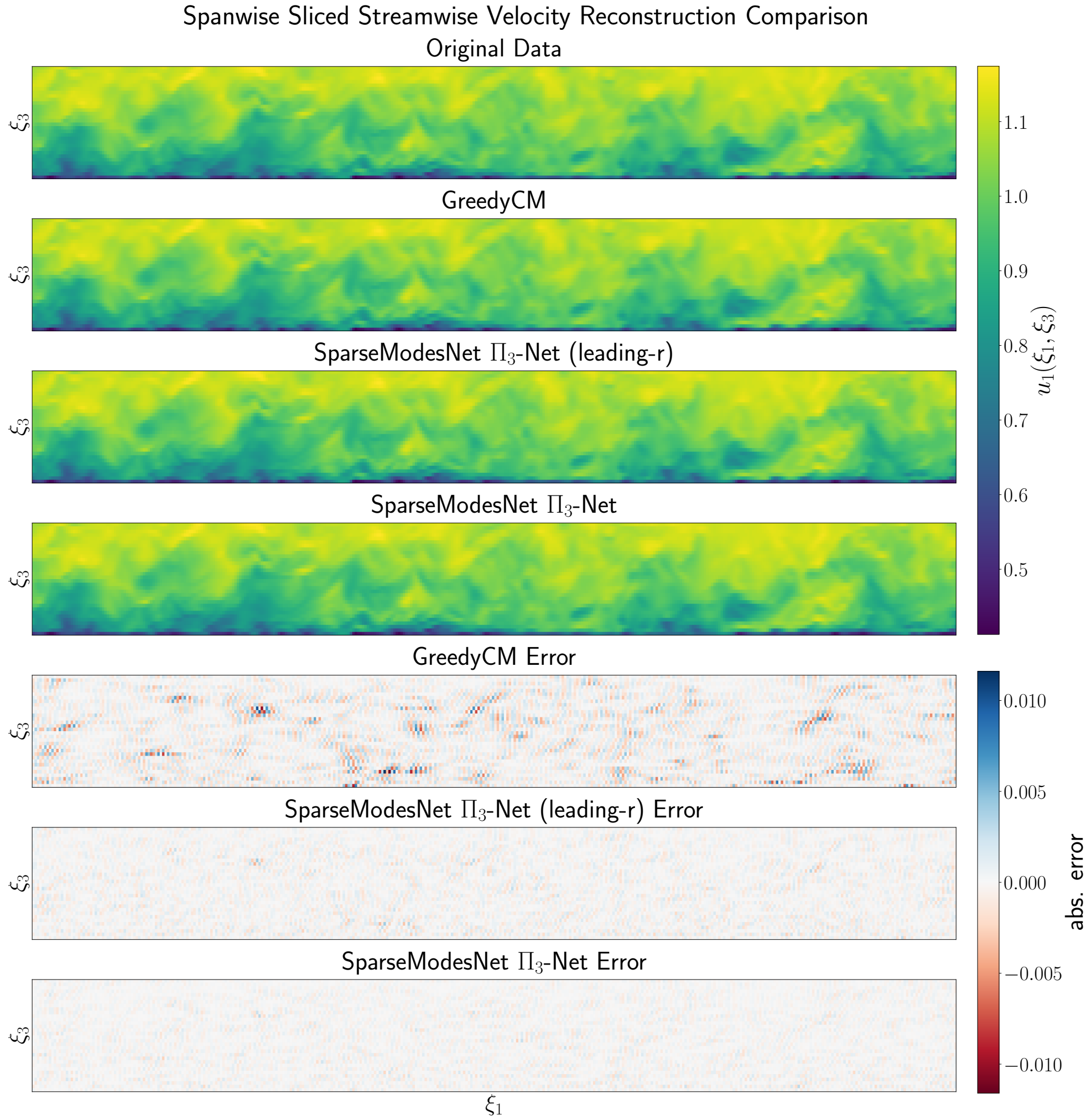


FIG. SM10. *Spanwise sliced streamwise velocity field reconstructions at a certain time $t$ using different decoder structures compared to the Greedy cubic manifold (GreedyCM) method.*

wall-normal component, which is more advection-dominated and exhibits stronger fluctuations than the streamwise component.

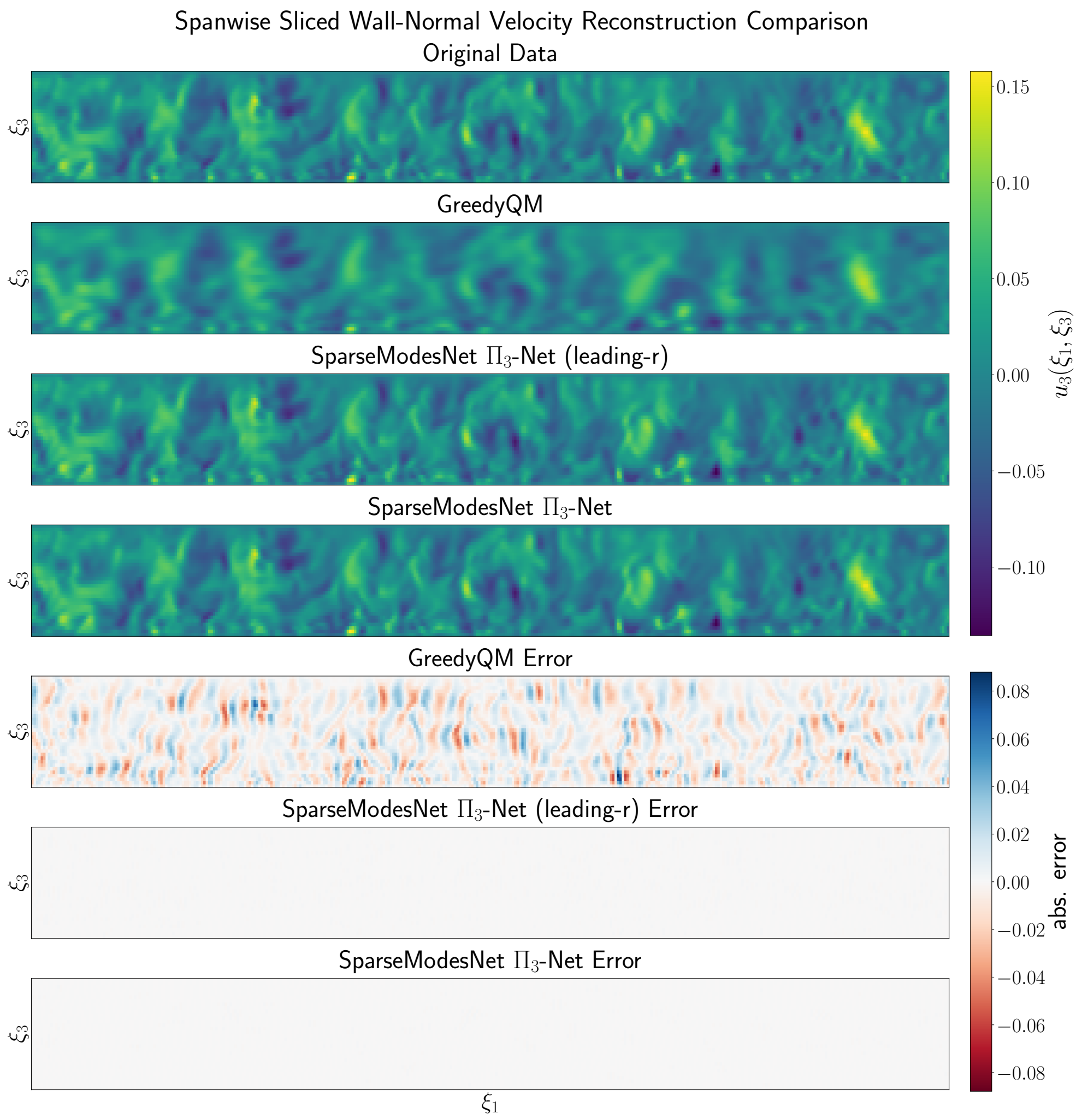


Fig. SM11. *Spanwise sliced wall-normal velocity field reconstructions at a certain time $t$ using different decoder structures compared to the Greedy quadratic manifold (GreedyQM) method.*

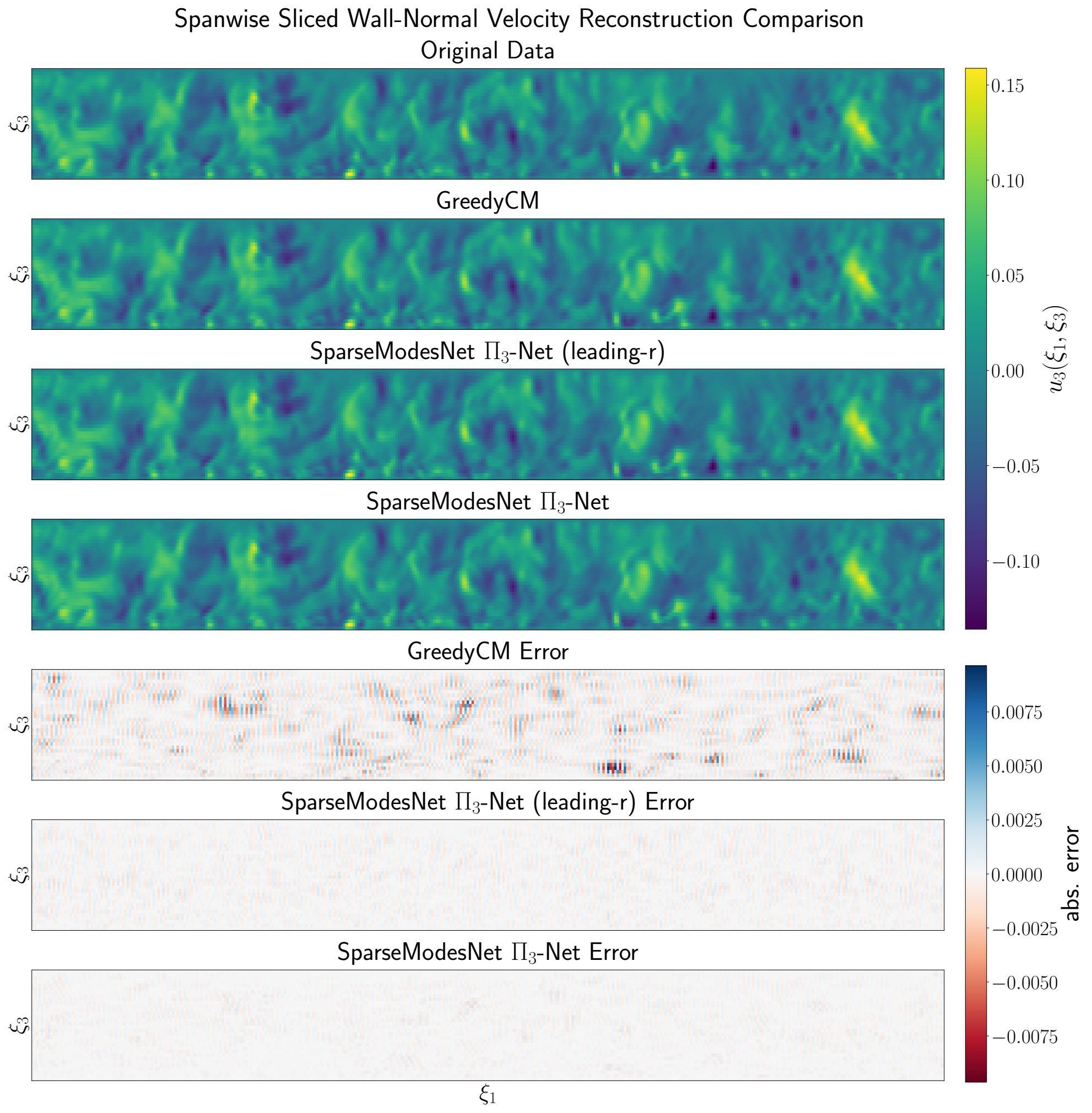


Fig. SM12. *Spanwise sliced wall-normal velocity field reconstructions at a certain time $t$ using different decoder structures compared to the Greedy cubic manifold (GreedyCM) method.*


\fi
%==================================%

\end{document}